\documentclass[usenatbib]{mn2e}
\pdfoutput=1

\pdfoutput=1
\usepackage{epstopdf}
\usepackage{float}
\usepackage{graphicx}
\usepackage{natbibmnfix, times}
 \usepackage{subfigure}
\usepackage{color}
\usepackage[T1]{fontenc} 
\usepackage{aecompl}

\usepackage{amssymb,amsmath}
\usepackage{bm}

\newcommand{\apj}{ApJ}
\newcommand{\apjl}{ApJ}
\newcommand{\apjs}{ApJS}
\newcommand{\aap}{A \& A}

\newcommand{\aj}{AJ}
\newcommand{\mnras}{MNRAS}

\newcommand{\na}{New Astronomy Review}

\newcommand{\apss}{Ap\&SS}

\newcommand{\icarus}{Icarus}

\def\red#1 {\textcolor{red}{#1}\ }   
\def\blue#1 {\textcolor{blue}{#1}\ }   

\title[Planet-disc interaction on a moving mesh]{Planet-disc interaction on a freely moving mesh}
 \author[Mu\~noz et al.]{D. J. Mu\~noz,$^{1,2}$\thanks{E-mail:dmunoz@cfa.harvard.edu} 
 K. Kratter,$^{3,6}$\thanks{Hubble Fellow.}
 V. Springel$^{4,5}$
 and L. Hernquist$^{1}$\\
$^{1}$ Harvard Smithsonian Center for Astrophysics, 60 Garden Street, Cambridge, MA 02138\\
$^{2}$ Center for Space Research, Department of Astronomy, Cornell University, Ithaca, NY 14853, USA\\
$^{3}$ JILA, University of Colorado, 440 UCB, Boulder CO 80309-0440, USA \\
$^{4}$ Heidelberg Institute for Theoretical Studies, Schloss-Wolfsbrunnenweg 35, 69118 Heidelberg, Germany \\
$^{5}$ Zentrum f\"{u}r Astronomie der Universit\"{a}t Heidelberg, ARI, M\"onchhofstr. 12-14, 69120 Heidelberg, Germany\\
$^{6}$ Steward Observatory, University of Arizona, 933 N Cherry Ave, Tucson, AZ, 85721, USA }

\begin{document}


\pagerange{\pageref{firstpage}--\pageref{lastpage}} \pubyear{2012}

\maketitle


\begin{abstract}
General-purpose, moving-mesh schemes for hydrodynamics have opened the
possibility of combining the accuracy of grid-based numerical methods
with the flexibility and automatic resolution adaptivity of
particle-based methods.  Due to their supersonic nature, Keplerian
accretion discs are in principle a very attractive system for applying
such freely moving mesh techniques.  However, the high degree of
symmetry of simple accretion disc models can be difficult to capture
accurately by these methods, due to the 
generation of geometric grid
noise and associated numerical diffusion, which is absent in polar
grids. 
To explore these and
other issues, in this work we study the idealized problem of
two-dimensional planet-disc interaction with the moving-mesh code
{\footnotesize AREPO}.  We explore the hydrodynamic evolution of discs
with planets through a series of numerical experiments that vary the
planet mass, the disc viscosity and the mesh resolution, and compare
the resulting surface density, vortensity field and tidal torque with
results from the literature. We find that the performance of the
moving-mesh code in this problem is in accordance with published
results, showing good consistency with grid codes written in polar
coordinates. We also conclude that grid noise and mesh distortions do
not introduce excessive numerical diffusion.  Finally, we show how the
moving-mesh approach can naturally increase resolution in regions of high density 
around planets and planetary wakes, while retaining the background flow at low
resolution. This provides an alternative to the difficult task of implementing
adaptive mesh refinement in conventional polar-coordinate codes.
\end{abstract}


\section{Introduction} 
Mesh-construction is a fundamental step in the process of solving sets
of partial differential equations numerically.  Although a mesh itself
is an ``extraneous" element to the underlying equations of
hydrodynamics (i.e., there is nothing in the theory that can tell us
the ``correct" way to discretize space), a given choice of mesh can
have a significant impact on the results of a numerical
simulation. This is of particular importance in simulations where the
hydrodynamical flows are under-resolved. In this case, the outcome on
large scales can be very sensitive to the numerical and physical
behavior at the resolution scale. The choice of the mesh can have as
much of an impact on the numerical result as other elements of the
computational methodology, such as the adequacy of the coordinate
frame used, the order-of-accuracy of the scheme, or even the
additional sub-resolution models implemented.  Given the degeneracy
between the mesh and other ``features" of a numerical code, the choice
of an optimal discretization approach (whether it is through a
grid-approach or a particle-based approach) tends to depend on the
problem being studied.

Recently, moving-mesh methods for computational hydrodynamics
(\citealp{spr10a,duf11}; but see also \citealp{bor87,tre88,duk89}), as
well as novel mesh-less approaches like that of \citet{mcn12a}, have
been shown to be an interesting and powerful tool for studying
high-mach-number, large-dynamical-range astrophysical flows.

High-mach-number flows are computationally challenging for several
reasons.  The most common complication is the so-called ``high-Mach
number problem" \citep{ryu93,bry95,fen04,tra04}. This problem
manifests itself when the kinetic energy density is much larger than
the internal/thermal energy density. Consequently, a small fractional
error in the velocity can translate into large fractional error in
temperature, eventually distorting the thermodynamic evolution of the
gas. Another problem is the strict limitation that the Courant
time-step condition imposes on high-velocity flows, which may extend
the computation time beyond practicality, with the additional peril of
excessive numerical noise accumulated over a large number of
integration time-steps.
 
Accretion discs are an example of high-Mach number flows that present
serious computational difficulties. In these systems, extremely small
Courant-limited time-steps are one of the greatest barriers to
high-resolution simulations. In addition, interesting phenomena in
these systems often evolve over many dynamical/orbital times, making
numerical studies extremely expensive.  The Mach number of a disc is
roughly the Keplerian speed divided by the local sound speed
$\mathcal{M}_\phi\sim v_K/c_s\sim 1/h$, where the aspect ratio $h$ of
a thin disc is $0.01-0.1$. Thus, Mach numbers can reach values of
10-100.

The importance and ubiquitousness of accretion discs in astrophysics
has propelled the development of several (magneto-) hydrodynamics
codes specifically written to handle global models of discs.  Global
disc simulations with grid codes over hundreds or thousands of orbits
became commonplace when the {\footnotesize FARGO} \citep[Fast
  Advection in Rotating Gaseous Objects;][]{mas00} algorithm was
introduced. In this scheme, the logic of the shearing-sheet
approximation is applied to global simulations; i.e., the Euler
equations are solved in a non-inertial rotating frame. To first order,
the disc is rotating at Keplerian speed, thus the equations can be
written in a local non-inertial frame rotating at rate
$\Omega_K=v_K/R$. As a result, the Courant criterion is thus based on
the deviation from the background velocity in this frame, and not on
the highly supersonic speed as measured in the inertial frame.
Besides the {\footnotesize FARGO} code, schemes like {\footnotesize
  RH2D} \citep{kle89}, {\footnotesize RODEO} \citep{paa06},
{\footnotesize RAPID} \citep{mud09}, and {\footnotesize DISCO}
\citep{duf13}, among others, have been tailored specifically for the
Euler equations in cylindrical/polar coordinates for supersonic
Keplerian flow.

Some of these cylindrical-coordinate codes (e.g., {\footnotesize
  RODEO}) were specifically designed to target the problem of
planet-disc or satellite-disc interaction, namely, the tidal
interaction between a disc and an embedded planet in Keplerian
rotation.  This gravitational coupling between a planet and a gas disc
has been studied in detail theoretically
(\citealp{gol79,gol80,lin79,lin86a,lin86b,war86,tak96,war97,tan02} and
more recently \citealp{raf12} and \citealp{pet12}) as well as
computationally
\citep[e.g.,][]{bat03,val06,dan08,dan10,don11a,don11b,duf13}.

Planet-disc interaction is particularly relevant for the early
dynamical evolution of protoplanets and proto-planetary systems, since
it is the mechanism behind disc-induced planet migration
\citep[e.g.][]{war86,war97}. The fully nonlinear integration of this
problem through numerical simulation is essential for understanding
the gravitational coupling between the planet and the surrounding
disc, especially if the planet is massive enough to open a gap.  Gaps
are not only important for the dynamics of planet migration, but they
may also provide morphological clues about the presence of young
planets around stars in observations of circumstellar discs. Evidence
from imaging and infrared spectroscopy showing central cavities
\citep{calv02,and11} and even narrow gaps \citep[e.g.,][]{esp07} in
gas discs opens the tantalizing possibility to infer the existence of
embedded planets in gas discs \citep[see, for example][for ongoing
  efforts in direct detection of these hypothetical planets]{krau12a}.

Although the process of gap-opening by a single planet on a circular
orbit embedded in a two-dimensional, isothermal disc is far too
idealized to be a realistic model for planet-disc interaction, it
remains an important basic test for hydrodynamic codes. The complexity
of the non-linear interaction between a planet and a disc does not
allow for traditional code {\it verification} (ithere is no known
exact solution), but the number of numerical experiments available in
the literature allow for the very important task of code {\it
  benchmarking}.  Most notably, \citet{val06} carried out an extensive
comparison of different grid-based and particle-based codes,
identifying similarities and disagreements between different numerical
schemes of widespread use in computational astrophysics. Along the
lines of this code comparison project, we apply a new numerical scheme
to the problem of planet-disc interaction in the gap-opening regime.

In this work, we use the code {\footnotesize AREPO} \citep{spr10a} for
the first time in numerical experiments of planet-disc
interaction. This problem has been studied in the low-planet mass case
by \citet{duf12} using the structured moving-mesh code {\footnotesize
  DISCO}. To our knowledge, an entirely freely moving mesh based on a
Voronoi tessellation has however not been applied successfully to this
problem before.

Moving-mesh methods share some of the spirit of the {\footnotesize
  FARGO} scheme, which is to solve the Euler equations in a frame
moving with the local flow in order to bypass the restrictions imposed
by the Courant criterion. However, one important difference is that
{\footnotesize AREPO} does not assume a priori a certain geometry and
magnitude of the underlying density and velocity fields. Instead, the
dynamics of the mesh automatically respond to whatever evolution
those fields follow.  The code enables cells to move self-consistently
and freely with the flow by moving a set of mesh-generating points
with the local gas velocity, with the mesh defined at every instance
as the Voronoi tessellation of the points.  This allows for a
quasi-Lagrangian interpretation of the tessellation: at high
resolution, each cell is a parcel of gas that follows the Lagrangian
trajectories of the flow with a minimal change of its gas content.

Recently, concerns about the use of a freely-moving mesh for the
planet-disc interaction problem have been raised \citep{duf12}.  However,
the observed limitations may well depend on details of the specific
hydrodynamical scheme explored and may not necessarily appear in other
tessellation codes. Indeed, in this work we find that {\footnotesize
  AREPO} is qualitatively competitive with polar grid codes in
simulations of planet-disc interaction, and that there is no reason to
consider the moving-mesh approach as fundamentally ill-suited to
accurately capture differentially rotating flows.

We begin in Section~\ref{sec:num_exp} by describing the numerical
models and explaining the primary features of the moving-mesh approach
applied to circumstellar disc dynamics. In Section~\ref{sec:results},
we present the main properties of our planet-disc interaction
simulations, describing the results for the evolution of the density
field, the vortensity and the net gravitational torque. We also
compare our results to the literature. In
Section~\ref{sec:discussion}, we discuss the implications of our
implementation with respect to concerns raised in the
literature about applying the moving-mesh approach to this
problem. Finally, in Section~\ref{sec:summary}, we summarize our
findings and present our conclusions.

\section{Numerical Experiments}\label{sec:num_exp}
 \begin{figure*}
\begin{tabular}{llll}
\includegraphics[width=4.8cm]{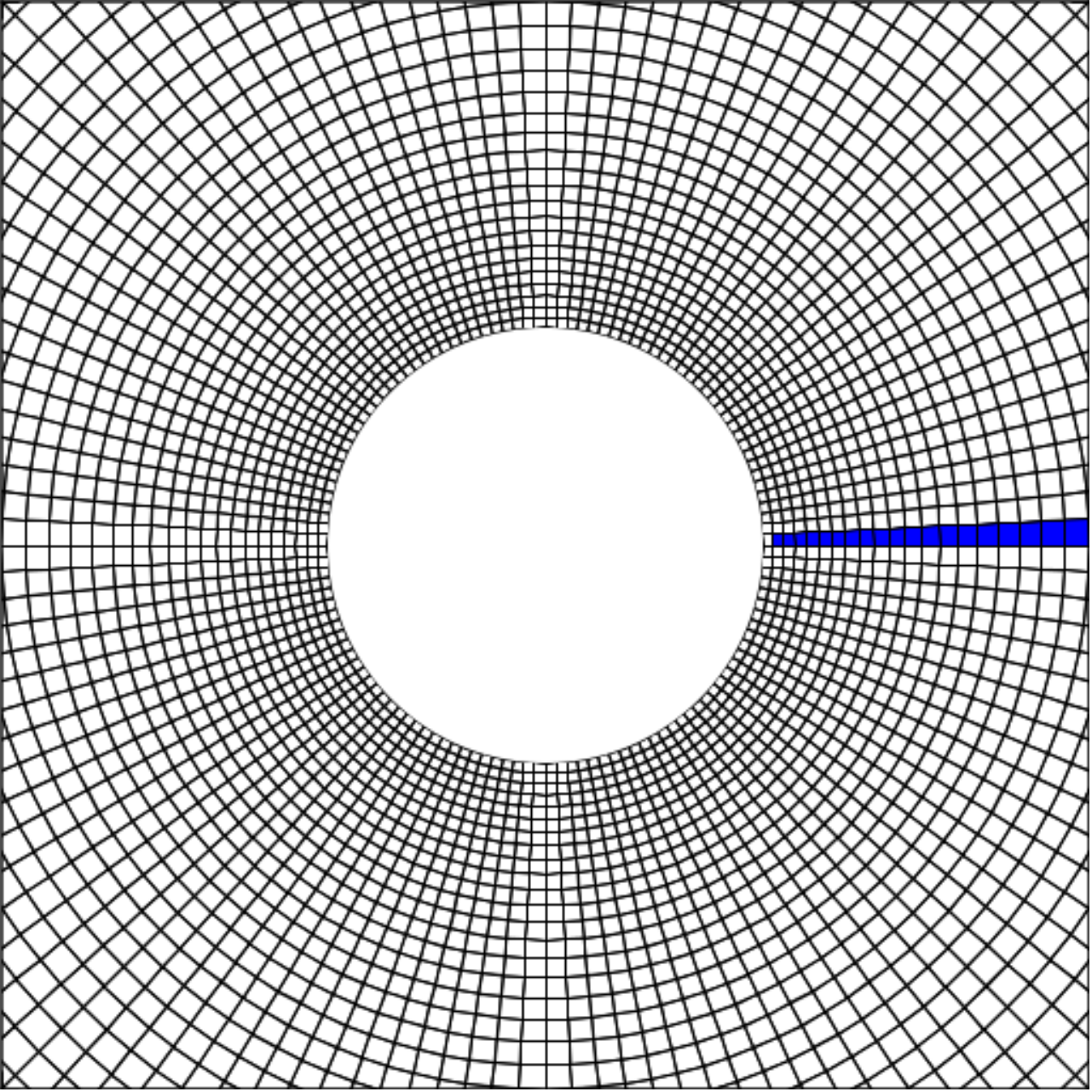}&
\includegraphics[width=4.8cm]{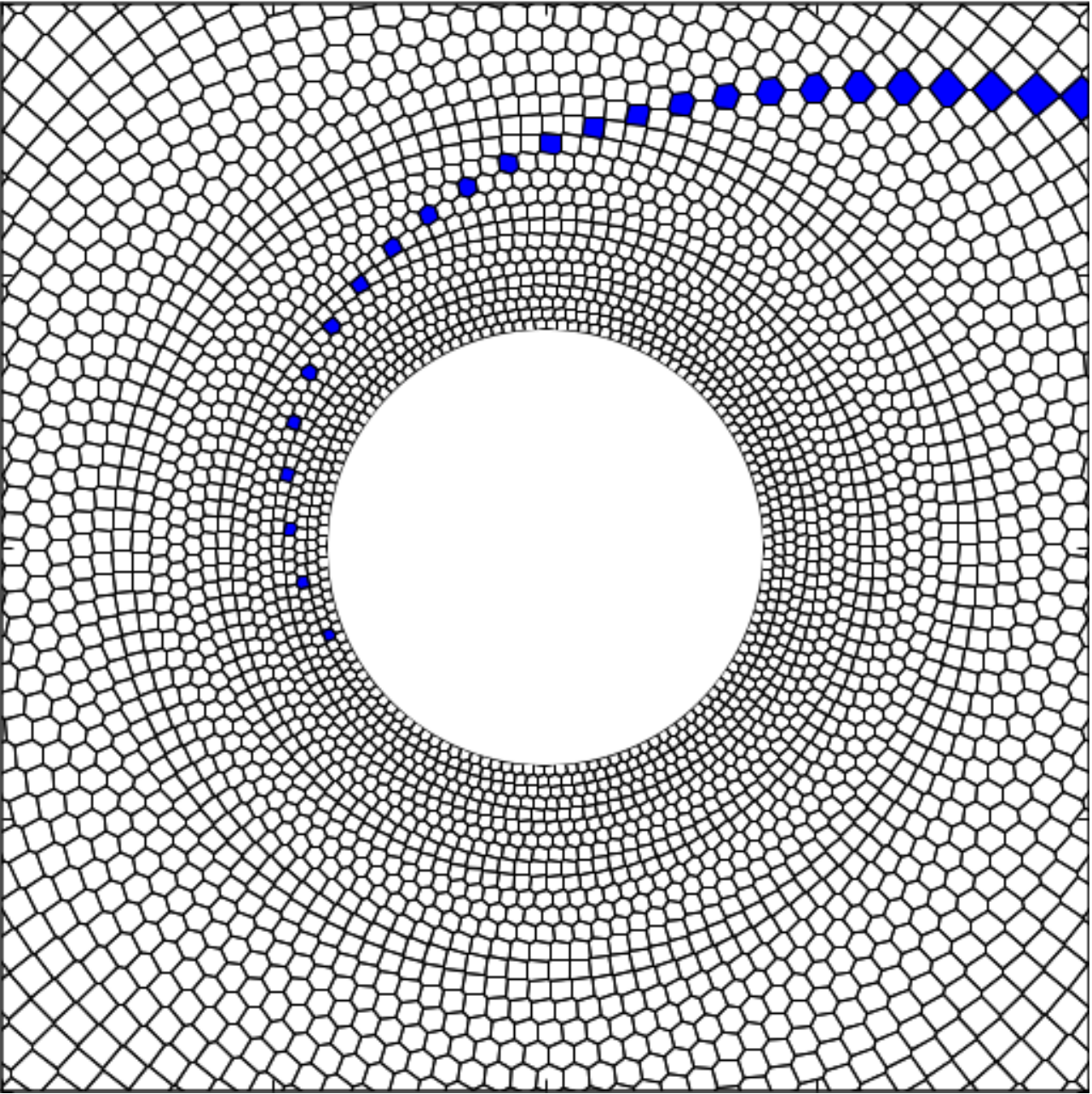}&
\includegraphics[width=4.8cm]{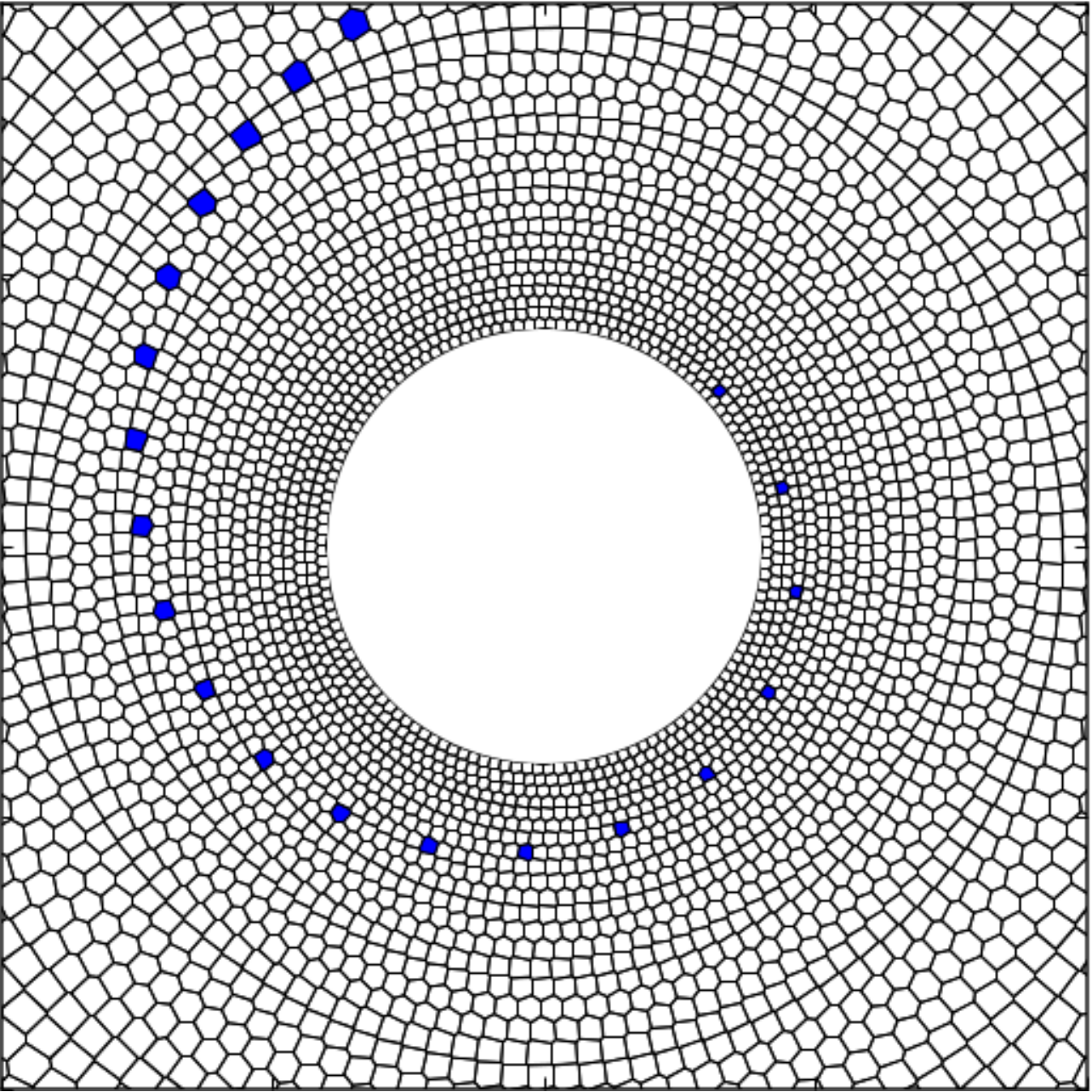}\\
\includegraphics[width=4.8cm]{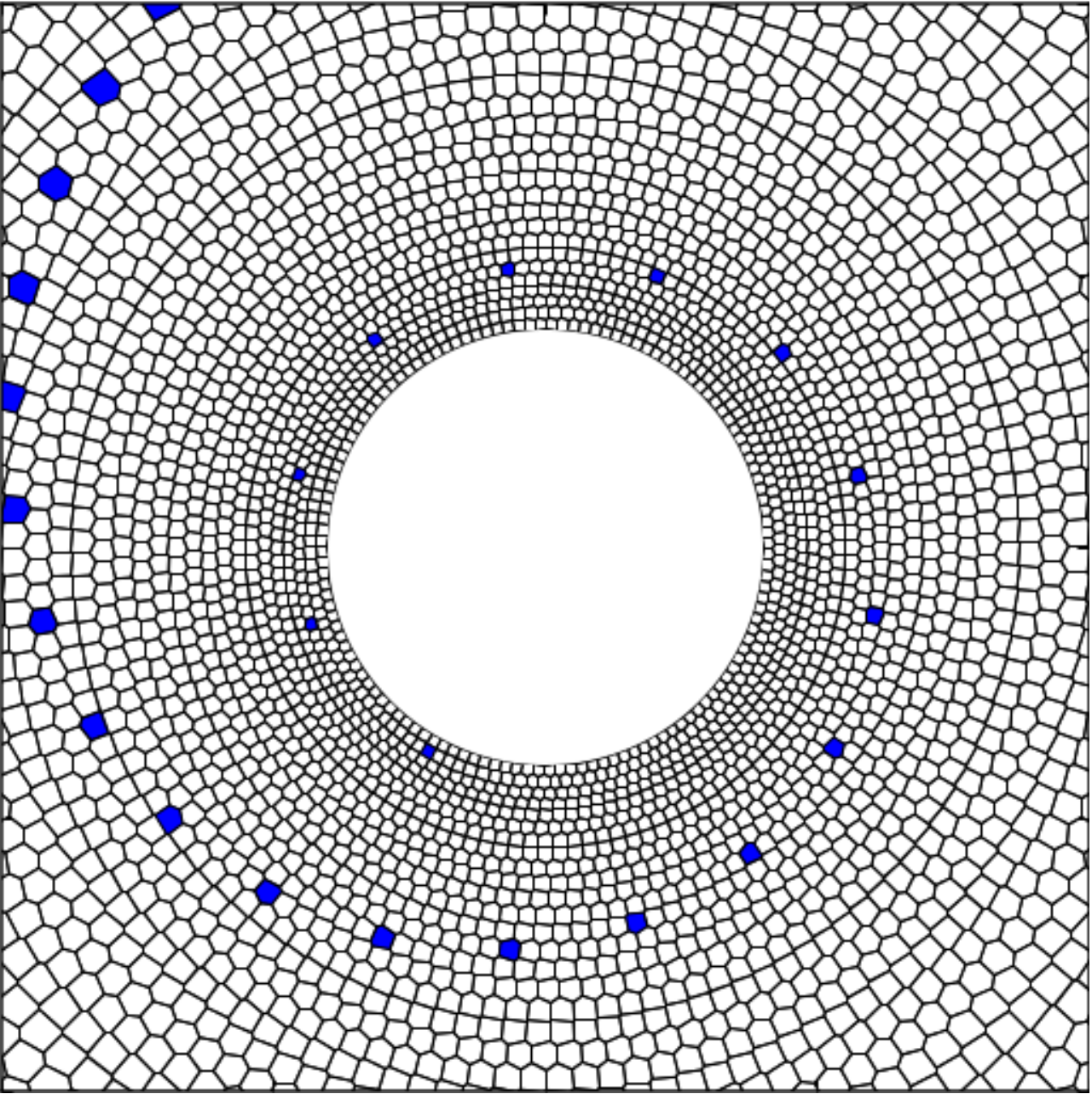}&
\includegraphics[width=4.8cm]{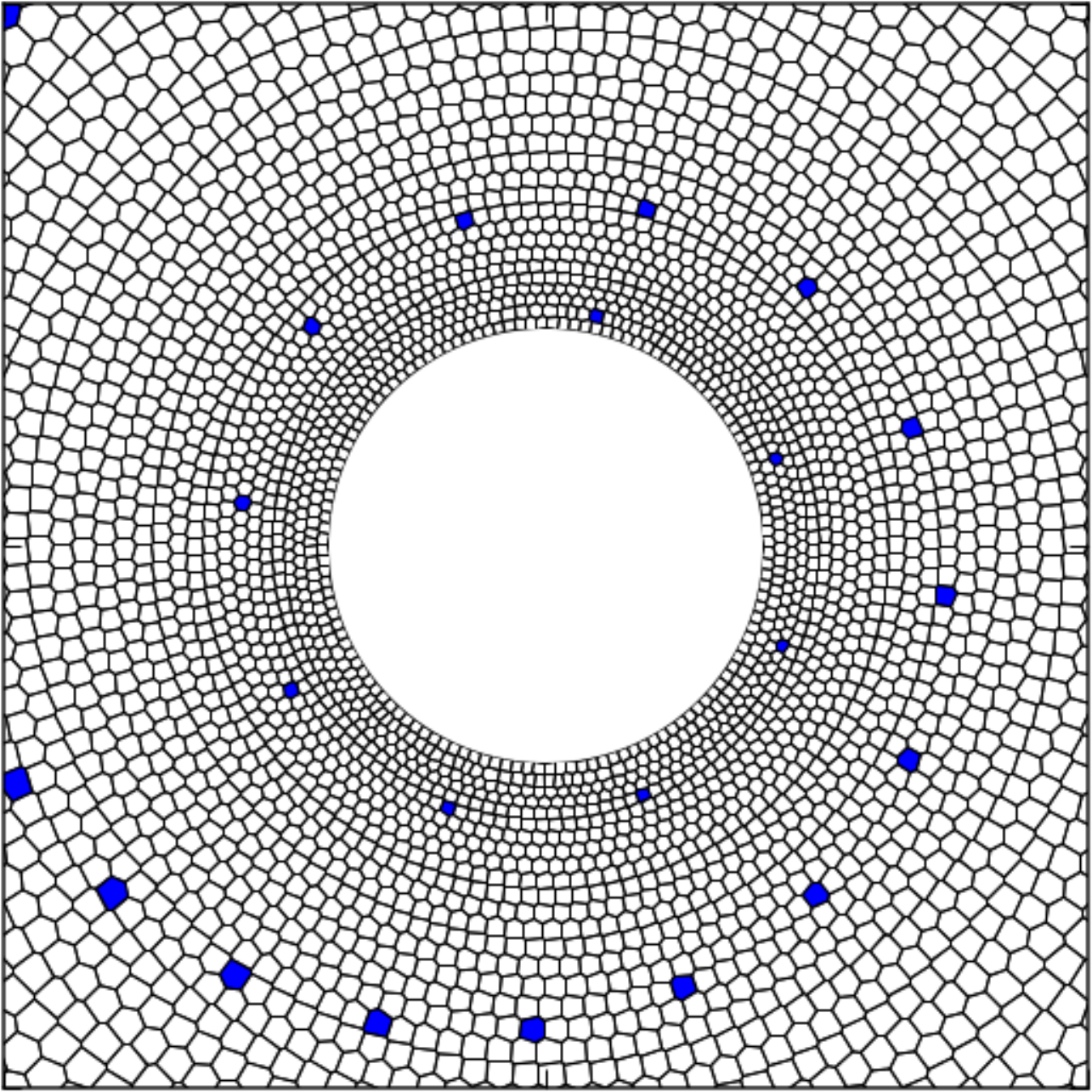}&
\includegraphics[width=4.8cm]{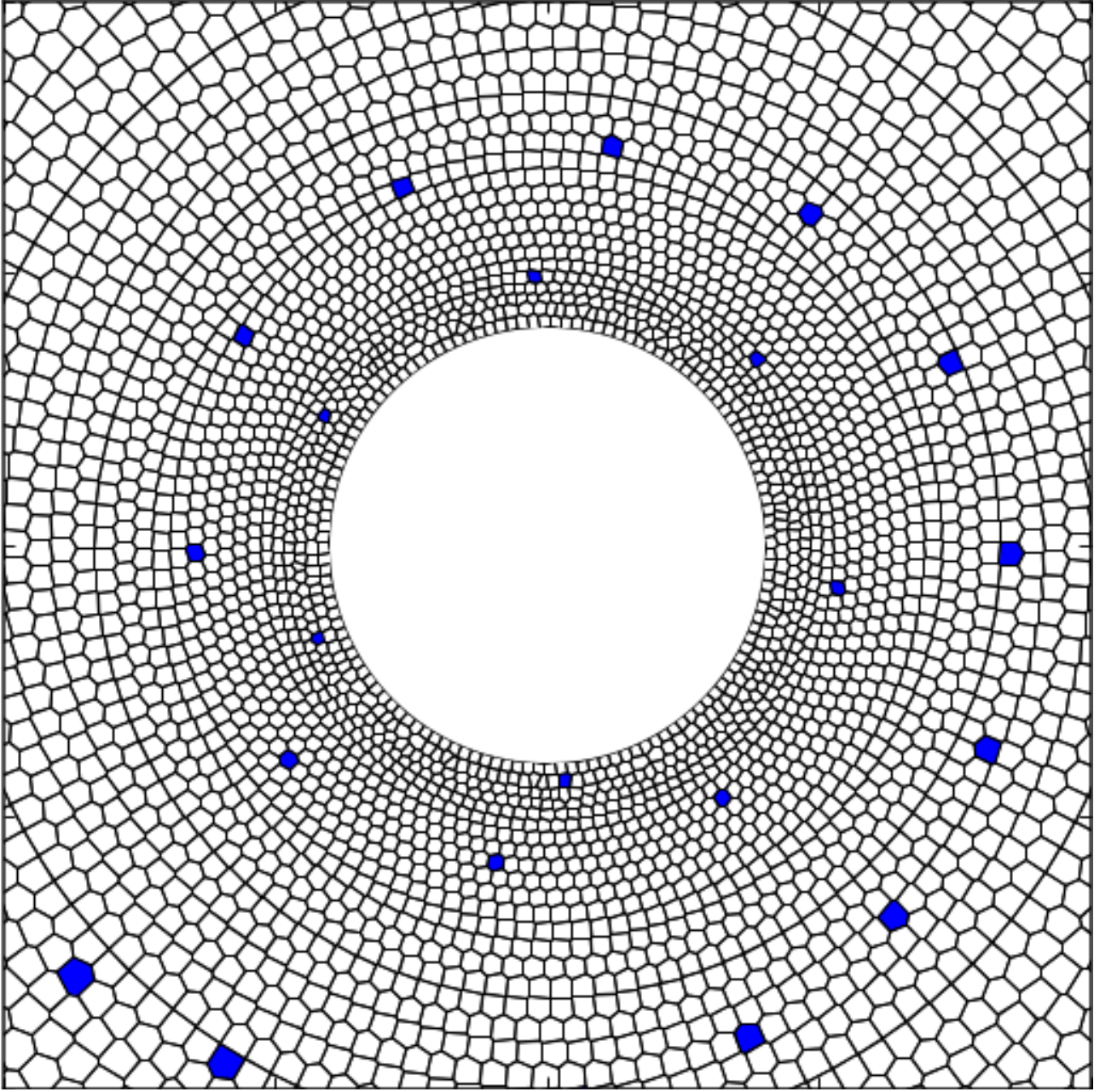}
\end{tabular}
\caption{Evolution of a Voronoi mesh under differential rotation
  supported by a Keplerian potential (time increases left-to-right and
  top-to-bottom).  The mesh-generating points are initially positioned
  in a polar distribution (logarithmic spacing in radius), which is
  roughly maintained.  The color-filled cells correspond to a set of
  cells tagged according to ID number at time $t=0$ (top left panel) and
  subsequently followed in time. The spatial distribution of the
  tagged cells highlights the ``quasi-Lagrangian" nature \citep{vog12}
  of the moving-mesh approach in the case of Keplerian shear.
\label{fig:sheared_mesh}}
\end{figure*}

\subsection{Problem Setup}

\paragraph*{Hydrodynamic Equations.} In this work we focus on the
solution of the Euler equations in two dimensions. In conservation-law
form, these equations are:
\begin{subequations}
\label{eq:euler}
\begin{align}
\label{eq:density}
\frac{\partial}{\partial t}\Sigma+\nabla\cdot(\Sigma\mathbf{v})&=0 ,\\
\label{eq:momentum}
\frac{\partial}{\partial t}\left(\Sigma \mathbf{v} \right)+ \nabla\cdot\left(\Sigma \mathbf{v}\otimes\mathbf{v} + P\mathbf{I} \right) &= 
-\Sigma\frac{\partial \Phi}{\partial \mathbf{r}}.
\end{align}
\end{subequations}
Note that the gravitational potential on the right hand side of
Equation~(\ref{eq:momentum}) is included as a source term. This is one
of the differences to some of the grid-based codes written in
cylindrical coordinates. In those coordinates, the Keplerian term can
be included directly into the conservation laws since the radial
gradient can be written as part of the divergence term in the
hyperbolic equations \citep[see][]{kle98,paa06}.
 
The fact that the gravity force is not included in a manifestly
conservative formulation into the Euler equations (however, see
\citealp{spr10a} and \citealp{jia13} for alternative approaches to
enforce a ``flux-based" description of gravity) implies that a
gravitational time-step criterion must be considered in addition to
the Courant-criterion time step. This acceleration must be taken into
account for cells to follow accurate orbital trajectories
(Figure~\ref{fig:sheared_mesh}). The bulk orbital motion of the cells
is carried out by a conventional kick-drift-kick (KDK) integrator
\citep[e.g.][]{sah92,pre99} which ``brackets" the hydrodynamical
finite-volume operation \citep{spr10a}.  Although the concept of
symplectic integration, as realized in the KDK leapfrog, offers little
advantage when applied to irreversible dynamics, it is nonetheless
true that, in practice, time-symmetric integrators show superior
performance compared with non-symmetric schemes (in terms of energy
and angular momentum conservation), even when integrating orbits of
SPH particles \citep{spr05c}.

For KDK integrators of particles in a Keplerian potential, about
50-100 time-steps should suffice to capture the orbit accurately. As a result,
{\footnotesize AREPO} could be more computationally expensive than the
classic {\footnotesize FARGO} scheme, because time-steps are allowed
to become shorter than the fluid-frame Courant time-step, and because
the motion of the mesh needs to be {\it solved for} instead of being
prescribed. In addition, the mesh needs to be re-tessellated at every
time-step. Therefore the concerns raised by \citet{don11b} \citep[see also][]{kle12}
about the use of the fluid-frame Courant time-step in {\footnotesize
 FARGO}, ignoring the gravitational influence of the planet, should
not be an issue in our case.

{The time-step for a cell of size $s_i$ is computed as 
$\Delta t_i=\min\{\Delta t_{\mathrm{grav},i},\Delta t_{\mathrm{CFL},i}\}$.
The gravitational time-step is defined as 
$\Delta t_{\mathrm{grav},i}=C_\mathrm{grav}\sqrt{ s_i / |\mathbf{f}|}$,
where $\mathbf{f}$ is the gravitational acceleration at the center of the cell 
and $C_\mathrm{grav}$ is a Courant-like
factor, which is typically chosen in the range of $10^{-2}-10^{-1}$. The moving-frame
Courant-Friedrichs-Lewy time-step is simply
$\Delta t_{\mathrm{CFL},I}=C_\mathrm{CFL} R_i/c_{s,i}$ where the Courant number
$C_\mathrm{CFL}$ is chosen in the range $0.2-0.3$ for 2D integrations and $c_{s,i}$ is the local
sound speed (constant everywhere for globally isothermal runs). In our low
resolution runs (see Table~\ref{tab:simulations} below), $\Delta t_{\mathrm{grav},i}\sim5\times10^{-3}/(2\pi)$ orbits, while  
$\Delta t_{\mathrm{CFL},i}\sim8\times10^{-2}/(2\pi)$ orbits, which implies that
 the gravitational time-step will be shorter than the fluid
frame Courant time step, and that we expect these
runs to be an order of magnitude more costly in CPU time than, say,
a run that uses a time-step based on $\Delta t_{\mathrm{CFL},i}$ alone.
For the disc properties adopted here (see below), the time-steps and $\Delta t_{\mathrm{CFL},i}$
and $\Delta t_{\mathrm{CFL},i}$ become roughly equal for a cell size of 
$s_i\sim \sim2.5\times10^{-3}$
in units where the planet semimajor axis is one, which corresponds to increasing
the number of radial zones by a factor of seven from our default number. Thus, at high resolution,
the Courant-time step typically dominates. At that point, the extra CPU cost of using {\footnotesize AREPO} should be mostly
limited to the tessellation routines, with a CPU usage that comprises roughly $40\%$ of the total.}

An important advantage of an entirely 
adaptable mesh is that
the complexity of the flow geometry does not impose any particular
restrictions on the computation, allowing for example simulations of
discs with distorted geometries. In the present example, the presence
of the planetary potential affects the evolution of the mesh, letting
it deviate from the nearly axially symmetric setup of
Figure~\ref{fig:sheared_mesh} to one that follows to the density
evolution of the disc, concentrating cells near spiral density wakes
and around the planet, and removing cells from gaps where gas is being
evacuated.

\paragraph*{Gravitational Potential.}
We represent the star-planet system by an
external, time varying potential:
\begin{equation}\label{eq:potential}
\Phi(\mathbf{r},t)= -\frac{GM_*}{|\mathbf{r}|}-\frac{GM_p}{|\mathbf{r}-\mathbf{r}_p(t)|}
+\frac{GM_p}{|\mathbf{r}_p(t)|^3}\mathbf{r}\cdot\mathbf{r}_p(t),
\end{equation}
where the third term on the right hand side corresponds to the
indirect term that results from choosing the coordinate system to be
fixed at the central star. The planet's position vector is
\begin{equation}
\mathbf{r}_p(t) = a_p \cos\left(2\pi\,
       {t}/{P_p}\right)\hat{\mathbf{x}}+a_p \sin\left(2\pi\,
       {t}/{P_p}\right)\hat{\mathbf{y}},
\end{equation}
(with $a_p=P_p=1$); i.e. the planet moves in a circular orbit around
the star.

The direct term corresponding to the planet potential (second term on
the RHS of Equation~\ref{eq:potential}) must be softened. We have
chosen a spline-type gravitational softening \citep{her89}
for the planet potential
as is usually done in {\footnotesize GADGET} \citep{spr01}. The spline
softening ensures a smooth transition into the exact Newtonian
potential at a finite distance from the planet (2.8 times the
gravitational softening parameter). We use a gravitational softening
of $\epsilon=0.03$ (0.6 times the disc scaleheight at the planet's
position) in agreement with the general setup proposed by
\citet{val06} \citep[see][for a discussion on the different types of
  softening and their effects]{don11b}. As it stands, this softening
has a different physical meaning than conventional approaches, and we
do not reduce it as we increase the resolution of our runs. This might
have some negative effects on convergence, since, as we see below
(Section~\ref{sec:gap_opening}), this softening length might be
already too large to capture accurately the non-linear effects
involved in the opening of gaps.

\begin{figure*}
\centering
\includegraphics[width=0.9\textwidth]{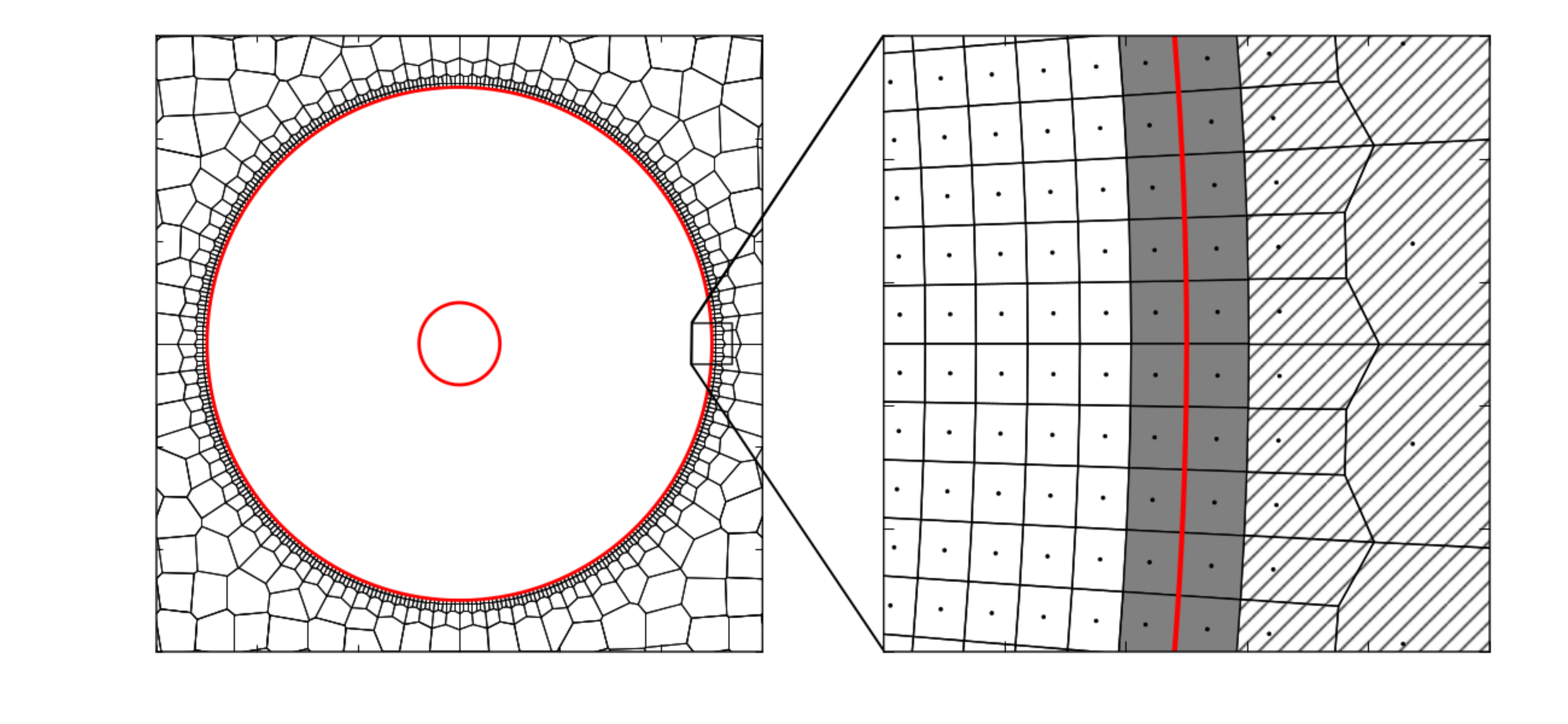}
\caption{Inner and outer boundaries (in red) for a two-dimensional
  circumstellar disc simulation. The computationally active domain --
  contained within the two boundaries -- only represents a radial
  portion of the disc, which is assumed to extend inside the inner
  boundary and outside the outer boundary. An inactive background
  Voronoi mesh is added outside $R_\mathrm{out}$. The main purpose of
  these additional cells is to fill in the computational box. However,
  in this example, the background cells are ``dead" and are never
  updated. A zoomed-in portion of the mesh shows how each boundary is
  constructed using pairs of mesh-generating points placed following
  two concentric circumferences
  \citep[see][]{ser01,spr10a,mun13}. Locally, the boundary is just one
  Voronoi interface and is, by definition, equidistant to both
  mesh-generating points. The left and right of this boundary define
  the ``interior" and ``exterior" domains, respectively. The last
  interior cell and the first exterior cell are shown in grey. These
  cells are referred to as ``inside" cell and ``outside" cell,
  respectively (see text), upon which boundary conditions are imposed
  \citep{mun13}. The background cells (cross-hatched region), although
  never updated, can be considered to contain quantity values ``at
  infinity" (i.e., as ``matching layer" boundaries) and are used to solve
  for the hydrodynamics quantities of the outside cells, in case these
  values cannot be fully determined from the interior domain (as in
  the case of reflective boundaries).
\label{fig:disk_boundaries}}
\end{figure*}

\paragraph*{Boundary Conditions.} 
One advantage of the moving-mesh approach is the high flexibility for
including moving boundaries {\it within} the computational domain
\citep{spr10a,mun13}. The inner and outer boundaries of a
circumstellar disc can be constructed from collections of
mesh-generating points describing concentric cylinders/circles. These
mesh-generating points can be made to move collectively, keeping the
overall shape of the boundary as the tessellation is updated. This
type of concentric boundary was already implemented in two- and
three-dimensional simulations of a Couette flow between concentric
cylinders by \citet{mun13}.

At the inner ($R_\mathrm{in}=0.25$) and outer ($R_\mathrm{out}=2.5$)
boundaries we impose reflective boundary conditions (see
Figure~\ref{fig:disk_boundaries} for a detailed description). In this
case, the primitive variables in an ``outside" cell (effectively
treated as a ghost cell, although it exists within the computational
domain) are copied from the adjacent ``inside" cell, except for the
velocity normal to the surface, which is reverted. The velocity
gradients are kept the same, thus the velocity normal to the surface
goes continuously through zero at the face, as required by reflective
boundaries. The initial-value Riemann problem at this interface is
solved in the same way as in the rest of the domain. To minimize wave
reflections off the boundaries, we impose an absorbing layer or
``wave-killing region" \citep{val06} that extends from the inner
radius up to $\widetilde{R}_\mathrm{in}=0.5$ and from the outer radius
down to $\widetilde{R}_\mathrm{out}=2.1$ thus reducing the
self-consistent computational domain to the region between those
radii. The absorbing region is implemented by adding a special source
term to the equations of motion, of the form
\begin{equation}
\frac{{\rm d}X}{{\rm d}t}=-\frac{X-X_0}{\tau}\Theta(R)~,~~
\text{with}~~\Theta(R)=\left[\frac{R-\widetilde{R}_\mathrm{in/out}}{R_\mathrm{in/out}-\widetilde{R}_\mathrm{in/out}}\right]^2~~,
\end{equation}
where $X$ represents each primitive or conservation variable, $X_0$ is
the reference value (the initial condition) and $\Theta(R)$ is a
parabolic ``ramp function" that vanishes at the edge of the absorbing
region.

This approach is similar in spirit to the perfectly matched layer of
\citet{ber94}, in which the wave damping is obtained by modifying the
equations of motions with frequency-dependent terms. The evanescence
of the waves is enforced by introducing an artificial complex quantity
into the dispersion relation of propagating waves, thus causing
exponential decay of their amplitude when needed.

\paragraph*{Shear Viscosity.}
A novel approach to physical viscosity has been recently developed by
\citet{mun13}. This approach uses a hybrid double-linear
reconstruction scheme to cope with the truncation errors of the
complex Voronoi cells while aiming to preserve the second order
accuracy of the scheme in time and space. The approach uses estimates
for higher-order derivatives to capture the spatial variability of the
shear to overcome the difficulties that can arise when mesh cells have
irregular shapes and an arbitrary number of neighbors. With this
approach, the time-centred diffusion fluxes can be estimated at each
interface.

\paragraph*{Equation of state} 
Circumstellar discs are often assumed to have an irradiation-dominated
temperature structure \citep[e.g.][]{chi97}, and thus a non-evolving
temperature specified by $T=T_0(R/R_0)^{-l}$.  This is referred to as
the ``locally isothermal" approximation, and $l =1$ is a common choice
(see \citealt{kra11} for a discussion of the validity of this
assumption).  The disc aspect ratio varies with radius as
$h=h_p(R/R_p)^{(1-l)/2}$, where $R_p$ is the radius of the planet's
orbit and $h_p$ is the disc aspect ratio at that location. Thus, discs
with $l=1$ have constant aspect ratio and discs with shallow
temperature profiles flare up with radius.

In shock-capturing Godunov schemes, this type of temperature structure
requires the use of an isothermal Riemann solver,
\citep[e.g.][]{bal94}, although a common shortcut is to run an
adiabatic simulation with $\gamma=1+\epsilon$ with $\epsilon$ small.
 In this work, we make use
of an exact (iterative) isothermal Riemann solver. For simplicity, we
set the disc to have the same temperature {\it globally} (i.e. $l=0$)
in all our simulations unless stated otherwise (see Section~\ref{sec:stockholm}).
This implies that the
disc aspect ratio ($h\sim c_s/v_k$) is not constant, in contrast with
several models found in the literature that choose $l=1$ and have,
as a consequnce $h=$constant. In all our runs,
the aspect ratio of the disc at the planet position is fixed to be
$h_p=0.05$, and a global sound speed value follows from this fixed
parameter.

\begin{table*}
\begin{center}
\begin{tabular}{ r | c | c | c | c | r | r | r  | c| r | c }
\hline
{Numerical experiment}
 & & \multicolumn{4}{c}{Physical parameters} &\multicolumn{3}{c}{Effective azimuthal resolution}\\
(label/radial zones) & &  $\mu$& &$\nu$&&$N_\phi(R_\mathrm{in})$&$N_\phi(R_p)$&$N_\phi(R_\mathrm{out})$\\ 
 \noalign{\smallskip}
\cline{1-1} \cline{3-5} \cline{7-9} 
\cline{1-1} \cline{3-5} \cline{7-9} 
\noalign{\smallskip}
{\bf \texttt{JUP\_128}}&& $10^{-3}$&&--                             & &      $\sim90$ & $\sim360$  & $\sim890$\\
\texttt{JUP-VISC-A\_128}&& $10^{-3}$&&$10^{-4}$       &    & "   & " & " \\  
{\bf \texttt{JUP-VISC-B\_128}}&& $10^{-3}$&&$10^{-5}$& & "   & " & "\\  
\texttt{JUP-VISC-C\_128}&& $10^{-3}$&&$10^{-6}$      &    & "   & " & "\\  
\texttt{JUP\_256}&& $10^{-3}$&&--                                    &    &      $\sim180$ & $\sim720$  & $\sim1780$\\ 
\texttt{JUP-VISC-A\_256}&& $10^{-3}$&&$10^{-4}$      &    & "   & " & "\\  
\texttt{JUP-VISC-B\_256}&& $10^{-3}$&&$10^{-5}$      &    & "   & " & "\\  
\texttt{JUP-VISC-C\_256}&& $10^{-3}$&&$10^{-6}$      &    & "   & " & "\\  
\texttt{JUP\_512}&& $10^{-3}$&&--                                    &     &      $\sim360$ & $\sim1440$  & $\sim3560$\\  
\texttt{JUP-VISC-A\_512}&& $10^{-3}$&&$10^{-4}$      &    & "   & " & "\\  
\texttt{JUP-VISC-B\_512}&& $10^{-3}$&&$10^{-5}$      &    & "   & " & "\\  
\texttt{JUP-VISC-C\_512}&& $10^{-3}$&&$10^{-6}$      &    & "   & " & "\\  
\\
{\bf \texttt{NEP\_128}}&& $10^{-4}$&&--                            &    &      $\sim90$ & $\sim360$  & $\sim890$\\
\texttt{NEP-VISC-A\_128}&& $10^{-4}$&&$10^{-4}$      &     & "   & " & " \\  
{\bf \texttt{NEP-VISC-B\_128}}&& $10^{-4}$&&$10^{-5}$&  & "   & " & "\\  
\texttt{NEP-VISC-C\_128}&& $10^{-4}$&&$10^{-6}$      &   & "   & " & "\\  
\texttt{NEP\_256}&& $10^{-4}$&&--                                    &   &      $\sim180$ & $\sim720$  & $\sim1780$\\ 
\texttt{NEP-VISC-A\_256}&& $10^{-4}$&&$10^{-4}$      &  & "   & " & "\\  
\texttt{NEP-VISC-B\_256}&& $10^{-4}$&&$10^{-5}$      &    & "   & " & "\\  
\texttt{NEP-VISC-C\_256}&& $10^{-4}$&&$10^{-6}$      &    & "   & " & "\\  
\texttt{NEP\_512}&& $10^{-4}$&&--                                    &     &      $\sim360$ & $\sim1440$  & $\sim3560$\\  
\texttt{NEP-VISC-A\_512}&& $10^{-4}$&&$10^{-4}$      &    & "   & " & "\\  
\texttt{NEP-VISC-B\_512}&& $10^{-4}$&&$10^{-5}$      &    & "   & " & "\\  
\texttt{NEP-VISC-C\_512}&& $10^{-4}$&&$10^{-6}$      &    & "   & " & "\\  
\hline
\end{tabular}
\end{center}
\caption{Simulation parameters. For each planet mass ($\mu=10^{-3}$
  and $\mu=10^{-4}$) we vary the resolution (in terms of the number of
  radial zones) and the viscosity coefficient. The different
  resolutions are $N_R=128$, $256$ and $512$ radial zones. The
  different viscosities are $\nu=10^{-4}$, $10^{-5}$ and $10^{-4}$ as
  well as runs not including explicit viscous terms. The runs in bold
  type-face denote the base of ``fiducial" runs: an inviscid and a
  viscous ($\nu=10^{-5}$) run for each mass ratio at a resolution of
  $N_R=128$.\label{tab:simulations}}
\end{table*}

\subsection{Initial Conditions}\label{sec:initial_conditions}
All the simulations presented in this paper are carried out in units
where $G(M_*+M_p)=a_p=1$ and thus $P_p=2\pi$, where $M_*$ is the mass
of the central object, $M_p$ is the planet mass, $a_p$ gives the
planet semi-major axis, and $P_p$ the planet's orbital period. The
planet-to-total mass ratio is $\mu\equiv M_p/(M_*+M_p)$. In addition,
we choose $G=1$, such that the star and the planet masses are $1-\mu$
and $\mu$, respectively. This choice of units reduces the relevant
physical parameters of our simulations to two: the mass ratio $\mu$
and the shearing viscosity $\nu$. In these units, the value of the
$\alpha$-viscosity coefficient \citep{sha73} at the planet's position
($R=1$) is $\alpha=\nu/h_p^2/\sqrt{1-\mu}$~~.
Table~\ref{tab:simulations} lists the different physical parameters
explored in this work, as well as the spatial resolution of each
numerical run.

\paragraph*{Surface Density Profile.}
Typically, the surface density in circumstellar discs is modelled with
a power-law profile
(i.e. $\Sigma(R)=\Sigma_0(R/R_0)^{-p}$~). We
choose our discs to have a constant surface density ($p=0$) of
$\Sigma=\Sigma_0=0.002 M_*/(\pi a_p^2)$, such that the enclosed mass
at the planet position is $\sim0.2\%$ of the mass of the star (the
disc is assumed to extend all the way to the star, beyond the inner
boundary). Note that real discs have steeper density profiles, ${-3/2
  < p < -1}$ \citep{and09}.

\paragraph*{Planet Mass}
In the units used here, the planet and star masses are determined by the
single parameter $\mu$.  We have explored two values of $\mu$:
$10^{-3}$ and $10^{-4}$ (Table~\ref{tab:simulations}).  For a stellar
mass of $M_\odot$, these mass ratios correspond approximately to the
planet masses of Jupiter and Neptune, respectively. Since simulations
have been shown to develop numerical artifacts if planets are added
impulsively into the disc, we increase the planet mass from zero up to
$M_p$ slowly in time as \citep[e.g.][]{val06}:
 \begin{equation}
 M_p(t)=\left\{
 \begin{array}{lr}M_p\,\sin^2\left(\frac{\pi t}{10 P_p}\right) & t\leq 5P_p\\ M_p  & t> 5P_p .\end{array}
 \right.
 \end{equation}
\paragraph*{Orbital Velocity Profile}
Our disc simulations are started from centrifugal equilibrium and
axial symmetry. From the Euler equations in cylindrical coordinates, the
equilibrium and axisymmetry condition implies an azimuthal velocity that satisfies 
\begin{displaymath}
\frac{v_\phi^2}{R}=\frac{\partial \Phi}{\partial R}+\frac{1}{\Sigma}\frac{\partial P}{\partial R}~~,
\end{displaymath}
where the Keplerian term $R({\partial \Phi}/{\partial R})=v_K^2=\sqrt{GM_*/R}$ is modified by a ``pressure
buffer" term, resulting in an orbital velocity that is slightly sub-Keplerian (for $c_s^2\ll v_K^2$)
\begin{equation}
v_\phi^2=v_K^2-c_s^2(l+p)~~,
\end{equation}
where the pressure buffer term comes from the initial temperature and
density gradients: $\partial P/\partial R=(\partial P/\partial c_s^2)
({\rm d} c_s^2/{\rm d}R)+(\partial P/\partial \Sigma) ({\rm
  d}\Sigma/{\rm d}R)$.  Because in this work we use an initially
constant surface density and a globally isothermal equation of state,
$p=l=0$, the initial rotation curve is strictly Keplerian.

\paragraph*{Initial Mesh} 
The setup of the initial mesh is flexible and can be chosen according
to the needs of the problem being simulated. In this work, we position
the mesh-generating points as a series of concentric, unaligned rings,
with constant separation $\Delta R$ in the radial direction. The same
separation is imposed between points in the azimuthal direction (i.e.,
$\Delta \phi\sim\Delta R/R$), thus keeping the cell size and cell
aspect radio nearly constant throughout the computational
domain. Unlike polar-like meshes (e.g. Figure~\ref{fig:sheared_mesh}),
which have constant azimuthal resolution at all radii, a constant cell
size implies a varying azimuthal resolution with radius. The effective
resolution at the inner and outer radius, and at the planet location
$R_p=1$ are listed in Table~\ref{tab:simulations}. The Hill radii of
the planets are $r_\mathrm{H}=3.2\times10^{-2}$ and $6.9\times10^{-2}$
for $\mu=10^{-4}$ and $10^{-3}$, respectively. This means that, for
the three resolutions explored here, the Hill sphere of the
Neptune-mass planet is 1.8, 3.7 and 7.4 cells across, while for the
Jupiter-mass planet, the Hill sphere is 4, 8 and 16 cells across.

\section{Results}\label{sec:results}
%

\subsection{Surface Density Field}\label{sec:density_field}
%

\subsubsection{General features}
The perturbed surface density profile of the disc provides a
qualitative means to asses the relative performance of different
numerical schemes.  The key features on which we base our comparison
are the overall shape of the tidal wake launched by the planet, its
location with respect to the predictions from linear theory
\citep{ogi02}, and how far from the planet the wake is damped. For the
Jupiter-mass simulations, the shape of the gap carved by the local
deposition of angular momentum is an important diagnostic for the
accuracy and numerical diffusivity of the code. 
Our analysis is guided by the code comparison project of \citet{val06}, although
their model setup is not entirely identical to the one presented in  
Section~\ref{sec:num_exp}. In said comparison project, the different numerical 
schemes differed on the sharpness of the gap, its degree of axisymmetry, its depth, the
smoothness of the remaining material, and the amount of gas retained
around the orbit's Lagrange points after 100 or more orbital periods.

\begin{figure*}
\centering
\includegraphics[width=0.49\textwidth]{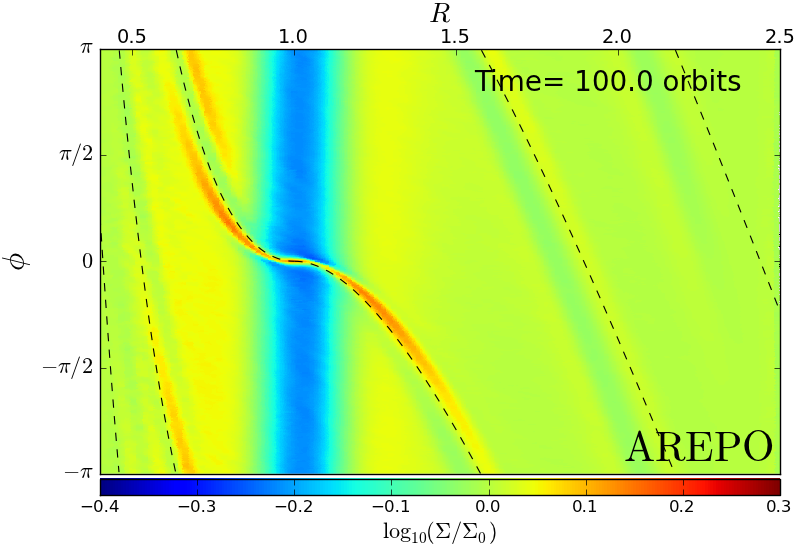}
\includegraphics[width=0.49\textwidth]{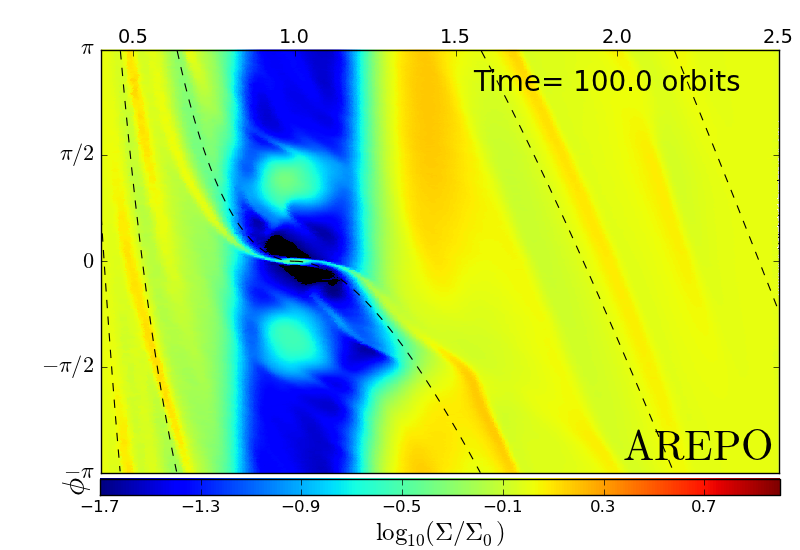}
\includegraphics[width=0.49\textwidth]{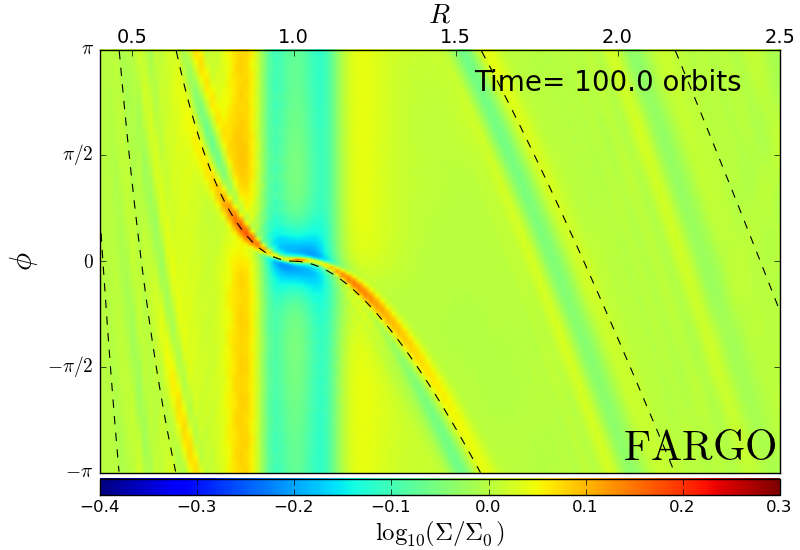}
\includegraphics[width=0.49\textwidth]{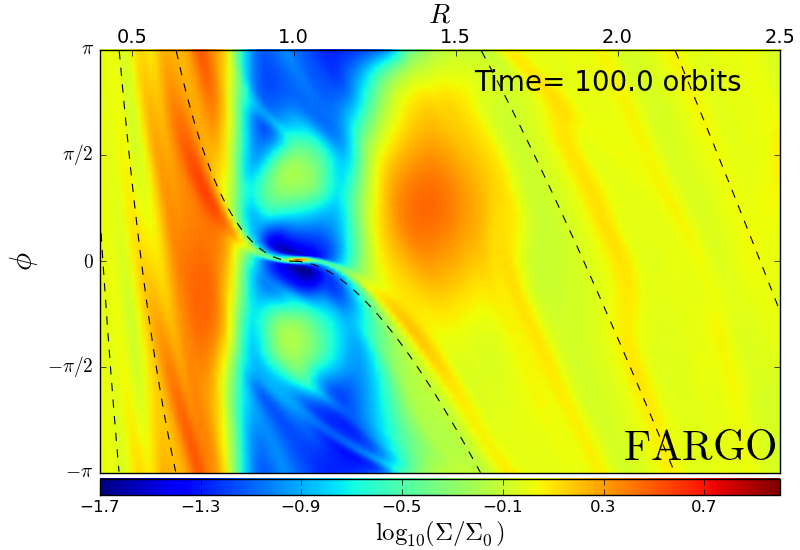}
\caption{Top panels: {\scriptsize AREPO} results of the
  surface density field at $t=$100 orbits for the fiducial inviscid simulations 
  \texttt{NEP\_128}  (top left panel; $\mu=10^{-4}$)  and \texttt{JUP\_128}
  (top right panel; $\mu=10^{-3}$).  For the Neptune-mass case (left panels), the
  color-coded density range is $-0.4<\log_{10}(\Sigma/\Sigma_0)<0.3$,
  while for the Jupiter-mass case (right panels), it is
  $-1.7<\log_{10}(\Sigma/\Sigma_0)<1.0$.  Bottom panels:
  same simulation setup as above, this time using the {\scriptsize FARGO}
  code \citep{mas00}. All four simulations are for {\it globally} isothermal
  (i.e., flared) discs. For the two mass ratios tested, {\scriptsize AREPO} gaps
  appear narrower and deeper than those obtained by {\scriptsize FARGO}. In
  addition, mass pile-up outside the gap edges is significantly
  less pronounced in the {\scriptsize AREPO} runs.
  In all frames, the dashed lines represent the
  location of the wake in the linear regime \citep{ogi02}.  Note that
  in the globally isothermal case, the location of the Lindblad
  resonances are different from those in a disc with a radial
  temperature gradient; in the present case, the spiral wake is less
  tightly wound in the outer disc and more tightly wound in the inner
  disc than the case with a $T\propto R^{-1}$ temperature profile.
\label{fig:density_polar}}
\end{figure*}

Figure~\ref{fig:density_polar} (top panels) shows the surface density field of the
two inviscid fiducial simulations (128 radial zones) for mass ratios
of $\mu=10^{-4}$ (Neptune-mass planet) and $\mu=10^{-3}$ (Jupiter-mass
planet) after 100 orbits. For easier comparison with earlier work, we
have applied a coordinate transformation to the cell coordinates and
converted the density field from the $x$-$y$ plane into the $R$-$\phi$
plane. The surface density field also includes an overlay of
the shape and location of the planetary wake derived in the linear regime by \citet{ogi02}\footnote{
The location of the planetary wake, for $R_p=1$, is given by
\begin{equation}
\nonumber
\phi(R)=\left\{
\begin{array}{lc}
2\pi ({t}/{P_p}) - \cfrac{2}{3}h^{-1}\left(R^{3/2}-\cfrac{3}{2}\ln R-1\right)+\pi  & R<1~~\\
2\pi ({t}/{P_p}) + \cfrac{2}{3}h^{-1}\left(R^{3/2}-\cfrac{3}{2}\ln R-1\right)+\pi & R>1 , 
\end{array}
\right.
\end{equation}
with $h=h_p(R/R_p)^{1/2}$, where $h_p=0.05$.\\
}.  {In addition, in Figure~\ref{fig:density_polar} (bottom panels)
we show equivalent runs using the {\footnotesize FARGO}
code \citep{mas00} with $N_R=128$ and $N_\phi=384$. 
Initial conditions and temperature structure (globally
isothermal) are the same for the  {\footnotesize AREPO} and the 
 {\footnotesize FARGO} runs, as well as the location of the inner and outer
 boundaries and the implementation of absorbing boundary conditions. Besides
 the obvious difference in the hydrodynamic solver, the  {\footnotesize AREPO} 
 and  {\footnotesize FARGO} runs differ in the distribution of resolution elements. 
 The {\footnotesize AREPO} cells are initially of roughly the same size, implying
 a radially varying azimuthal resolution (see Table~\ref{tab:simulations}.
The {\footnotesize FARGO} grid consists of linearly spaced radial zones and
 constant number azimuthal zones per radial bin, implying that cells do not have
 a constant aspect ratio and have area that
 grows proportionally  to $R$. At the location of the planet ($R=1$) both
 setups roughly contain the same number of azimuthal zones.}

{After 100 planetary orbits, both the Neptune-mass and Jupiter-mass simulations
have cleared  -- to different
degrees -- a gap in the coorbital region. {\footnotesize AREPO} 
and {\footnotesize FARGO} runs roughly agree on the width and depth of the gap,
although the {\footnotesize AREPO} results feature a sharper transition at the
edge of the gap and slightly deeper gaps (see Section~\ref{sec:gap_opening} below).
The Jupiter-mass runs retain similar amount 
of gas in the Lagrange  $L_4/L_5$ points, although the minimum density 
within the gap close the to planet is lower in the {\footnotesize AREPO} case.} 

{Outside the
gap, there is very good agreement between  {\footnotesize AREPO} 
and {\footnotesize FARGO} for the Neptune-mass run}. 
The background flow in this  case is solely disturbed
by the spiral wake launched by the planet; the averaged density field,
in turn, barely differs from the initial condition. The Jupiter-mass planet, on the other
hand, does alter the background mean flow significantly, in part owing
to the larger amplitude of the spiral wake (in the linear regime, the
amplitude of the wake is proportional to $M_p$) and to the density
bump caused by material being displaced from the coorbital
region. Indeed, a significant part of the mass in the coorbital region
is relocated right outside the gap edge. 
{This gas pile-up is more pronounced in the  {\footnotesize FARGO} simulations 
(very significantly in the Jupiter-mass case, but also to some degree in the Neptune-mass
case). }

{The deviations from a uniform density field 
in the $\mu=10^{-3}$ simulations are largest outside the edge of the gap, and exhibit
time variability (in both {\footnotesize FARGO} and
{\footnotesize AREPO}). This is in contrast with the $\mu=10^{-4}$ simulations,
in which, aside from a gradually carved shallow gap, the density field
is nearly stationary for 100 orbits. 
The Jupiter-mass {\footnotesize FARGO} run seems more violently unstable
than its {\footnotesize AREPO} equivalent. The large amplitude density
fluctuation in this case are attributed to ``edge instabilities"
and vortices in the disc  (e.g.,
\citealp{lov99}; see Section~\ref{sec:vortensity} below), which
have been previously  observed in numerical simulations by
\citet{kol03}, \citet{li05} and \citet{val06}.
We have confirmed that the appearance of these transient
features is favored by higher resolution, and eventually it even takes place
for the Neptune-mass simulation with both {\footnotesize AREPO} 
and {\footnotesize FARGO}. }

\subsubsection{Clearing of gaps  and evolution of the surface density profile}\label{sec:gap_opening}

The tidal wake launched in a disc is a result of 
waves excited by a perturber moving at supersonic speeds and its shape arises
as the superposition of all harmonics.  In the linear regime, the wake
shape asymptotically reaches a stationary profile
\citep{nar87,raf12} and its amplitude is proportional to the planet mass
\citep[e.g.][]{don11a,raf12}. These waves carry angular momentum,
and unless they remain non-dispersive (i.e. the in linear regime) throughout the computational
domain and  for the full extent of the integration time, some damping process
will couple these waves to the disc fluid, triggering the transfer of angular
momentum to the gas, resulting in the evolution of the background
density field. Well into the non-linear regime, the deposition of angular momentum
into the disc is expected to open a gap in the coorbital region if the planet is
massive enough.

One of the gap-opening criterions found in the literature is the ``thermal criterion"
of \citet{lin93}. This criterion states that a planet of mass $M_p$ will open
a gap in the disc if
\begin{equation}\label{eq:gap_opening}
M_p>M_\mathrm{th}\equiv\frac{c^3_{s,p}}{\Omega_pG}~~,
\end{equation}
which roughly corresponds to the case in which the planet's Hill radius
is comparable to the disc scale height. 

For the parameters used in the present work, the so-called thermal
mass $M_\mathrm{th}$ is $\sim1.25\times10^{-4}$, which is slightly
above the Neptune-mass planet ($M_p=10^{-4}$), and eight times smaller
than the Jupiter-mass planet ($M_p=10^{-3}$), thus our results in
shown in Figure~\ref{fig:density_polar} fall within expectations according
to this criterion. For a further discussion
of gap-opening and its dependence on resolution and viscosity
coefficient, see Section~\ref{sec:gap_opening} below.

Note that once there is a gap present, the gravitational coupling between the
disc and the planet cannot be calculated in the linear regime assuming
a uniform background density, since this will lead to erroneous
estimates of the tidal torque density \citep{pet12}. Not only does the
torque density profile in gapped systems differ from the original
calculation of \citet{gol79} but also the shape of the tidal wake
differs. While in the Neptune-mass simulations the gap is not deep
enough to significantly alter the shape of the wake compared to the
theoretical results in the linear regime
(Figure~\ref{fig:density_polar}, left panel), the Jupiter-mass runs
(Figure~\ref{fig:density_polar}, right panel) show an evident mismatch
between the theoretical wake position and the actual density maxima of
the wake.

\begin{figure}
\centering
\includegraphics[width=0.48\textwidth]{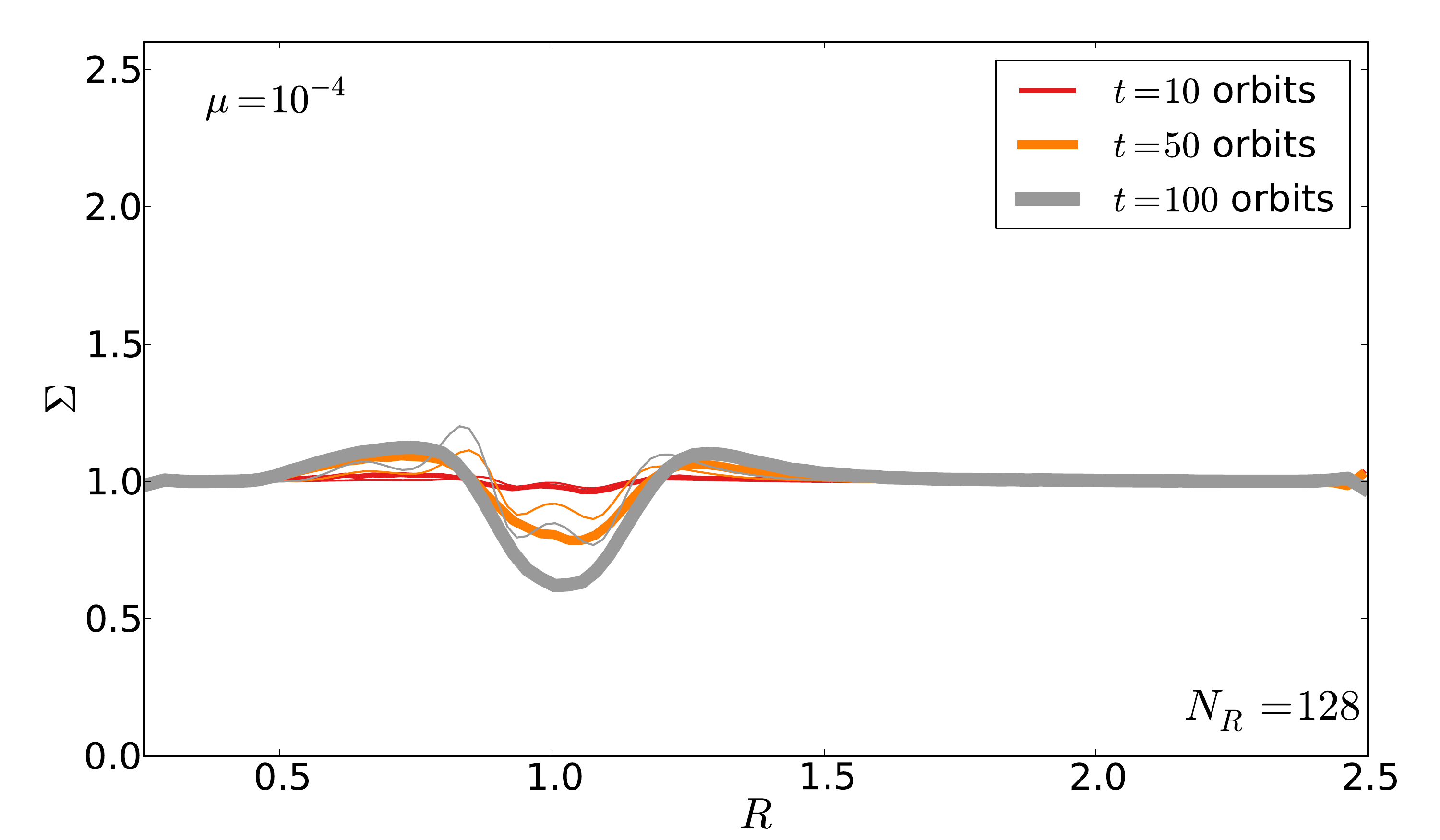}
\includegraphics[width=0.48\textwidth]{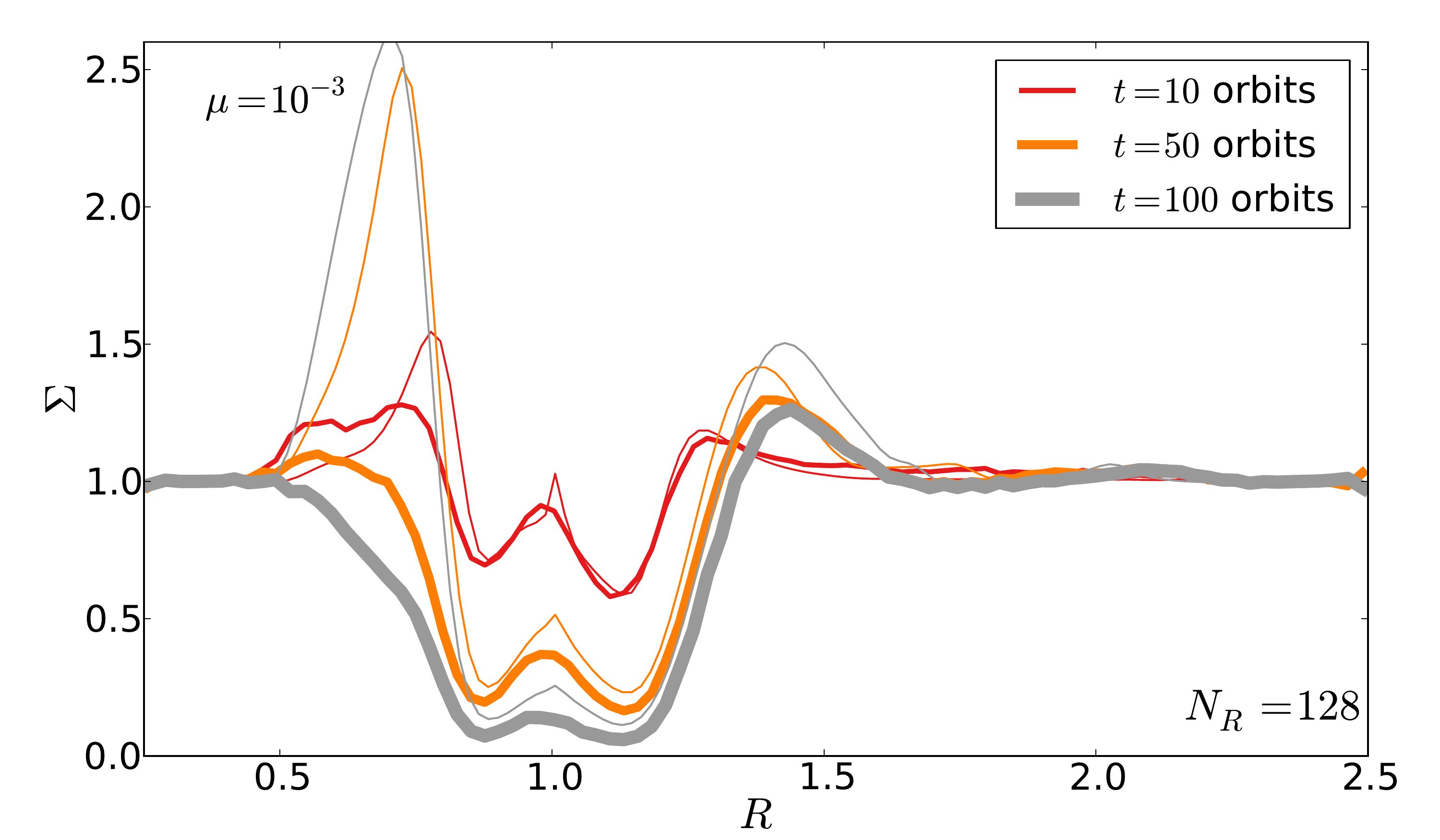}
\caption{Top panel: azimuthally averaged surface density profile for
  the inviscid fiducial simulation with $\mu=10^{-4}$ at three
  different times (10, 50 and 100 orbits). Thick lines correspond to
  the {\scriptsize AREPO} run (top left panel of Figure~\ref{fig:density_polar})
  and thin lines correspond to the equivalent  {\scriptsize FARGO} 
  (bottom left panel of Figure~\ref{fig:density_polar}).
   Bottom panel: same as top
  panel but with $\mu=10^{-3}$.
\label{fig:gap_profile}}
\end{figure}

Figure~\ref{fig:gap_profile} presents the evolution of the (azimuthally
averaged) surface density profile for the Neptune-mass (top panel) and
Jupiter-mass (bottom panel) planets at three different times for each
numerical integration,  showing how a shallow gap is
carved in the former case, and a deep gap in the latter.
 {Thick lines correspond to our fiducial
 {\footnotesize AREPO} runs, and thin lines to the equivalent  {\footnotesize FARGO}
 runs (same simulations as  in Figure~\ref{fig:density_polar}). }
 In the
Neptune-mass case, the density field features a dip within the coorbital
region of about $40\%$ of its initial value after 100 orbits. This deficit increases
to over $90\%$ in the Jupiter-mass case. {Although the width
and depth of the gap are roughy consistent between the {\footnotesize AREPO} 
and {\footnotesize FARGO} runs, these seem to diverge with respect to each other in time, 
as can be readily noticed for the $\mu=10^{-4}$ simulations at 100 orbits.}

{As mentioned above, the most striking difference between  {\footnotesize AREPO} 
and {\footnotesize FARGO}  is the magnitude of
the density bumps at the edges of the gap. This disagreement is more pronounced 
for the inner edge. The in/out asymmetry
suggests that resolution may be to blame because the number of azimuthal
zones increases with radius (Table~\ref{tab:simulations}). Figure~\ref{fig:gap_profile} demonstrates that the mass deficit in the inner
edge of the gap grows with time. This growth might be related to the lack of angular
momentum conservation or angular momentum diffusion\footnote{
Although the terms ``non-conservation"  and 
``numerical diffusion" are sometimes found in the literature as
interchangeable, they are different, as a numerical scheme can conserve a
quantity globally, but still transport it diffusively across interfaces owing to truncation error.
Diffusion is an unavoidable consequence of truncation and finite resolution, while
non-conservation is a consequence of the time-stepping algorithm used and can, in principle, be
avoided \citep[see,][for an example]{kle98}.
} combined with an absorbing boundary condition. As the inner
density bump starts to develop (e.g., red curves in Figure~\ref{fig:gap_profile}, bottom panel),
it is already wider in the {\footnotesize AREPO} case, ending at the edge
of the wave absorbing region. This can lead to 
a ``leakage" of mass through the boundary}, since
over-densities will be damped out, in turn violating mass conservation \citep[see][]{val06}.
{Thus, the disagreement in the density profile of the inner disc can only worsen with time}.
 Mass loss seems to take place in our low resolution Jupiter-mass run, 
but higher-resolution runs do reveal a local density maximum in the inner disc close to the edge of
the wave-absorbing region (see Section~\ref{sec:viscosity_resolution} below). 

{Local non-conservation of angular momentum resulting from certain type
of integrators can be the cause of spurious evolution of gaps in disc simulations
 \citep[see e.g.,][]{kle98,paa06}.
We know that {\footnotesize AREPO} does not conserve angular momentum
exactly because of two reasons: (1) owing to the solving of the Euler equations
in  a cartesian formulation, and (2) owing to the inclusion of the gravitational
acceleration as a source term evaluated at the cell centers. Since the cells 
bulk velocities are updated via a well-validated leapfrog algorithm, we expect
the second of these effects to non cumulative and  therefore  of lesser importance. The first effect however,
can produce artifacts at low resolution.}

{In practice, a cell's bulk angular 
momentum can be lost to the cell's inner circulation or ``cell spin",
which cannot be properly captured by the linear reconstruction of the velocity field.
This is obviously a function of resolution, but also a function of the shear rate
(and, to a lesser extent, the local density gradient). Since
our radial resolution is nearly constant, we expect local shear rate 
($\equiv (R/2)(d\Omega/dR)\propto R^{-3/2}$) to be responsible for the
inner region of the disc suffering from angular momentum loss most significantly.
The fact that the shear is set almost entirely
by the Keplerian potential, means that this effect has a constant sign and will
act cumulatively. This might explain why material in the inner region is smoothly
transported to the absorbing boundary (Figure~\ref{fig:gap_profile}). }

{\citet{kle98} discussed the numerical creation and destruction of angular momentum
for disc simulations using rotating-frame, cylindrical-coordinates schemes in
which the Coriolis term is included as a source term rather than as a a flux
divergence in the angular momentum conservation equation. Although an entirely
different numerical scheme, the numerical bias identified by \citet{kle98} in
some integration schemes bears some resemblance to what we have called
cell spin, in the sense that it arises due to the evaluation of  angular momentum
source terms at the center of cells, rather than calculating how much angular momentum
enters and leaves a cell through its interfaces. Although not considered within the
present work, we explore this effect in further depth in \citet{pak14}, suggesting alternatives
to better deal with angular momentum trapped as cell spin. For now, we confirm
that this effect is indeed resolution dependent, and that the deficit in the density
bump is significantly diminished by using moderately higher resolution
(Section~\ref{sec:viscosity_resolution}).}

Qualitatively, our results are in agreement with the thermal criterion of
Equation~(\ref{eq:gap_opening}), which states that our Jupiter-mass
simulations should open a gap while the Neptune-mass runs should
not. However, the thermal criterion has been questioned by recent
evidence of planets with masses smaller than the thermal mass being
able to open gaps \citep[e.g.][]{don11b,duf12,zhu13}.

Objections to the thermal criterion were raised by \citet{goo01}
\citep[see also][]{raf02a,raf02b}, {whose calculations showed that 
deposition of angular momentum from the launched waves onto the disc
leading to clearing of the gap can happen for planet mass
smaller than $M_\mathrm{th}$ if these waves
steepen into shocks at a finite distance from the planet $x_\mathrm{sh}$:}
\begin{equation}\label{eq:shock_distance}
|x_\mathrm{sh}|\approx0.93\left(\frac{\gamma+1}{12/5}\frac{M_p}{M_\mathrm{th}}\right)^{-2/5}H_p~.
\end{equation}
{After breaking, the wave's angular momentum is gradually transferred to the disc, which
in turns leads to density evolution and the possibility of clearing a gap, even for
planet masses below $M_\mathrm{th}$ \citep{raf02a,raf02b}}

Evidently, wave steepening depends on grid resolution, thus for inviscid
simulations, the gap shape and depth will depend on whether the
angular momentum is injected into the disc by shock dissipation or
by a viscous-like dissipation onto the computational grid.
This is why high-resolution simulations are
required \citep{don11a,duf12} for gaps to be seen at small values of
$\mu$ ($\lesssim10^{-5}$), since otherwise numerical diffusivity can
mimic the gap-filling consequence of physical viscosity.

For the isothermal ($\gamma=1$) examples shown here, and recalling
that the planet mass in the Neptune case is $M_p=4/5M_\mathrm{th}$,
the shock distance given by Equation~(\ref{eq:shock_distance}) is
$|x_\mathrm{sh}|\approx1.09H_p\sim0.055$. For the Jupiter case
($M_p=8M_\mathrm{th}$), we have
$|x_\mathrm{sh}|\approx0.44H_p\sim0.022$. Note that, in the vicinity
of the planet, the radial extent of a Voronoi cell at $t=0$ is
$1.8\times10^{-2}$, $9.0\times10^{-3}$ and $4.5\times10^{-3}$, for
$N_R=128,256$ and $512$, respectively. Therefore, the shock distance
is not resolved at all for the lowest resolution runs.  Furthermore,
the shock distance for the Jupiter-mass case is smaller than $\varepsilon=0.6H_p$, the
gravitational softening used in our simulations\footnote{
We chose a softening parameter of $\varepsilon=0.6H_p=0.03$ 
\citep[as in][]{val06} since this value gives 2D torque magnitudes 
closer to their corresponding 3D values \citep[see][]{tan02}.},
thus the precise gap-opening mechanism cannot be
adequately captured in these simulations.

\begin{figure}
\centering
\includegraphics[width=0.45\textwidth]{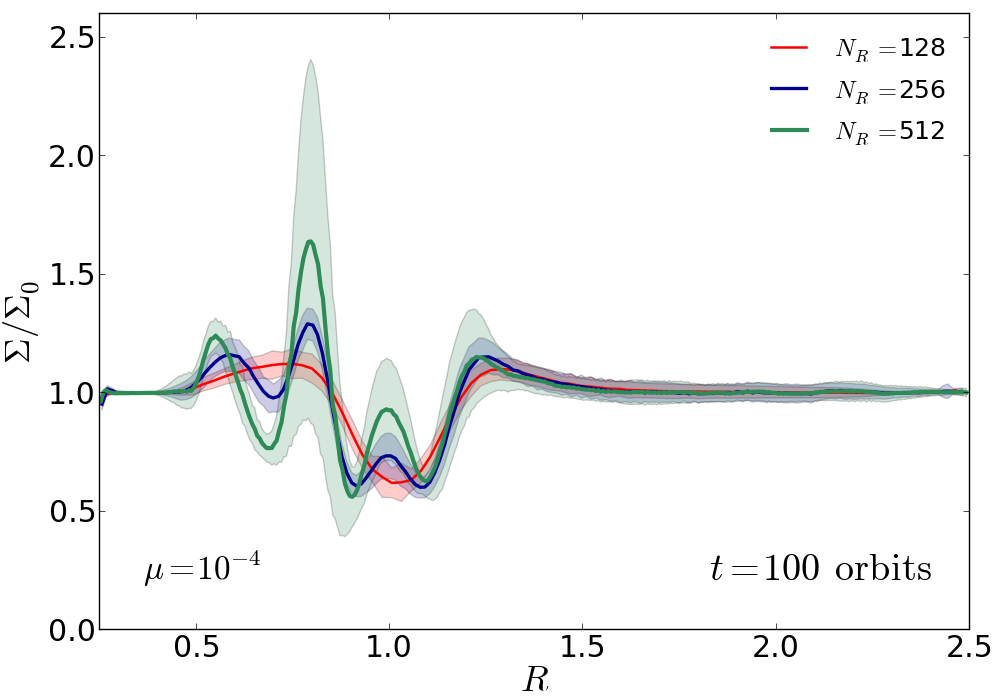}
\includegraphics[width=0.45\textwidth]{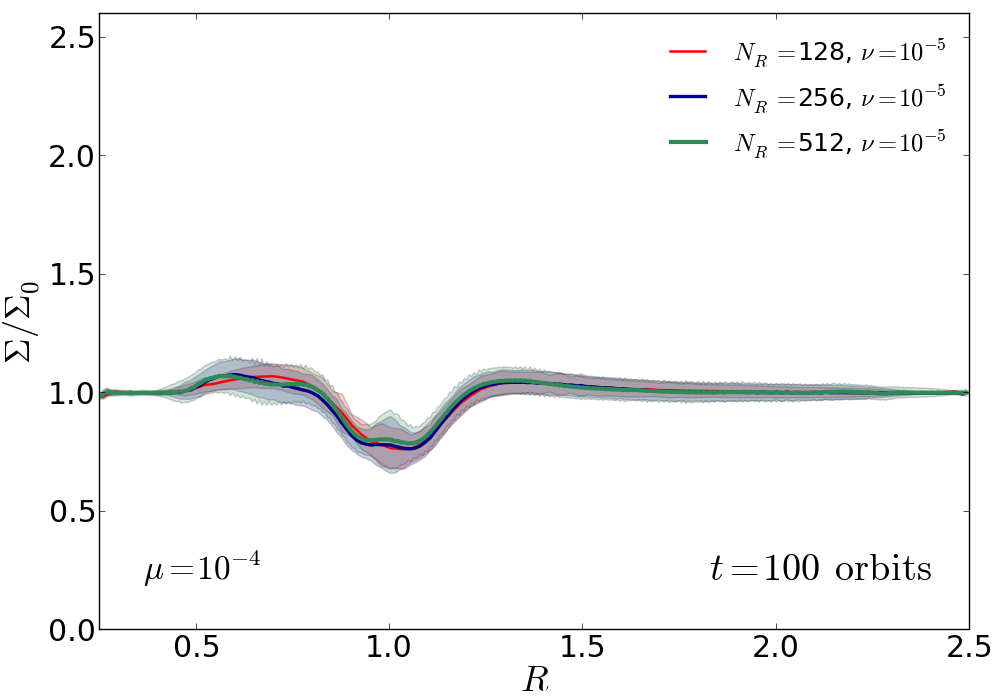}
\caption{Azimuthally-averaged gap profiles at $t=100$~orbits with
  $\mu=10^{-4}$.  In addition to the azimuthal means (solid curves),
  1-$\sigma$ (one standard deviation) shaded areas are added to
  quantify deviations from axial symmetry. Top panel: profiles for
  inviscid simulations at three different resolutions, with $N_R=128$
  (red), 256 (blue) and 512 (green). Bottom panel: profiles for
  viscous simulations ($\nu=10^{-5}$) at the same 3 resolutions.
\label{fig:NEP_gap_profile}}
\end{figure}

\begin{figure}
\centering
\includegraphics[width=0.45\textwidth]{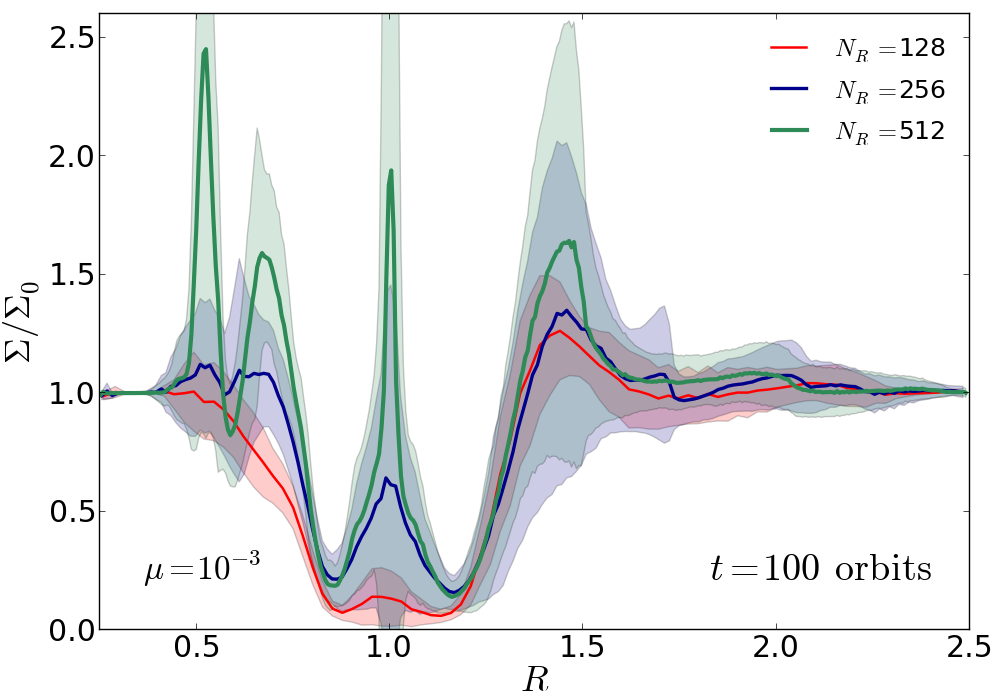}
\includegraphics[width=0.45\textwidth]{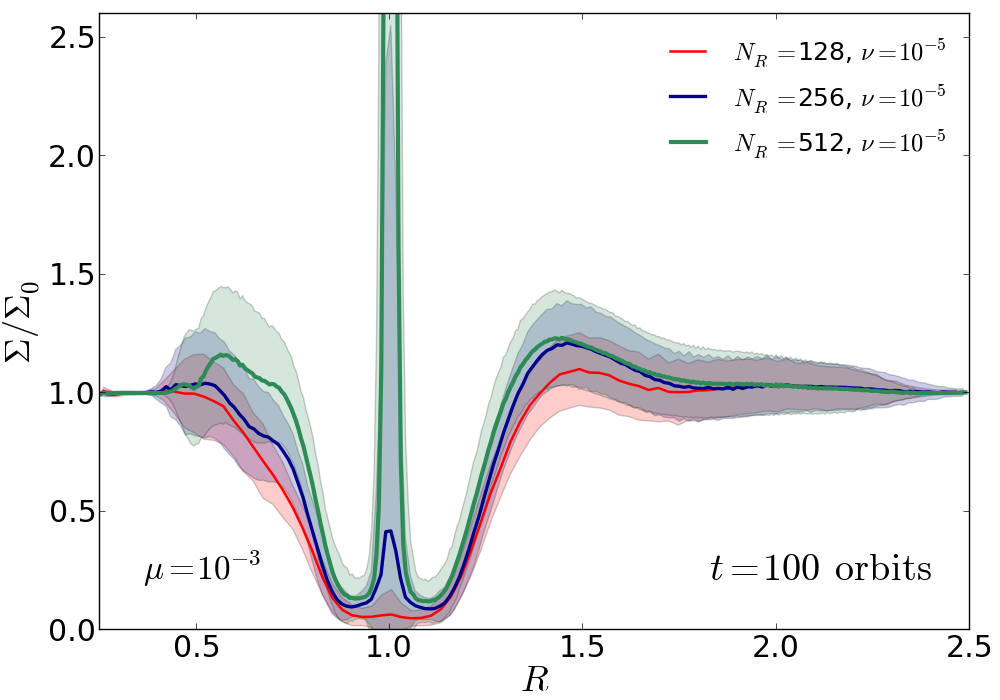}
\caption{Same as Figure~\ref{fig:NEP_gap_profile}  but for a mass ratio of
$\mu=10^{-3}$.
\label{fig:JUP_gap_profile}}
\end{figure}

\subsubsection{Development of edge instabilities}\label{sec:edge_instabilities}
The surface density field in polar coordinates emphasizes the
departure from axial symmetry of the density gap cleared by the
Jupiter-mass planet (Figure~\ref{fig:density_polar}, right
panel). This is something to be aware of in moving-mesh codes, which
are known to develop morphological asymmetries somewhat faster than
fixed grid codes \citep[c.f.][]{spr10a,spr11,mcn12b}.  The early
development of asymmetries during the growth of instabilities can be
seeded by ``grid noise" arising even in initially symmetric initial
conditions.  In moving-mesh codes, grid noise tends to have a greater
influence because it is not only composed of round-off errors in the
flux calculation in different directions, but also of truncation
errors in the mesh-drifting algorithm. The finite accuracy with which
the locations of the mesh-generating points are evolved propagates to
the geometry calculation of cell face centers, ultimately affecting
the evolution of the fluid quantities. This problem is absent in
structured mesh codes, especially since they often implement
``symmetric sweeping" of the grid faces to preserve the symmetries of
the flow to greater accuracy (cf.~the examples of oblique 2-D shock
tubes in {\footnotesize ATHENA}; see also \citealp{sij12}).  
For discussions on symmetric sweeping we
refer the reader to \citet{alo99} and \citet{sut10}.

Although the amplification of asymmetries in the flow can be favored
by a moving-mesh, they are {\it not generated} by the mesh itself in
our planet-disc interaction simulations.  To test this, we compare the
density field in our fiducial inviscid simulations ($\mu=10^{-4}$ and
$\mu=10^{-3}$) at different times, looking for transients.  Aside from
a gradually carved shallow gap, the background density in the
Neptune-mass case is nearly stationary over a period of 100 orbits,
with no sign of instabilities or transient over-densities. In the
Jupiter-mass simulation, on the other hand, the density field outside
the gap shows variations that can be of the order of the density peaks
associated with the spiral wake.  Thus, the development of strong
density bumps and vortices outside the gap in the Jupiter-mass case is
not a feature of {\footnotesize AREPO} but a consequence of the large
gap carved out by the planet, corresponding to the ``edge
instabilities" previously observed in numerical simulations by
\citet{kol03}, \citet{li05} and \citet{val06}.  These edge effects are
usually associated with vortices right outside the gap
\citep{val07,lyr09}, which are present in our Jupiter-mass
simulations, but absent in the Neptune-mass runs (see
Section~\ref{sec:vortensity} below).

{As seen from the right panels of Figure~\ref{fig:density_polar}, the Jupiter-mass 
{\footnotesize AREPO} runs are at most as unstable as the 
{\footnotesize FARGO} runs. We have confirmed that this effect increases with resolution and
eventually happens even for the Neptune-mass simulation with
both {\footnotesize AREPO} and {\footnotesize FARGO}. 
Therefore,} we conclude that the sensitivity to the development of asymmetries
does not arise spontaneously in {\footnotesize AREPO}, but only in
hydrodynamically unstable regimes. The importance of the preservation
of symmetries (which do not occur in nature) well into the non-linear
regime in an unstable configuration is a matter of debate among
researchers \citep[e.g.,][]{mcn12b}.  Although we do not believe it is
a critical component of a hydrodynamical scheme, future work should
address pathological cases in which grid noise might affect the
convergence rate in a code like {\footnotesize AREPO}.

\begin{figure*} 
\centering
\includegraphics[width=0.32\textwidth]{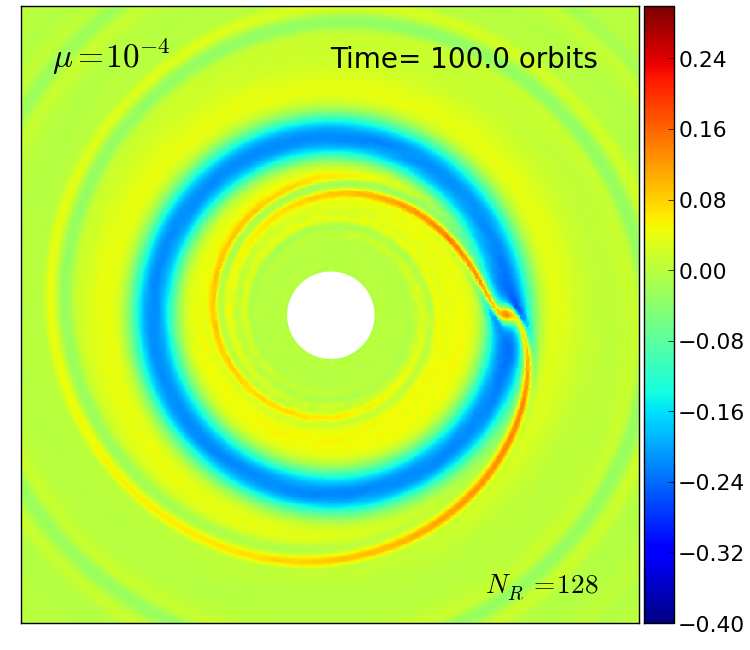} 
\includegraphics[width=0.32\textwidth]{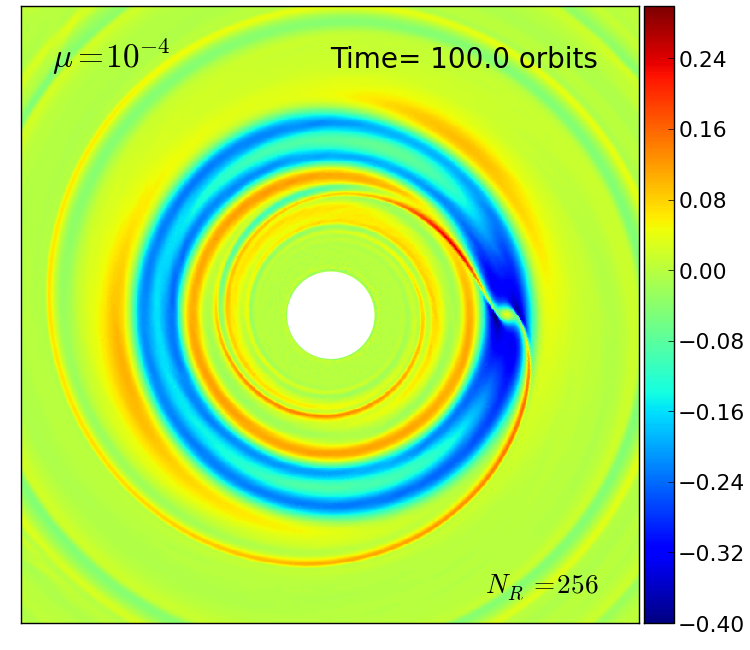}
\includegraphics[width=0.32\textwidth]{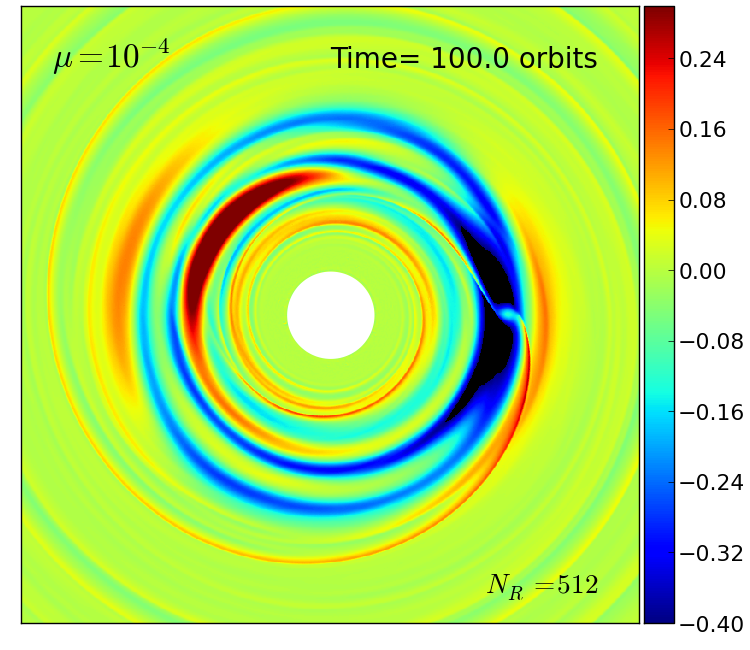}\\
\includegraphics[width=0.32\textwidth]{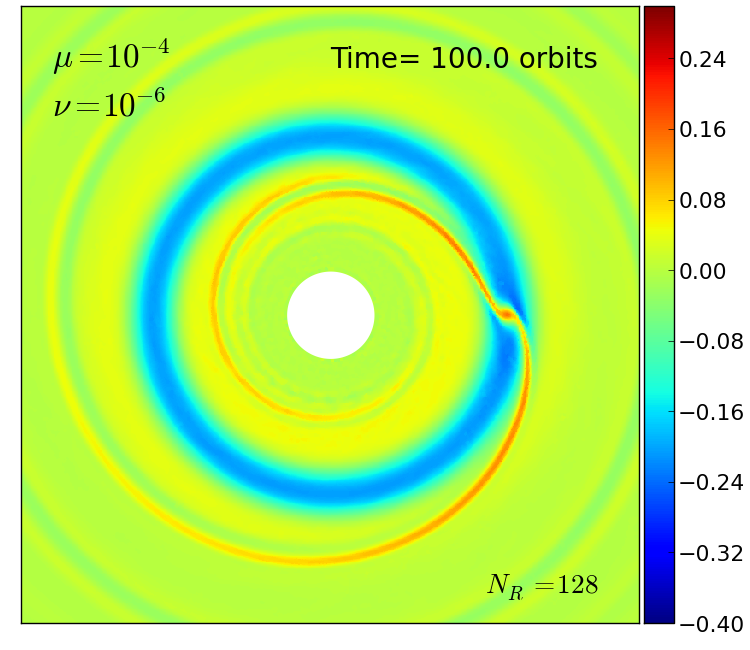}
\includegraphics[width=0.32\textwidth]{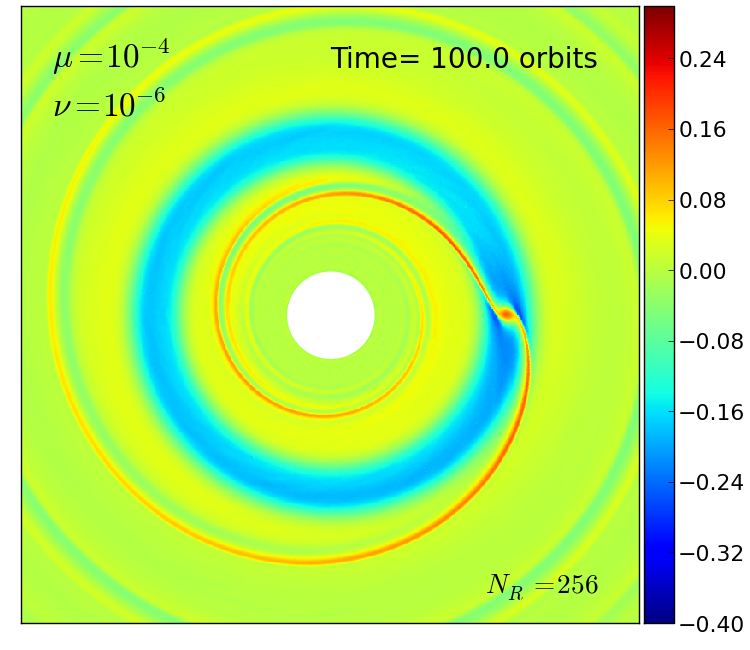}
\includegraphics[width=0.32\textwidth]{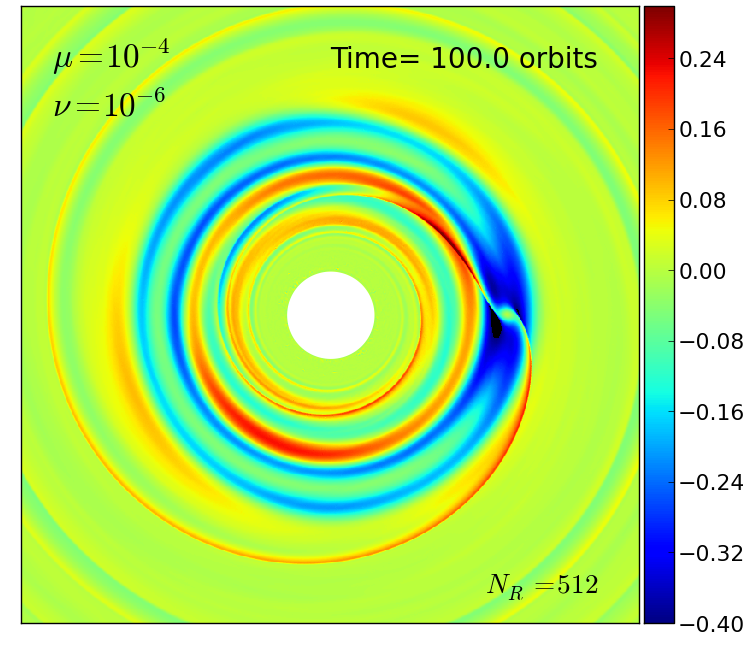}\\
\includegraphics[width=0.32\textwidth]{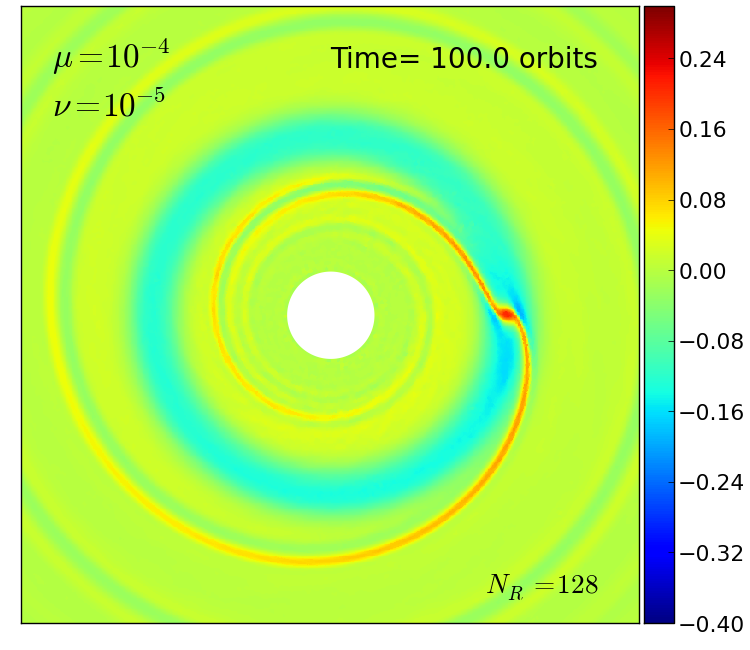}
\includegraphics[width=0.32\textwidth]{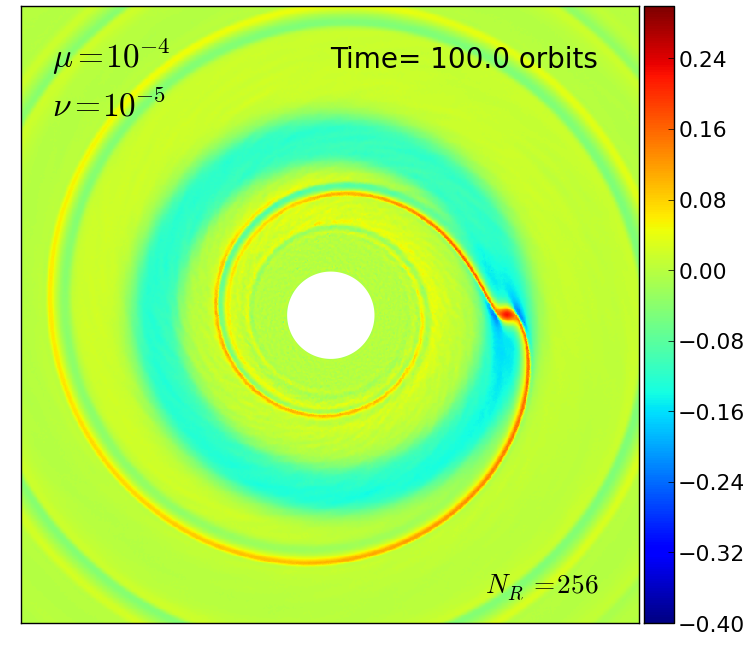}
\includegraphics[width=0.32\textwidth]{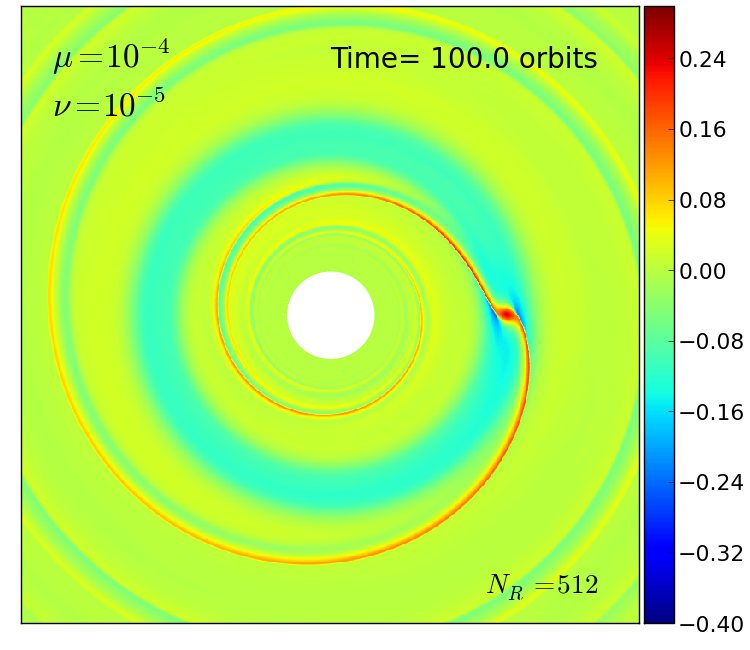}\\
\includegraphics[width=0.32\textwidth]{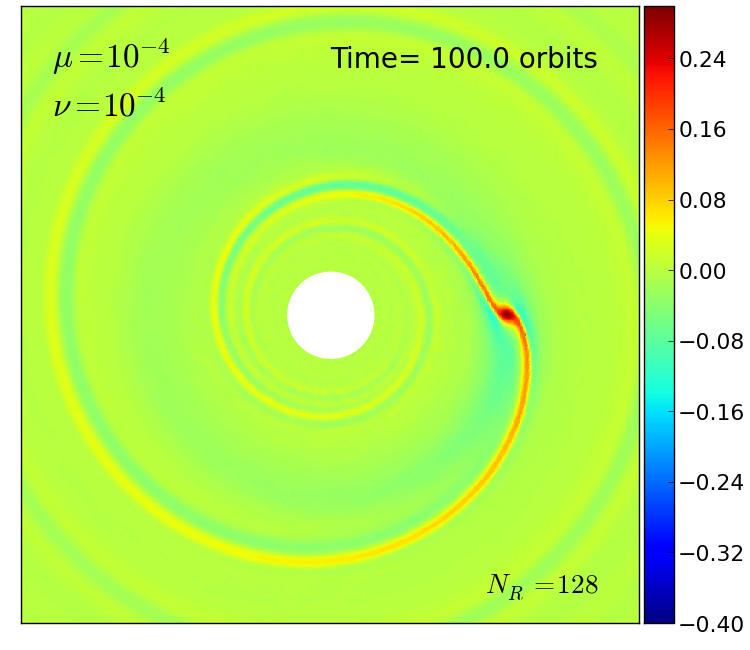}
\includegraphics[width=0.32\textwidth]{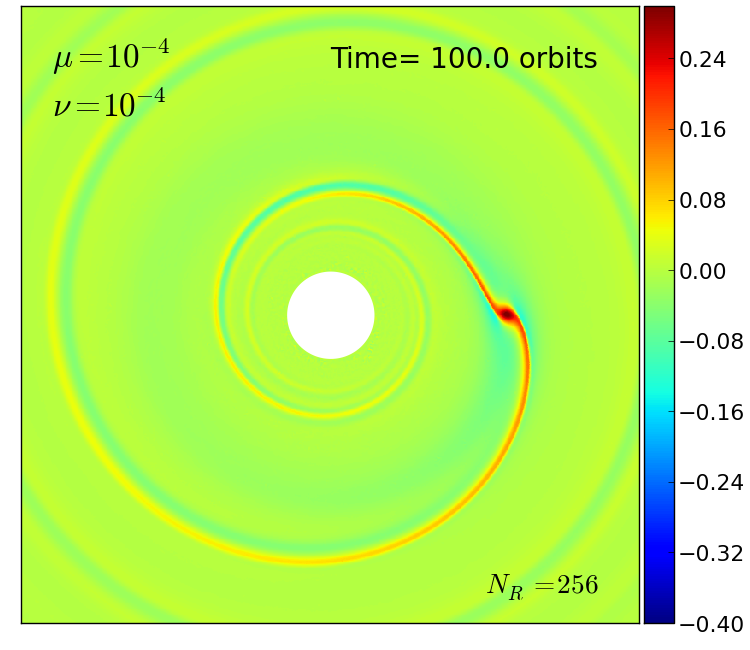}
\includegraphics[width=0.32\textwidth]{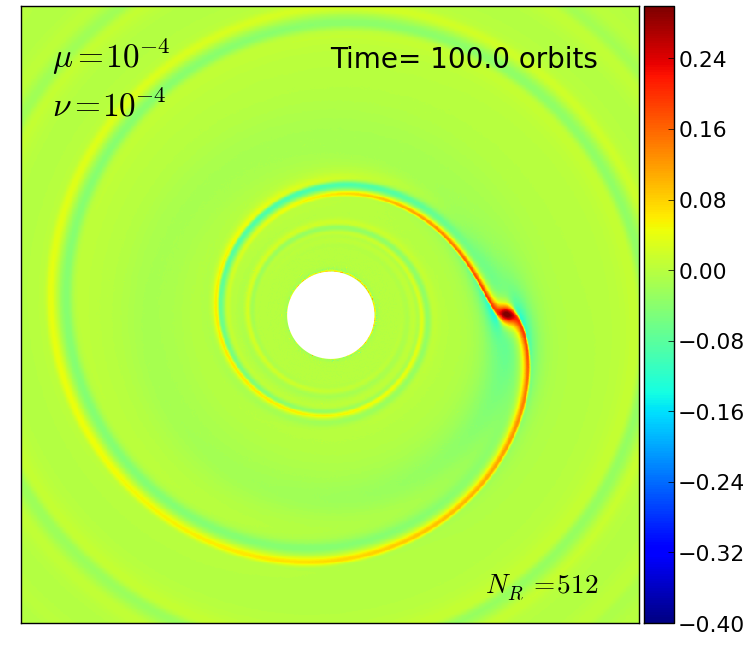}
\caption{Surface density fields for Neptune-mass ($\mu=10^{-4}$)
  simulations with different resolutions (for different columns) and
  physical viscosity coefficients (for different rows). The first,
  second and third column have 128, 256 and 512 radial zones,
  respectively. The first row corresponds to inviscid (i.e., no
  explicit viscosity terms) runs; the second, third and fourth rows
  have viscosity coefficients of $\nu=10^{-7}$, $10^{-6}$ and
  $10^{-4}$, respectively.
\label{fig:neptune_density_frames}}
\end{figure*}

\begin{figure*}
\centering
\includegraphics[width=0.32\textwidth]{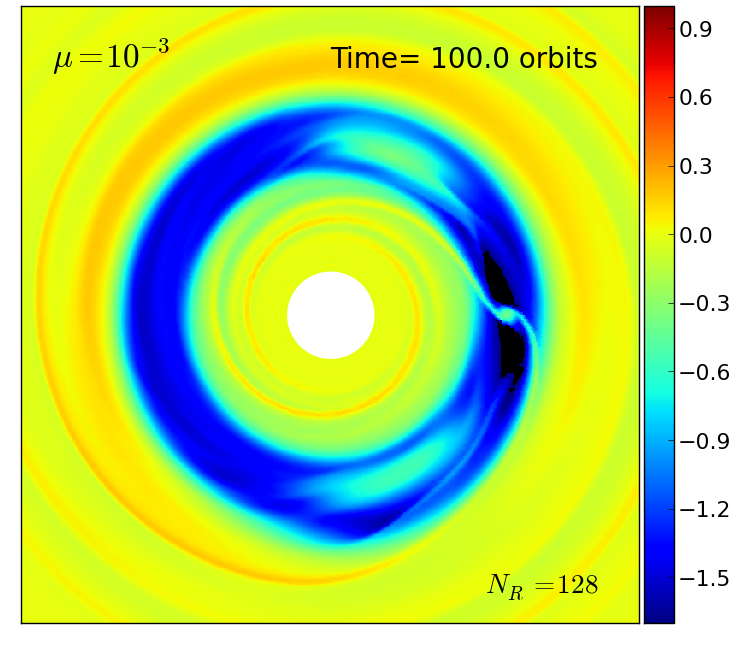} 
\includegraphics[width=0.32\textwidth]{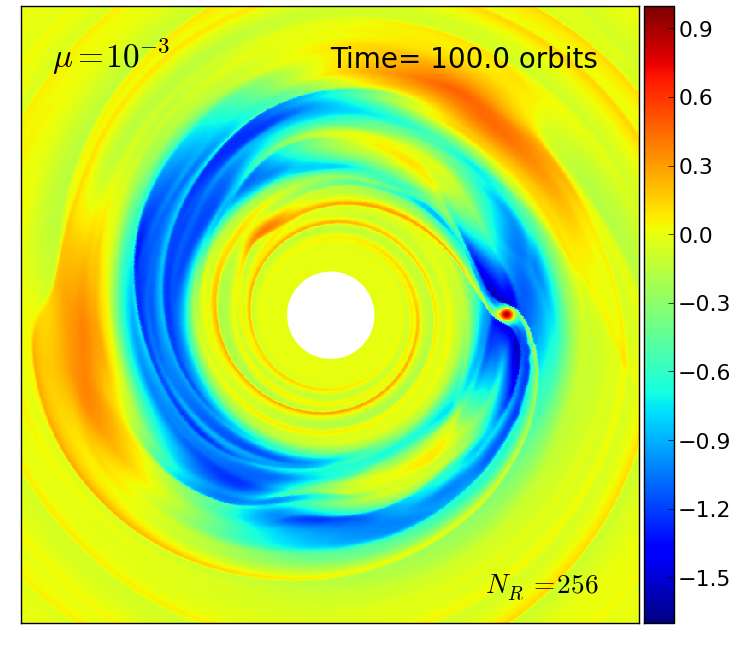}
\includegraphics[width=0.32\textwidth]{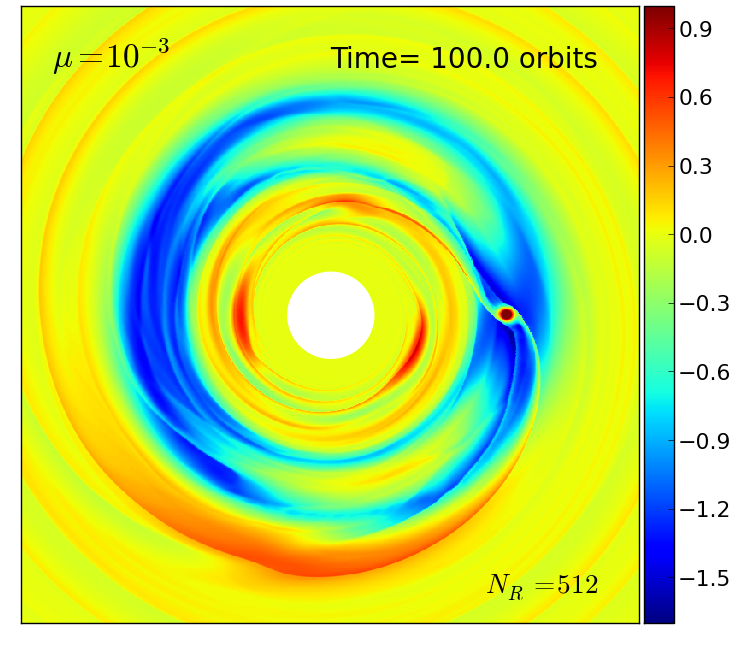}\\
\includegraphics[width=0.32\textwidth]{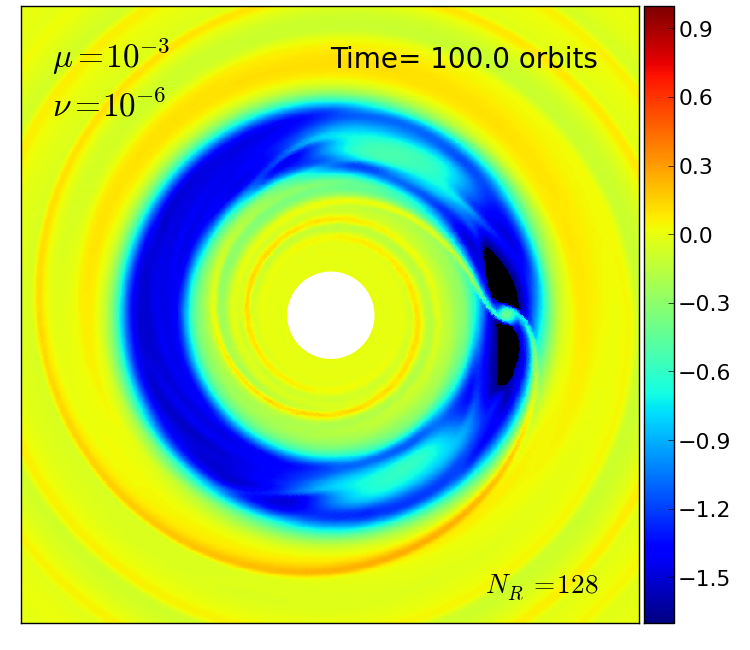}
\includegraphics[width=0.32\textwidth]{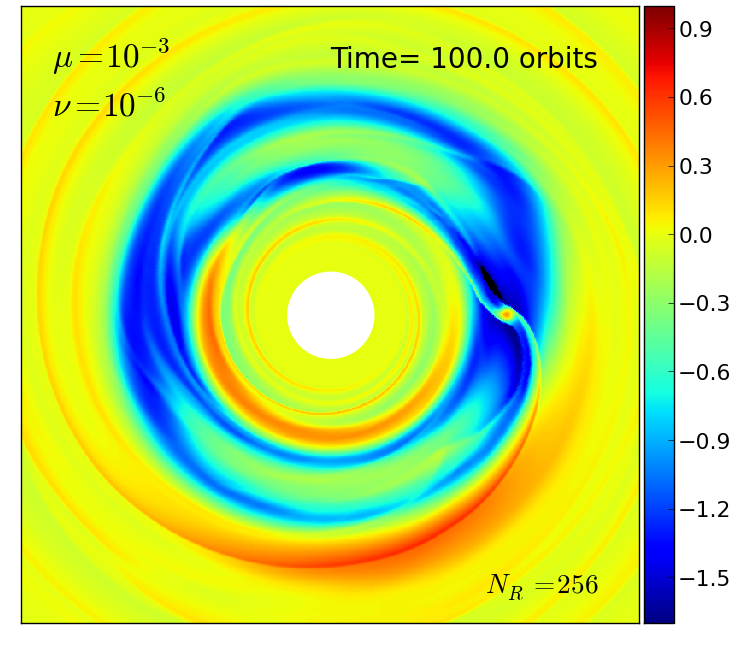}
\includegraphics[width=0.32\textwidth]{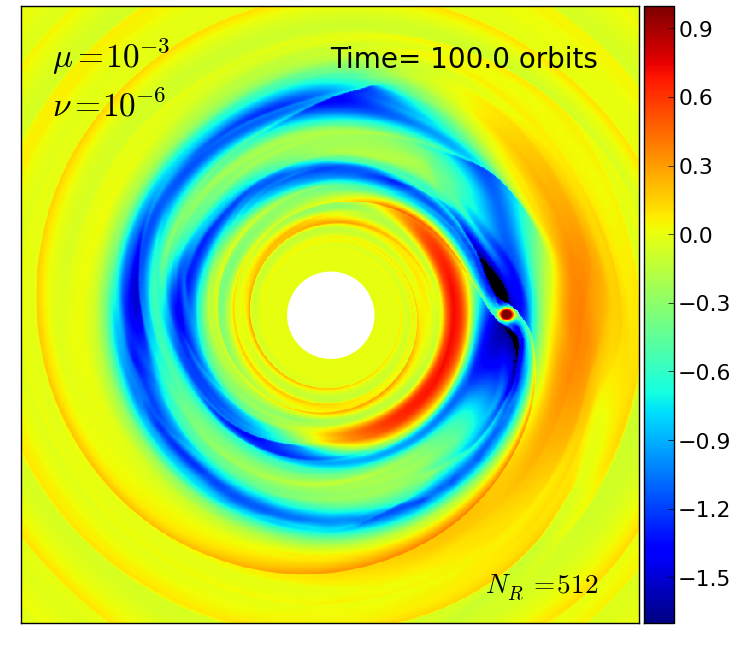}\\
\includegraphics[width=0.32\textwidth]{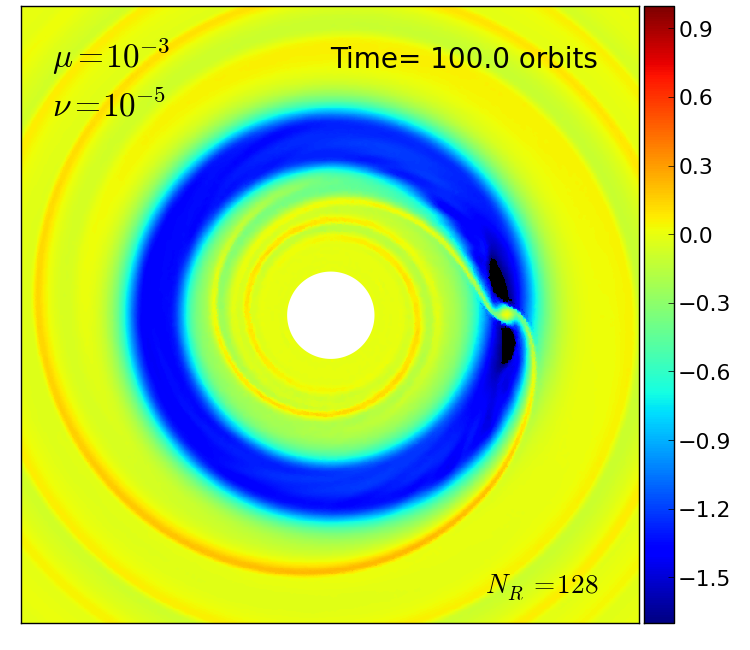}
\includegraphics[width=0.32\textwidth]{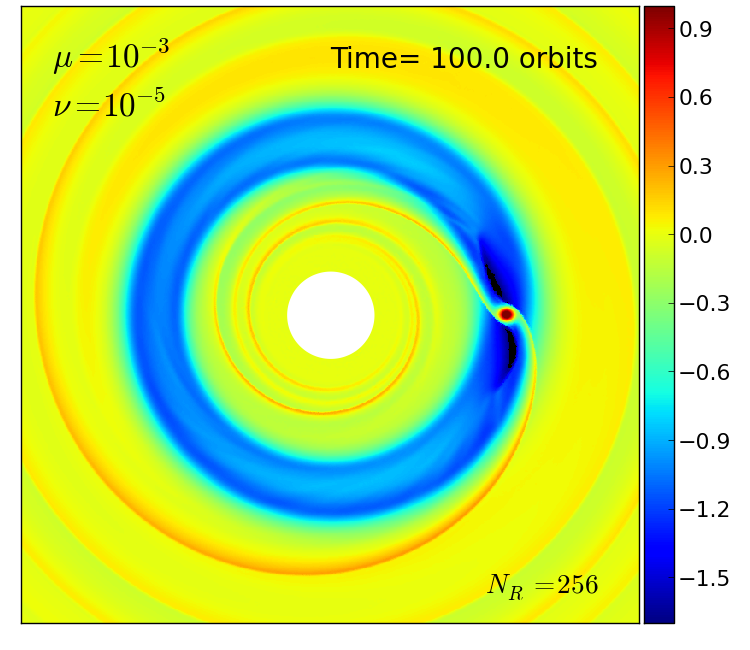}
\includegraphics[width=0.32\textwidth]{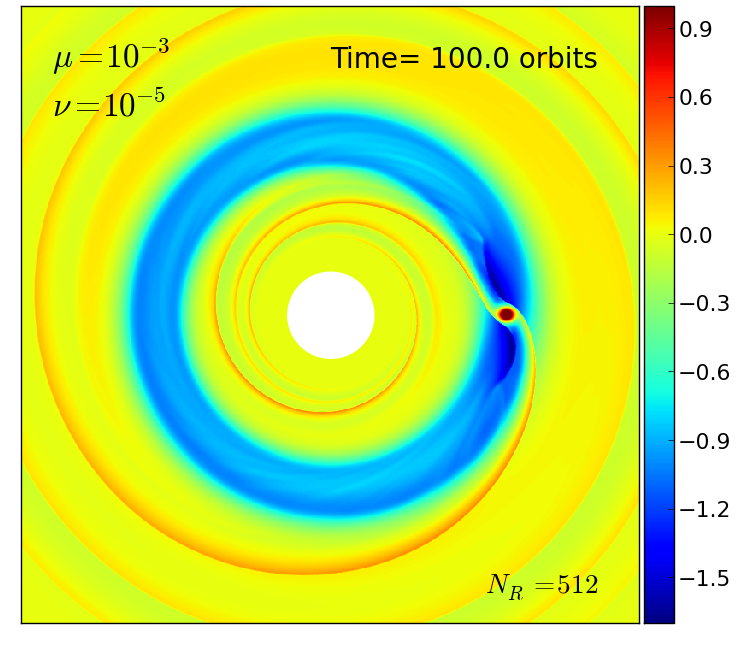}\\
\includegraphics[width=0.32\textwidth]{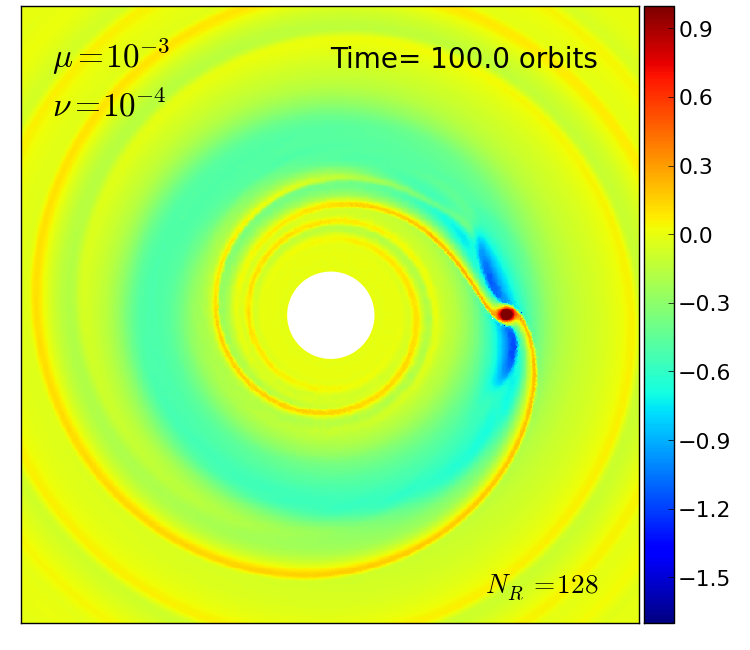}
\includegraphics[width=0.32\textwidth]{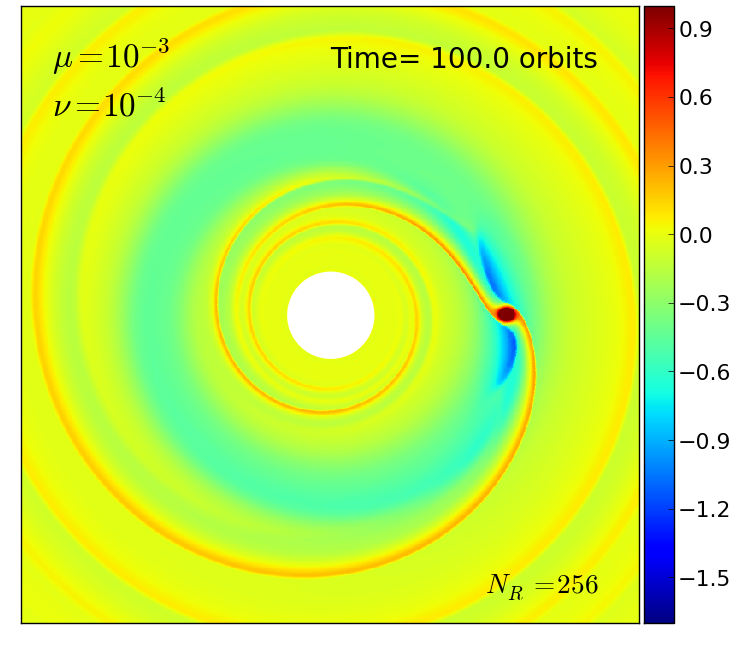}
\includegraphics[width=0.32\textwidth]{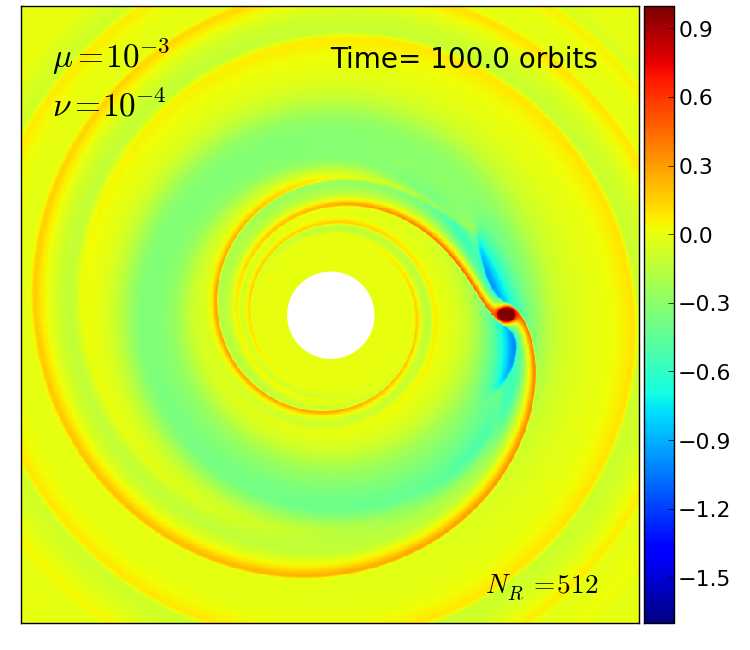}
\caption{Same as Figure~\ref{fig:neptune_density_frames} but for a Jupiter-mass
planet ($\mu=10^{-3}$).
\label{fig:jupiter_density_frames}}
\end{figure*}

\subsection{Effects of viscosity and higher resolution}\label{sec:viscosity_resolution}
Viscosity (or viscosity-like numerical diffusivity) not only affects
the mechanism by which gaps form, but also the long term balance
between the tidal torque and the viscous torque
\citep{cri06,cri07}. Balancing the angular momentum flux of these two
competing effects, \citet{zhu13} find a ``viscous criterion" for gap formation
of the form
\begin{equation}\label{eq:viscous_criterion}
\left(\frac{M_p}{M_*}\right)^2\gtrsim \frac{10\nu H_p^3}{R_p^5\Omega_p}~~,
\end{equation}
which, for the disc parameters of this work, reduces to
${M_p}/{M_*}\gtrsim \sqrt{1.25\times10^{-3}\nu}$.  This inequality
means that the critical mass ratio $\mu\approx{M_p}/{M_*}$ should be
above $\sim1.11\times10^{-4}$ for a viscosity coefficient of
$\nu=10^{-5}$ ($\alpha\approx4\times10^{-3}$ at the planet position);
i.e., the Neptune-mass case is again border-line capable of opening a
gap.

Figure~\ref{fig:NEP_gap_profile} shows the azimuthally-averaged
surface density field for the inviscid (top panel) and the viscous
(bottom panel) Neptune simulations after 100 orbits for three
different resolutions: $N_R=128$, 256 and 512. The surface density
profiles include a 1-$\sigma$ contour to quantify the degree of
axisymmetry in the disc. There is
no convergence in the inviscid simulations, while very good
consistency is found in the viscous runs across the different
resolutions, in addition to uniformity in the azimuthal direction.

Similarly, Figure~\ref{fig:JUP_gap_profile} shows the
azimuthally-averaged surface density profile for the Jupiter-mass
simulations for three different resolutions. 
In the inviscid case, the edge instabilities cause much
wilder variations than in the Neptune analog, and grow in amplitude
as resolution is increased, eventually affecting both inner and outer
edges of the gap. In the
viscous case, the flow is smoother and deviations from the azimuthal
mean are smaller, although the 1-$\sigma$ contours still show broader
scatter than in the Neptune-mass case.  This lack of a converging trend
is probably due to the fact that the edge instabilities are not entirely suppressed by a viscosity
of $\nu=10^{-5}$, and that the gap is deep enough to make the
instability effective.

A grid of models with varying resolution ($N_R=128$, 256 and 512) and viscosity
coefficient ($\nu=0$, $10^{-6}$,  $10^{-5}$ and $10^{-4}$) is shown in
Figures~\ref{fig:neptune_density_frames} and~\ref{fig:jupiter_density_frames}.
According to the viscous gap-opening criterion of Equation~(\ref{eq:viscous_criterion}),
we find that the critical mass ratio $M_p/M_*$ is $3.5\times10^{-4}$ and $3.5\times10^{-5}$,
for  $\nu=10^{-4}$ and  $\nu=10^{-6}$, respectively. Therefore, with $\nu=10^{-4}$
($\alpha\approx4\times10^{-2}$ at $R=1$), gap formation should be suppressed
for the Neptune-mass planet and significantly affected for the Jupiter-mass planet. 
These very viscous runs (bottom rows of Figures~\ref{fig:neptune_density_frames} 
and~\ref{fig:jupiter_density_frames}) show a high degree of smoothness,
and no evident dependence on resolution, even for the Jupiter-mass case.

For $\nu=10^{-6}$
($\alpha\approx4\times10^{-4}$  at $R=1$), both the Neptune-mass and the
Jupiter-mass planets should be able to open partial or full gaps,
provided numerical diffusion is less important than physical viscosity.  This is easily
satisfied by the Jupiter-mass run at all explored resolutions (second row
from top in Figure~\ref{fig:jupiter_density_frames}), but not so clearly for the
Neptune-mass
case, for which the $N_R=128$ run is indistinguishable from the inviscid run,
while the $N_R=512$ version is able to open a sharp gap.

\begin{figure}
\includegraphics[width=0.48\textwidth]{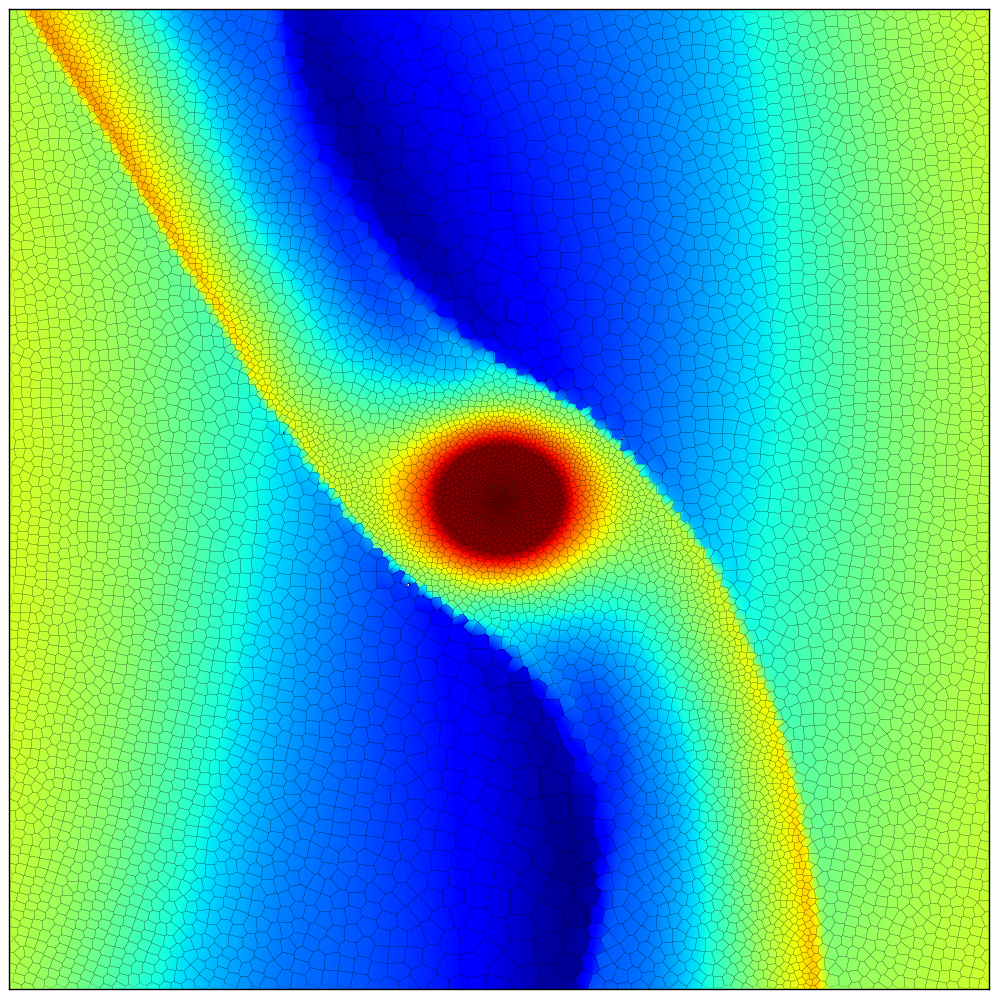}
\includegraphics[width=0.48\textwidth]{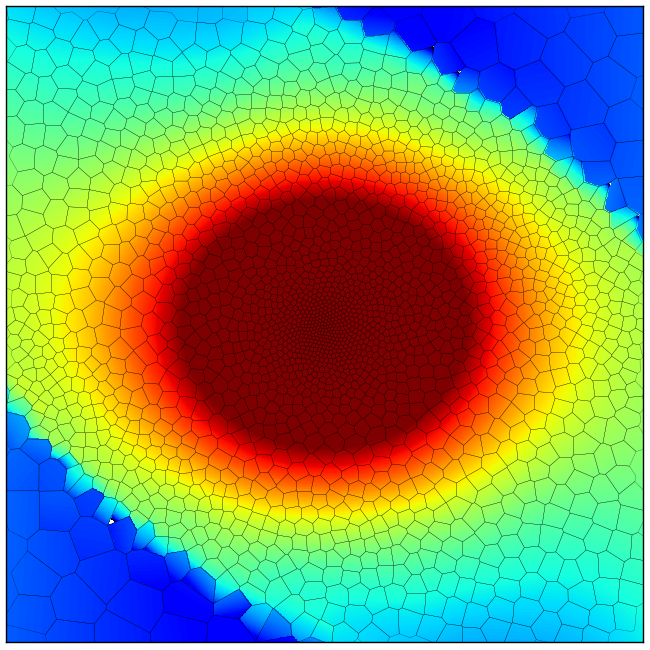}
\caption{Density field and mesh structure of a viscous Jupiter run
  (simulation \texttt{JUP-VISC-B\_512}) close to the planet. Top
  panel: the visualized region covers a square area of side length
  equal to $0.5$ (in units where $R_p=1$), centered on the planet
  position at $t=100$ orbits. Bottom panel: zoomed-in region of side
  length equal to $0.14$, roughly twice the planet's Hill sphere. The
  concentration of cells toward the potential well minimum increases
  the resolution within the Hill sphere roughly by a factor of $\sim
  20^2$-$30^2$ with respect to the initial cell spacing. In this case,
  the Hill radius is $(\mu/3)^{1/3}\sim0.07$, corresponding to 4, 8
  and 16 cells across at $t=0$ for the runs with $N_R=128$, 256 and
  512, respectively.  However, after 100 orbits, the Hill sphere can
  have up to 200 cells across the disc in the viscous run with
  $\nu=10^{-5}$ and $N_R=512$. If the entire simulation had been run
  with a globally uniform cell size like that obtained within the Hill
  sphere, the number of radial zones would have been $N_R=6500$, a
  resolution comparable to the one used by \citet{don11a,don11b} and
  \citet{duf12}.
\label{fig:planet_wmesh}}
\end{figure}

The quasi-Lagrangian nature
of our scheme can be seen in the very high density of gas at the
planet position (Figure~\ref{fig:JUP_gap_profile}).  This effect
appears to be enhanced with increasing viscosity. Since high-viscosity
(e.g. second row from the top in
Figure~\ref{fig:jupiter_density_frames}) suppresses the formation of
deep gaps, the gas supply onto the planet is not halted and thus
sustained accretion can proceed.  This accretion of gas translates to
a high concentration of cells within the planet's Hill sphere
(Figure~\ref{fig:planet_wmesh}).

Interestingly, this property of moving-mesh schemes opens up new
possibilities to overcome the strict resolution requirements imposed
by the detailed hydrodynamics of gap clearing.  For example,
Figure~\ref{fig:planet_wmesh} shows a high concentration of cells
close to the planet, which allows for the study of circumplanetary
discs within {\it global} circumstellar disc simulations, combining a
coarse large-scale disc with a very high resolution Hill sphere
region, thus concentrating all the computational power in the regions
of interest; this provides an alternative to attempts to implement adaptive
mesh refinement in cylindrical coordinates, which has resulted to be very difficult 
in the past. {Recently, adaptive refinement has been
successfully implemented in cylindrical codes by \citet{gre13},
allowing for unprecedentedly detailed, high-resolution studies of circumplanetary flow.
Of course, a refined polar grid at the scales of the planet's Hill sphere
is still nearly cartesian, a restriction that does not exist for the moving mesh.}

Similarly, Figure~\ref{fig:planet_wmesh} shows how the spiral wakes
also concentrate a larger number of cells than the background
flow. This property of moving-mesh methods can provide an alternative
to the extremely high resolution studies that have been carried out by
\citet{don11a,don11b} and \citet{duf12}, concentrating the resolution
elements specifically on the wake.

\begin{figure*}
\centering
\includegraphics[width=0.33\textwidth]{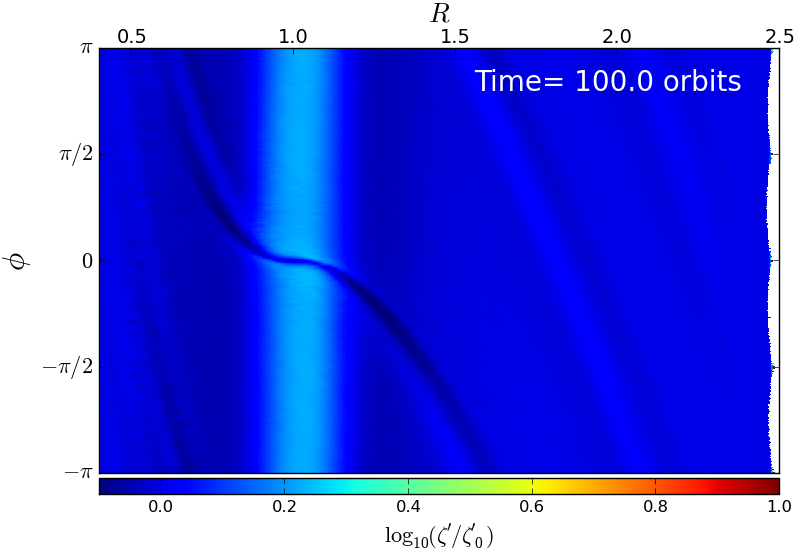}
\includegraphics[width=0.33\textwidth]{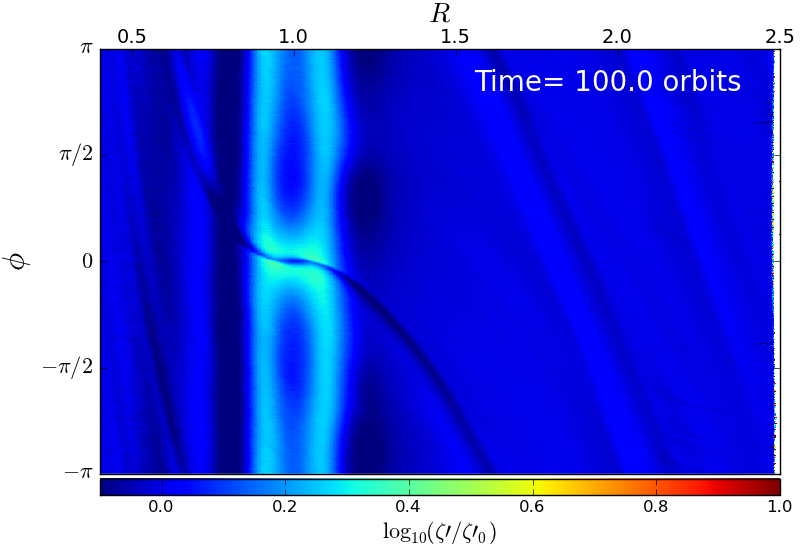}
\includegraphics[width=0.33\textwidth]{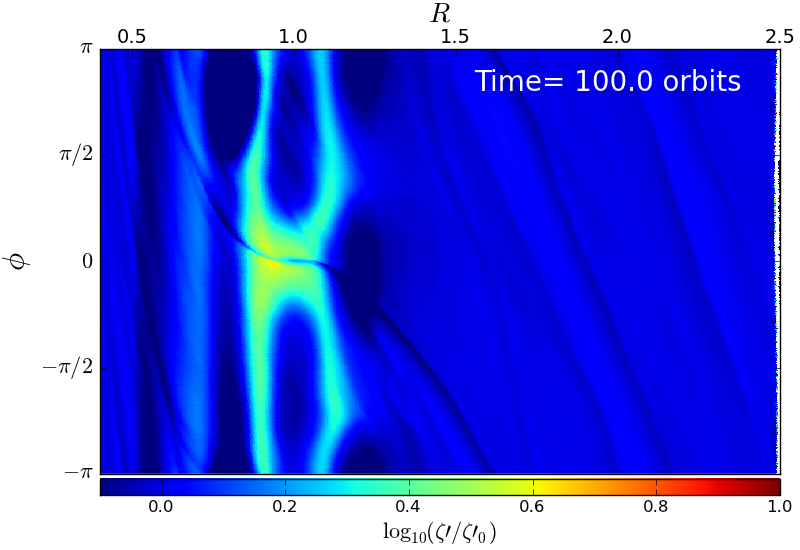}
\includegraphics[width=0.33\textwidth]{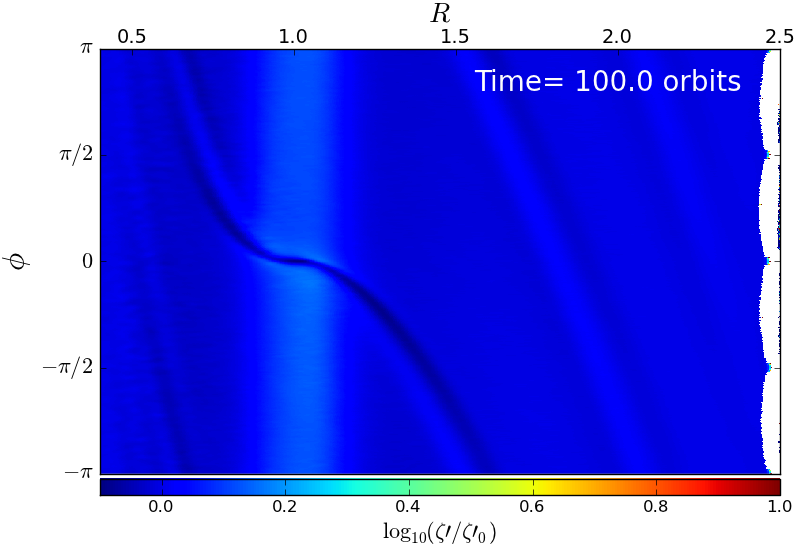}
\includegraphics[width=0.33\textwidth]{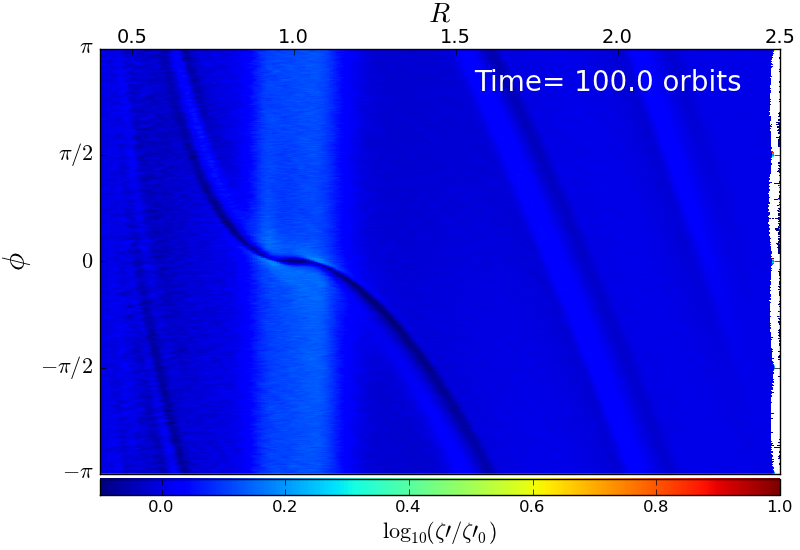}
\includegraphics[width=0.33\textwidth]{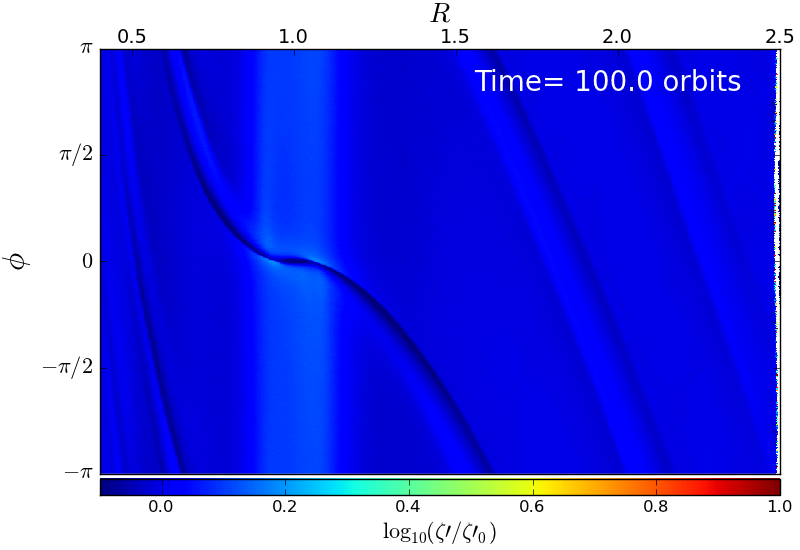}
\caption{Deviations in the vortensity field (time $t=100$ orbits) for
  the Neptune-mass simulations at three different resolutions
  including the inviscid and viscous cases. Top panels: inviscid
  Neptune runs with resolutions (left to right) $N_R=$128, 256 and
  512.  Bottom panels: viscous ($\nu=10^{-5}$) Neptune runs with
  resolutions (left to right) $N_R=$128, 256 and 512.
\label{fig:neptune_vortensity}}
\end{figure*}

\begin{figure*}
\centering
\includegraphics[width=0.33\textwidth]{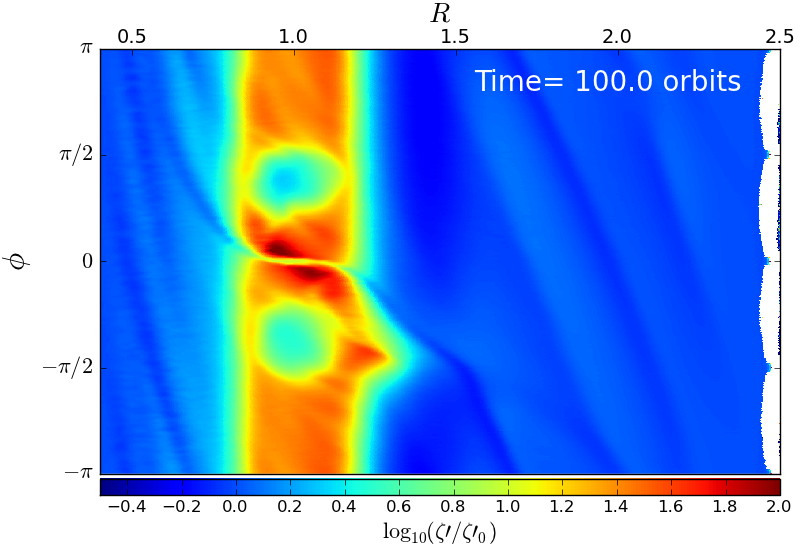}
\includegraphics[width=0.33\textwidth]{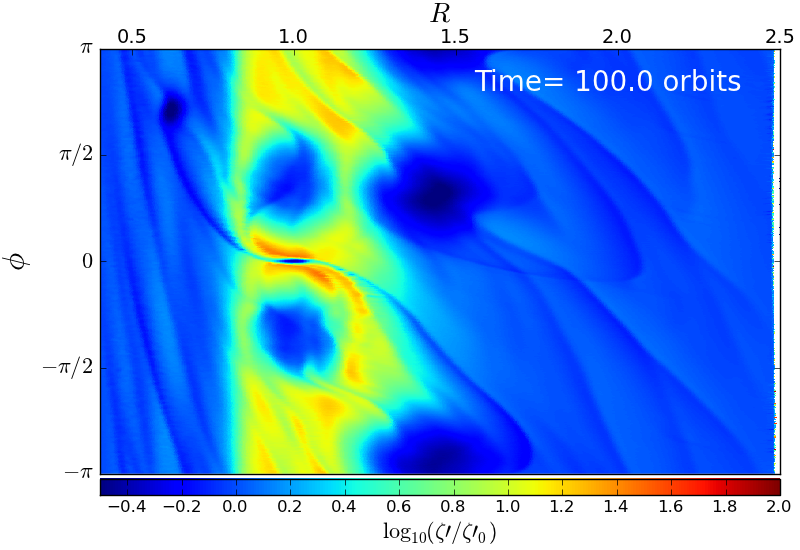}
\includegraphics[width=0.33\textwidth]{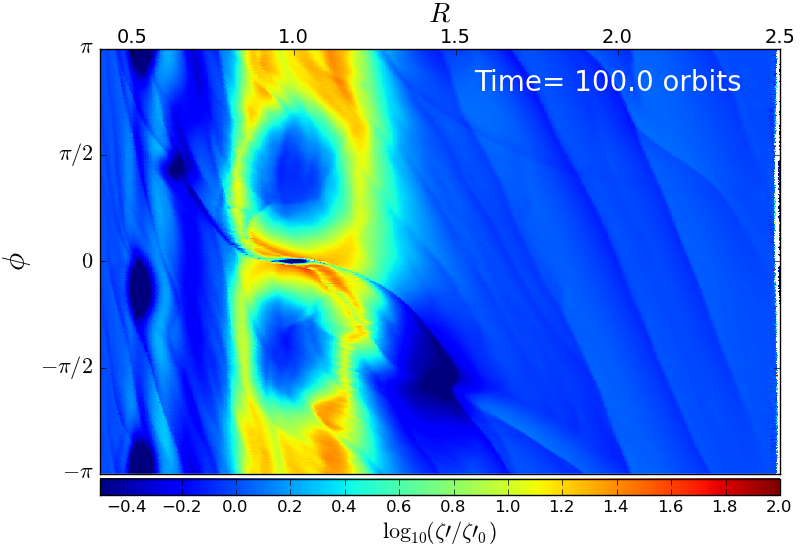}
\includegraphics[width=0.33\textwidth]{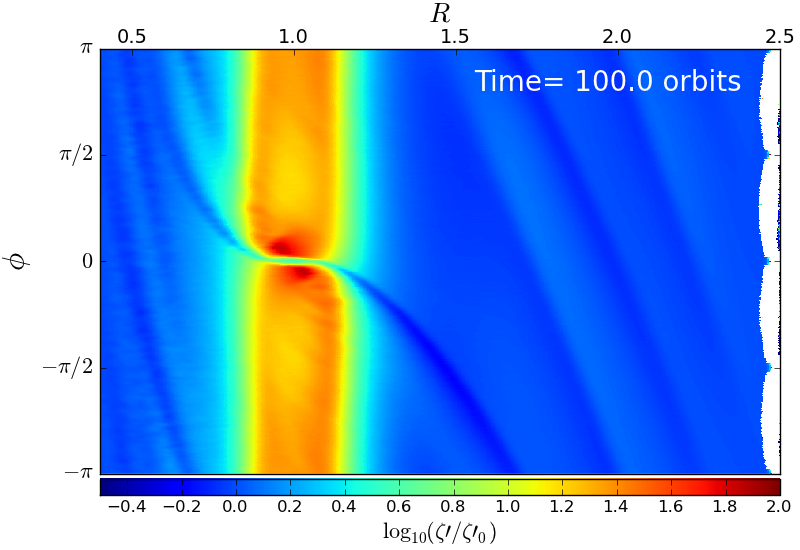}
\includegraphics[width=0.33\textwidth]{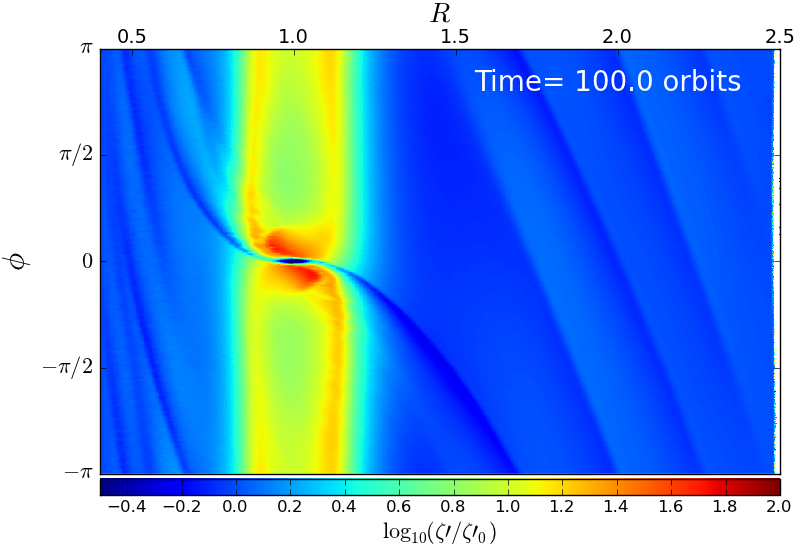}
\includegraphics[width=0.33\textwidth]{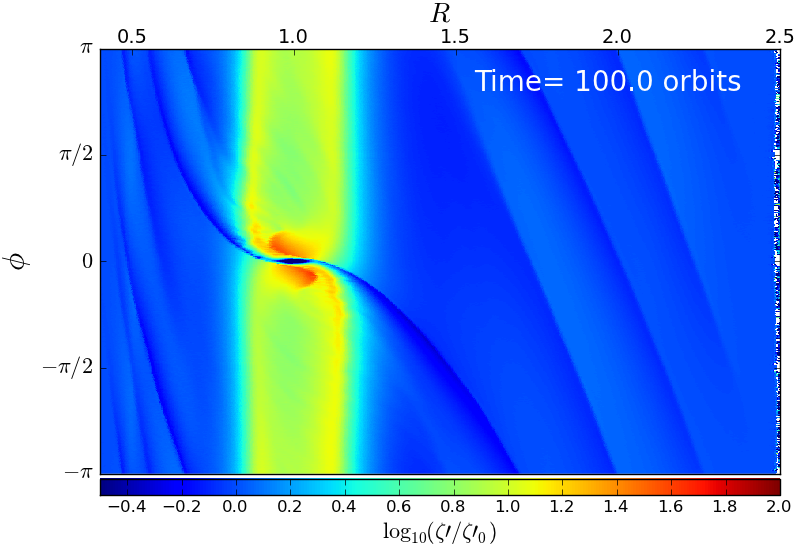}
\caption{Same as Figure~\ref{fig:neptune_vortensity}, but for the
  Jupiter-mass runs.  Top panels: inviscid Jupiter runs with
  resolutions (left to right) $N_R=$128, 256 and 512.  Bottom panels:
  viscous ($\nu=10^{-5}$) Jupiter runs with resolutions (left to
  right) $N_R=$128, 256 and 512.
\label{fig:jupiter_vortensity}}
\end{figure*}

\subsection{Vortensity Field}\label{sec:vortensity}

{The vorticity field provides another powerful numerical diagnostic in planet-disc interaction simulations.
In particular, we are interested in the vortensity (or "potential vorticity") 
field $\zeta$, defined as $\zeta\equiv\omega_z/\Sigma$, where $\omega_z=|\nabla\times\mathbf{v}|$,
since the radial variability of this quantity is what is the source of edge instabilities
observed in gap-opening simulations. }

\citet{lov99} and \citet{li00} found that a Keplerian disc is unstable
to azimuthal perturbations when the following quantity reaches a local
extremum
 \begin{equation}
 \nonumber
 \mathcal{L}(R)\equiv\mathcal{F}(R)(P\Sigma^{-\gamma})^{2/\gamma}.
 \end{equation}
In a barotropic disc, $\mathcal{F}(R)=\Sigma\Omega\kappa^{-2}$, where
$\Omega$ is the orbital angular velocity and $\kappa=[R({\rm
    d}\Omega^2/{\rm d}R)+4\Omega^2]^{1/2}$ is the epicyclic frequency,
which can be related to the $z$-component of the vorticity for {\it
  axisymmetric flow} by $\omega_z=(1/R)\partial(R^2\Omega)/\partial
R=\kappa^2/(2\Omega)$.  Therefore, for the globally isothermal case
(i.e., $\gamma=1$),
 \begin{equation}\label{eq:rwi}
 \mathcal{L}(R)\propto\frac{\Sigma}{2\omega_z}=\frac{1}{2\zeta}~~.
 \end{equation}

Equation~(\ref{eq:rwi}) implies that local
extrema in the radial vortensity profile can trigger 
the Rossby wave instability \citep[][]{lov99,li00,li01,tag01},
which saturates into vortices that
are efficient at destroying the extremum in
vortensity \citep[e.g.][]{meh10}. However, the presence of a massive
planet can sustain the density bump, enabling a sustained production
of vortices \citep{val07,lyr09}. {Thus, a numerical scheme's capability
of producing steep vortensity gradients at the edge of a planet-carved
gap translates directly into the production of the edge instabilities observed
in the surface density field.}

Since our simulations are run using an
inertial reference frame, we obtain the {\it absolute} vorticity
$\nabla\times\mathbf{v}$ directly from the interpolated velocity
field. In order to obtain the {\it relative} vorticity (i.e., the
vorticity as seen in the frame rotating with the planet) we simply
subtract $2\Omega_p$, i.e., the vorticity associated with the
background flow. The relative vortensity $\zeta'$ is then
\begin{equation}
\zeta'=\frac{|\nabla\times\mathbf{v}'|}{\Sigma}=\frac{|\nabla\times\mathbf{v}-2\Omega_p\hat{\mathbf{k}}|}{\Sigma},
\end{equation}
which we normalize by $\zeta_0'\equiv\zeta'(t=0)$.

We calculate $\zeta'/\zeta'_0$ for
three different resolutions in the inviscid Neptune, the viscous
Neptune, the inviscid Jupiter and the viscous Jupiter cases. The
normalized vortensity fields are shown in
Figure~\ref{fig:neptune_vortensity} for Neptune, and
Figure~\ref{fig:jupiter_vortensity} for Jupiter. As discussed in
Section~\ref{sec:gap_opening}, the inviscid runs should be expected to
converge slowly, since higher resolution runs will be effectively less
diffusive, and the appearance of vortices should increase with
increasing resolution, as a consequence of sharper gaps. Indeed, both
inviscid examples (top rows in Figures~\ref{fig:neptune_vortensity}
and~\ref{fig:jupiter_vortensity}) show an increase in the number of
vortices every time the resolution is improved. The inviscid Neptune
runs at $N_R=256$ and $N_R=512$ show the vortensity rings described by
\citet{lin10}. The different vortensity fields for the viscous runs
are nearly indistinguishable from each other, confirming that a
viscosity of $\nu=10^{-5}$ is enough to suppress the shock-induced
generation of vortensity.  We do expect the viscous runs to show some
degree of convergence. Interestingly, there is consistency between the
$N_R=256$ and $N_R=512$ run, for both the Neptune and Jupiter cases,
but the $N_R=128$ case clearly stands out as unconverged. This
confirms our previous observation that 128 radial zones might not be
enough to capture the global flow properly, especially if vortices are
expected to develop.

Some of the sharp vortensity features obtained by other cylindrical
grid codes cannot be reproduced entirely here at the
fiducial resolution of $N_R=128$ (left panels in
Figures~\ref{fig:neptune_vortensity}
and~\ref{fig:jupiter_vortensity}). Indeed, the vortensity peaks found
near the edges of the partial gap in the Neptune simulations are
shallower in our example. In addition, we see less structure within
the gap. The Jupiter run, on the other hand, does show a sharp
transition in vortensity across the edge of the gap, and succeeds in
capturing the vortensity ``islands" at the $L_4/L_5$ Lagrange
points, which is not achieved by all codes 
\citep[see, for example,][]{val06}. Also, the vortensity field
is devoid of reflections from the boundaries, which shows that our
absorbing boundary condition is effective at eliminating such
artifacts. The vortensity features in inviscid runs with
$N_r=256$ (for both Neptune and Jupiter examples) are significantly
sharper and richer than those of \citet{val06}.  Therefore, we believe
that to achieve results comparable to those of well established codes for
planet-disc interaction, we would need a number of radial zones lower than 256
but higher than 128.  We also point out the effective azimuthal resolution at $R=R_p$
in our fiducial runs is $\sim360$, still below the $N_\phi=384$ used in the 
{\footnotesize FARGO} simulations we have included for comparison, as well
as in the simulations of \citet{val06}.

\begin{figure}
\centering
\includegraphics[width=0.38\textwidth]{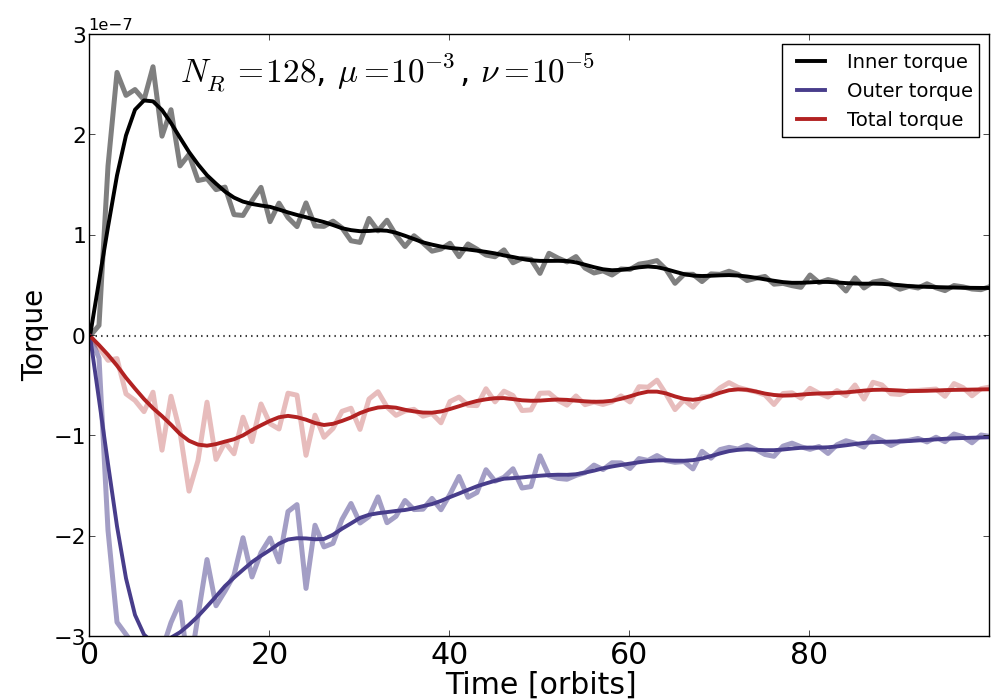}
\includegraphics[width=0.38\textwidth]{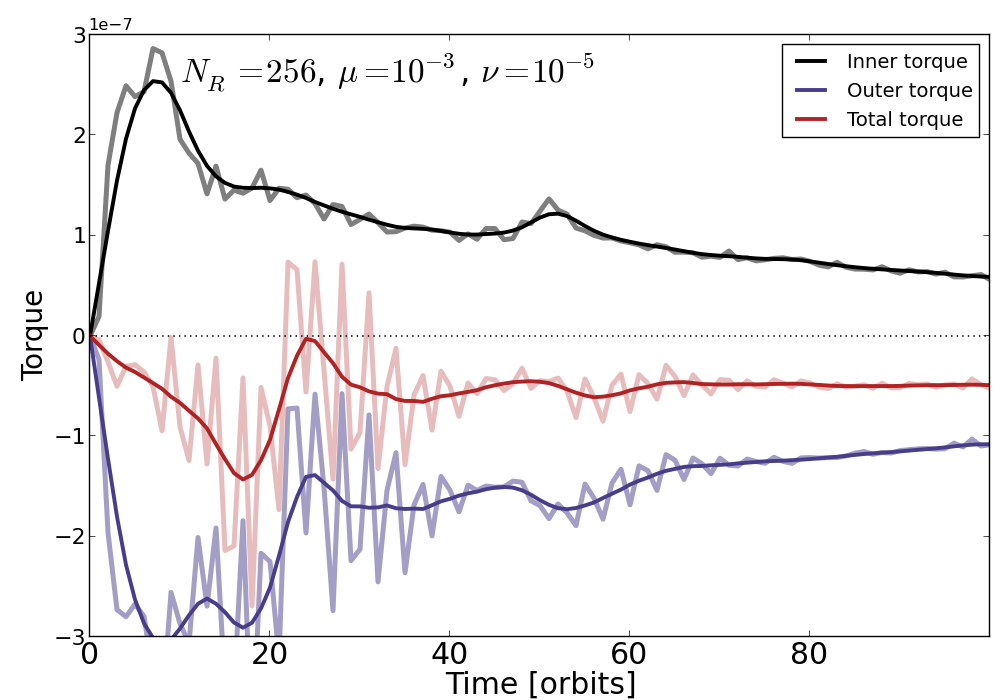}
\includegraphics[width=0.38\textwidth]{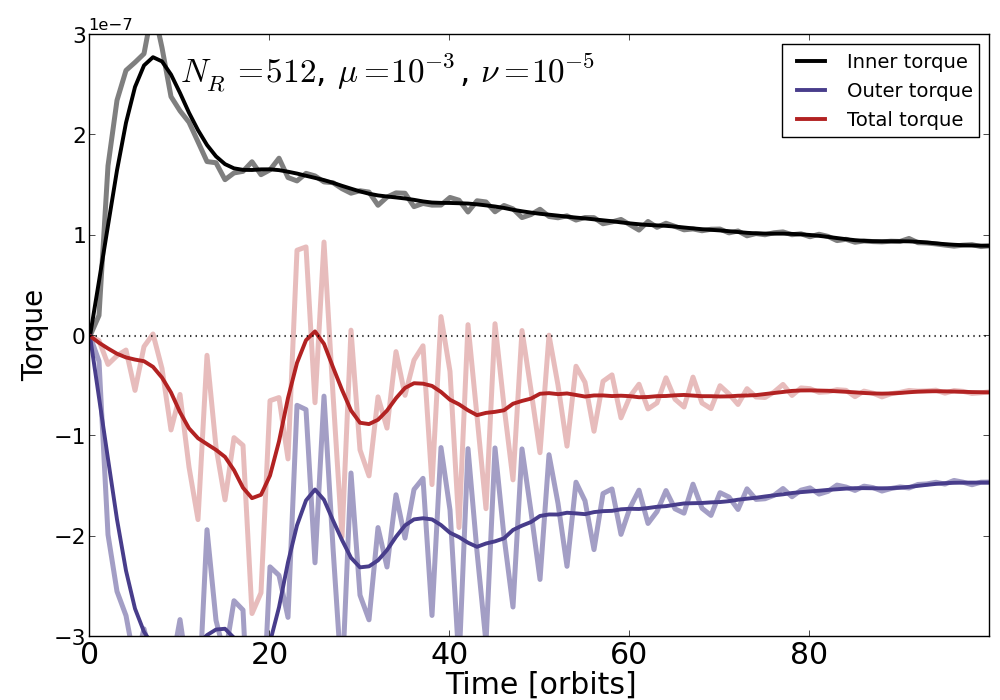}
\caption{Tidal torque evolution for a Jupiter-mass simulation with
  viscosity coefficient $\nu=10^{-5}$ at three different resolutions:
  $N_R=128$ (top panel), 256 (mid panel) and 512 (bottom panel),
  corresponding to runs \texttt{JUP-VISC-B\_128},
  \texttt{JUP-VISC-B\_256} and \texttt{JUP-VISC-B\_512}, respectively.
  The total torque at $t=100$ orbits is consistent across all
  resolutions, although the magnitude of the inner and outer torques
  of the highest resolution run differs from the rest. The transient
  stage of the torque calculation (up to $\sim50$ orbits) shows no
  hints of convergence.  Material within the Hill sphere of the planet
  has been excluded in the measurement.
\label{fig:jup-visc_torque}}
\end{figure}

\begin{figure}
\centering
\includegraphics[width=0.35\textwidth]{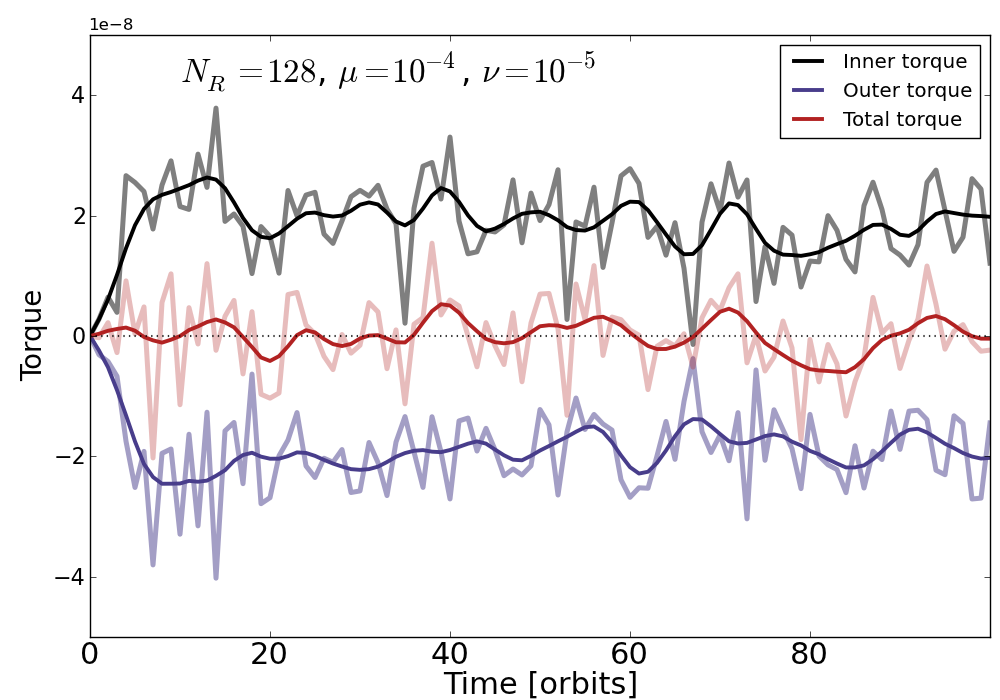}
\includegraphics[width=0.35\textwidth]{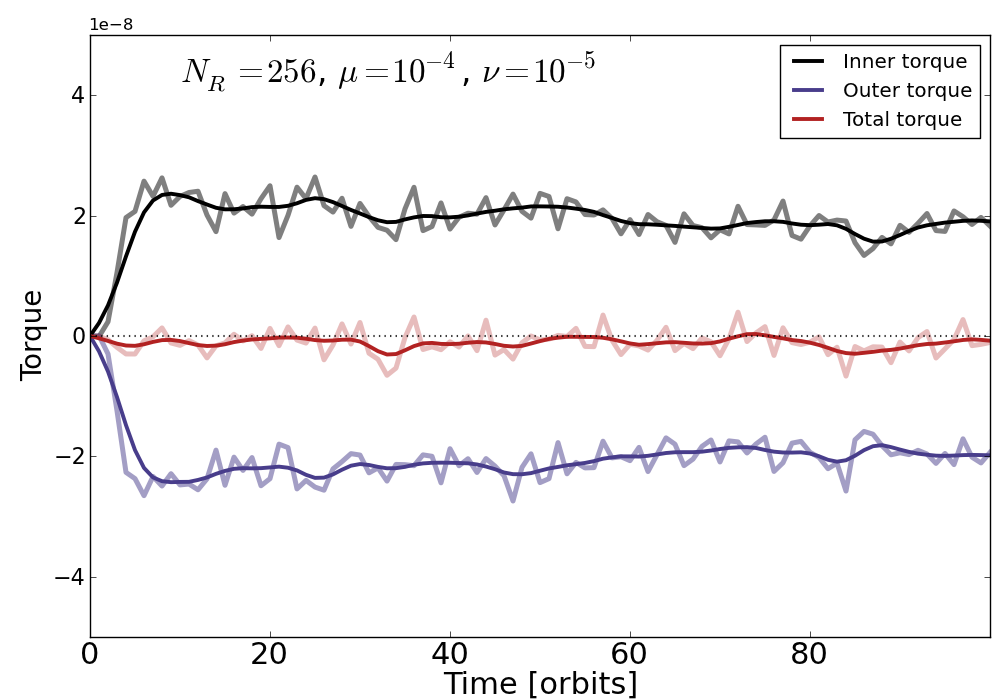}
\includegraphics[width=0.35\textwidth]{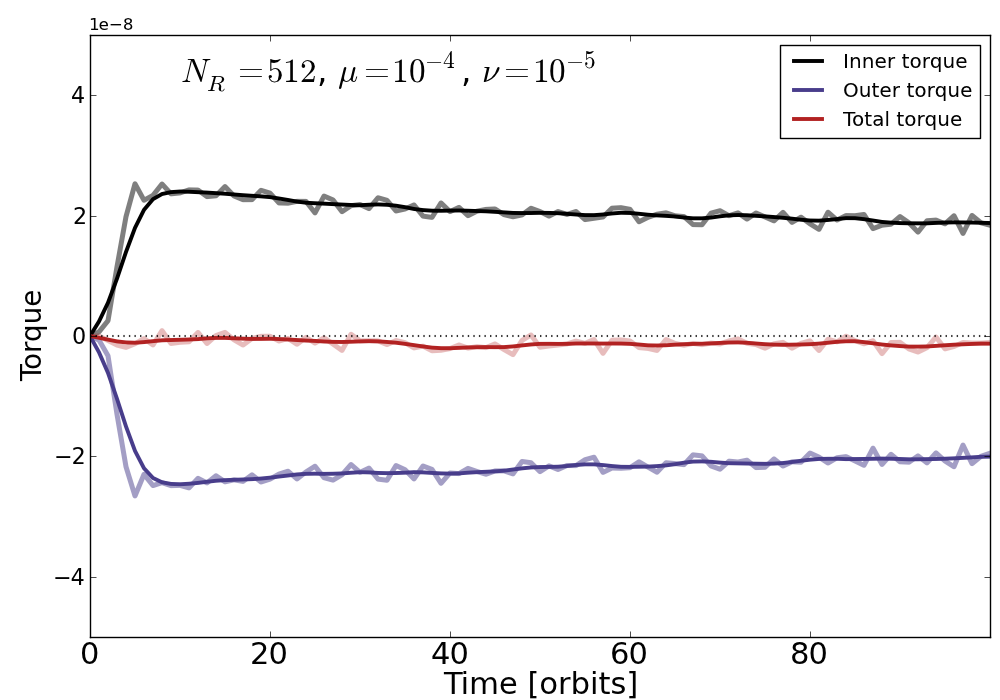}
\caption{Same as Figure~\ref{fig:jup-visc_torque} but for a
  Neptune-mass planet (runs \texttt{NEP-VISC-B\_128},
  \texttt{NEP-VISC-B\_256} and \texttt{JUP-VISC-B\_512}, from top to
  bottom). The net torque toward the end of the simulation is
  $\sim1.0\times10^{-9}$.  Convergence is achieved with resolution,
  albeit slowly for separate inner and outer torques. There is a
  trade-off in comparison to the Jupiter-mass case: since the density
  fluctuations due to the planet wake are weaker, the smaller the
  planet mass, the higher the resolution is needed to recover the
  theoretically expected negative net torque; on the upside, the
  smoother variation of this configuration (there is no gap opening)
  guarantees that a consistent solution can be reached with high
  enough resolution, as opposed to the transient-plagued torque
  evolution of Figure~\ref{fig:jup-visc_torque}.
\label{fig:nep-visc_torque}}
\end{figure}

\subsection{Total Torque Evolution}\label{sec:torques}
Finally, we consider the measurement of the tidal torques.
Ultimately, a successful planet-migration simulation will be
determined by the accurate calculation of the tidal torques, since
these determine the rate of change of the planet's angular
momentum. The tidal torque exerted by the disc onto the planet is
\begin{equation}\label{eq:torque_formula1}
\begin{split}
\mathbf{T}=& \mathbf{r}_p\times \int\limits_\mathrm{disc}\mathbf{f}\,  \Sigma(\mathrm{r})\,{\rm d}A\\
=&GM_p \,\mathbf{r}_p\times\left[\int\Sigma(\mathbf{r})\,g_1(|\mathbf{r}'|)\,\mathbf{r}'{\rm d}A \right],
\end{split}
\end{equation}
where $\mathbf{r}'=\mathbf{r}-\mathbf{r}_p$ and $g_1(y)$ is the
spline-softened gravitational force function \citep{spr01}, which
evaluates to the exact Keplerian value of $1/y^3$ for $y\geq
h=2.8\epsilon$, with $\epsilon$ being the traditional Plummer softening
parameter. We discretize Equation~(\ref{eq:torque_formula1}) as
\begin{equation}\label{eq:torque_formula2}
\mathbf{T}=GM_p \,\mathbf{r}_p\times\left[\sum_{i=0}^{N}M_i
 \, g_1(|\mathbf{r'}|)\,\mathbf{r'} \right],
\end{equation}
summing over all gas cells of masses $M_i$, and using
$\mathbf{r}'=\mathbf{r}_i-\mathbf{r}_p$, where $\mathbf{r_i}$ is the
cell centroid. The cells within the planet's Hill sphere, i.e.,
$|\mathbf{r}'|\lesssim(\mu/3)^{1/3}$, are excluded from the sum in
Equation~(\ref{eq:torque_formula2}).  For a globally isothermal disc
with constant background density profile, the linear theory net torque
(Lindblad plus corotation torques) in two dimensions is given by
\citep{tan02}
\begin{equation}\label{eq:torque_theory}
T\equiv\mathbf{T}\cdot\hat{\mathbf{z}}=-1.16\;T_0~~~\text{with}~~ T_0\equiv
\left(\frac{\mu}{1-\mu}\frac{1}{h_p}\right)^2\Sigma_0
R_p^4\Omega_p^2~~,
\end{equation}
which equals $-2.96\times10^{-7}$ for $\mu=10^{-3}$ and
$-2.95\times10^{-9}$ for $\mu=10^{-4}$, implying type-I migration
timescales $\tau\sim M_pR_p^2\Omega_p/T_0$ of $\sim6\times10^2P_p$ and
$\sim6\times10^3P_p$, respectively.

{We have computed the torque time series according to 
Equation~(\ref{eq:torque_formula2}) for all our simulations. To faciliate
the analysis of the torque over long timescales, we have created smoothed versions
of these time series by applying a Hamming filter with window length of
2 and 10 orbits.
 At the end of the simulations, the smoothed torque is
consistently slightly negative, consistent with an inward migration of the planet.}
The measured torque toward the end of the Jupiter-mass
run is $\sim-2\times10^{-8}$ in units where $P_p=2\pi$ and
$G=M_*+M_p=\Omega_p=1$, which is about one order of magnitude smaller
than the total torque value predicted by
Equation~(\ref{eq:torque_theory}). On the other hand, the torque value
for the Neptune-mass run is $\sim -4\times10^{-9}$, roughly
consistent with the
value from linear theory. The significant discrepancy between the measured torque and
linear theory in the $\mu=10^{-3}$ case (Equation~\ref{eq:torque_theory}
gives a torque value of $2.96\times10^{-7}$) is not
surprising due to the strong non-linearity of the simulation, but also
because the carved out gap is sufficiently deep to drastically change the
gravitational coupling to the disc \citep{pet12}, decreasing the
magnitude of the Lindblad torque. As discovered by \citet{pet12}, the total
Lindblad torque of a gapped disc with minimum gap density
$\Sigma_\mathrm{gap}\sim0.1\Sigma_0$ is closer to the Lindblad torque
of a density disc with background density $\Sigma_\mathrm{gap}$ rather
than $\Sigma_0$ in Equation~(\ref{eq:torque_theory}). This partially explains 
why the measured torque in our runs is an order of magnitude smaller 
than the value predicted by Equation~(\ref{eq:torque_theory}). 

Despite the consistently negative {\it smoothed} tidal torques measured 
toward the end of our simulations, {\it raw} torque time series may tell a different story.
For the Jupiter-mass simulations, the small scale asymmetric features observed in the density field
(Section~\ref{sec:density_field}) are large enough that they can alter
the torque onto the planet significantly, even to the point of
changing its sign on timescales comparable with the planet's orbital
period. These fluctuations, identified as vortices in our simulations,
are a phenomenon observed previously by \citet{kol03}, \citet{li05}
and \citet{val07}. In some cases, the planet-disc torque can be
dominated by planet-vortex scattering \citep{lin10} triggering a
runaway migration sometimes referred to as type III migration
\citep[see also][]{mas03,art94,pap05}. Since this type of migration
depends sensitively on the vortensity field within the coorbital
region, the exclusion of the Hill sphere from the torque calculation
might produce drastically different results compared to a fully
consistent, self-gravitating planet-disc interaction
\citep[e.g.][]{cri09a}. {Although this behavior is not observed
in the fiducial inviscid Neptune-mass run, edge instabilities
do appear as resolution is increased  (see density fields in the top panels of
Figure~\ref{fig:neptune_density_frames}), a result that we have confirmed
with {\footnotesize FARGO} runs. Thus, torque calculations as a mean
to assess numerical convergence may not produce conclusive results,
unless the variability can be suppressed by a physical mechanism
such as viscosity.}  

Another source of fluctuations in the measured torque is provided if
the Hill sphere material is included in the calculation of Equation~\ref{eq:torque_formula2}.
The fluctuating density in this region can produce significant noise in the torque measurement
for the Jupiter-mass simulations., On the other hand, the inclusion
of the Hill sphere has little effect on the measured torque
in the Neptune-mass simulations; in fact,
its inclusion results in smaller fluctuations on the net torque
calculation. 

In order to assess the reliability of the torque calculation, we
measure $\mathbf{T}$ at three different resolutions ($N_R=128$, 256
and 512; see Table~\ref{tab:simulations}). Since we do not expect the
inviscid runs to converge given the high sensitivity of this problem
to diffusion (whether it is of numerical or physical origin), we
explore the convergence of the torque evolution for the viscous runs
with $\nu=10^{-5}$. Figure~\ref{fig:jup-visc_torque} shows the torque
evolution for the viscous Jupiter-mass runs at the three different
resolutions listed in Table~\ref{tab:simulations}.  Although the
qualitative torque evolution is very consistent across the different
resolutions, the transient period ($t\lesssim50$~orbits) shows a
non-converging behavior. Note that the location and shape of this
variability, which does not vanish after applying a time-series
smoothing, is not entirely inconsistent across the different
panels. However, the amplitude of these transients grows with
resolution, indicating that a viscosity of $\nu=10^{-5}$ is not high
enough to suppress vorticity generation near the coorbital region that
can generate chaotic variations in the net torque. Note that the net
torque at $t=100$~orbits is very consistent across all resolutions,
showing a robust convergence of the planet's migration rate. In the
near-stationary regime ($t\gtrsim50$~orbits), the individual
components of the torque (the inner and outer contributions) are
independently consistent with each other for the $N_R=128$ and
$N_R=256$ runs. However, and despite a consistent net torque, the
individual components show a discrepancy at $N_R=512$. Still higher
resolution runs are hence required to settle whether these simulations
are converged in all relevant aspects.

Figure~\ref{fig:nep-visc_torque} shows better convergence
than Figure~\ref{fig:jup-visc_torque}.  The smaller perturbation
exerted on the disc by a Neptune-mass planet implies that the torque
magnitude can be more easily ``buried" by the noise than in the
Jupiter-mass case since now the outer and inner torques are
fractionally much closer in value than in the Jupiter-mass cases. The
separation between inner and outer torques might prove ambiguous in a
moving-mesh code, since cells centers can switch from the inner to the
outer region and vice-versa from one time step to the next.
Similarly, cells can enter and leave the
Hill sphere, introducing fluctuations in the inferred gravitational
torque if it is computed by treating cells as point masses.  This
explains in great part the fluctuations in the torque calculation.  As
opposed to the wildly varying density field in the Jupiter-mass
calculation, the quasi-stationarity of the Neptune case provides more
reassurance that convergence can be reached. Indeed, although a
resolution of $N_R=512$ is needed to beat down the fluctuations,
convergence can be observed in Figure~\ref{fig:nep-visc_torque},
showing that a slight asymmetry between the inner and outer torques
will cause an inward migration of the planet.

\subsection{FARGO and the Stockholm Code Comparison Project}\label{sec:stockholm}

\begin{figure}
\begin{center}
\includegraphics[width=0.48\textwidth]{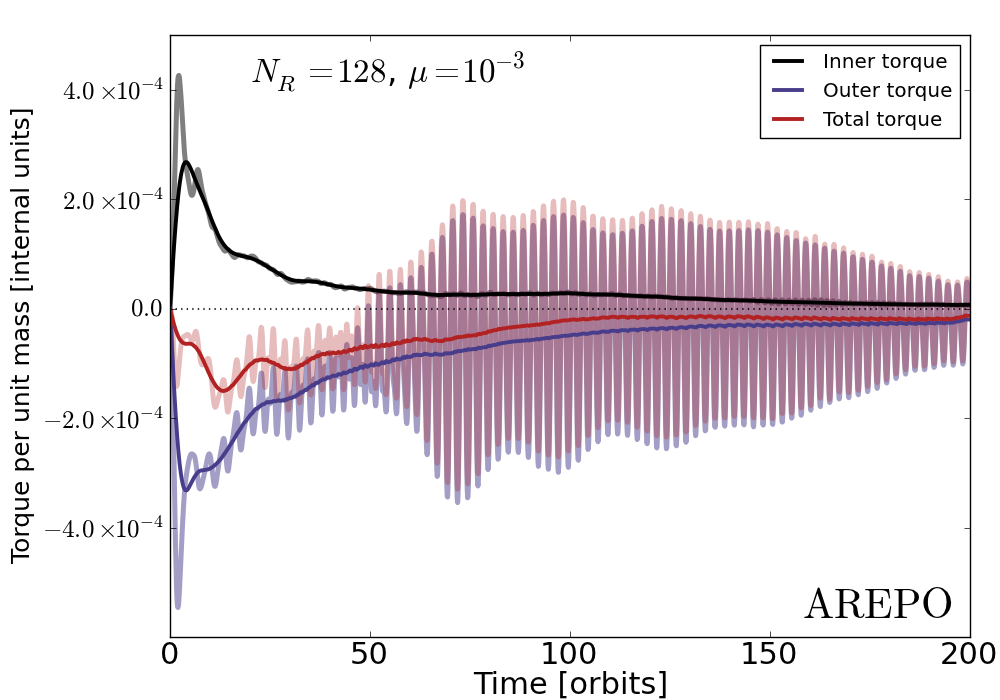}
\includegraphics[width=0.48\textwidth]{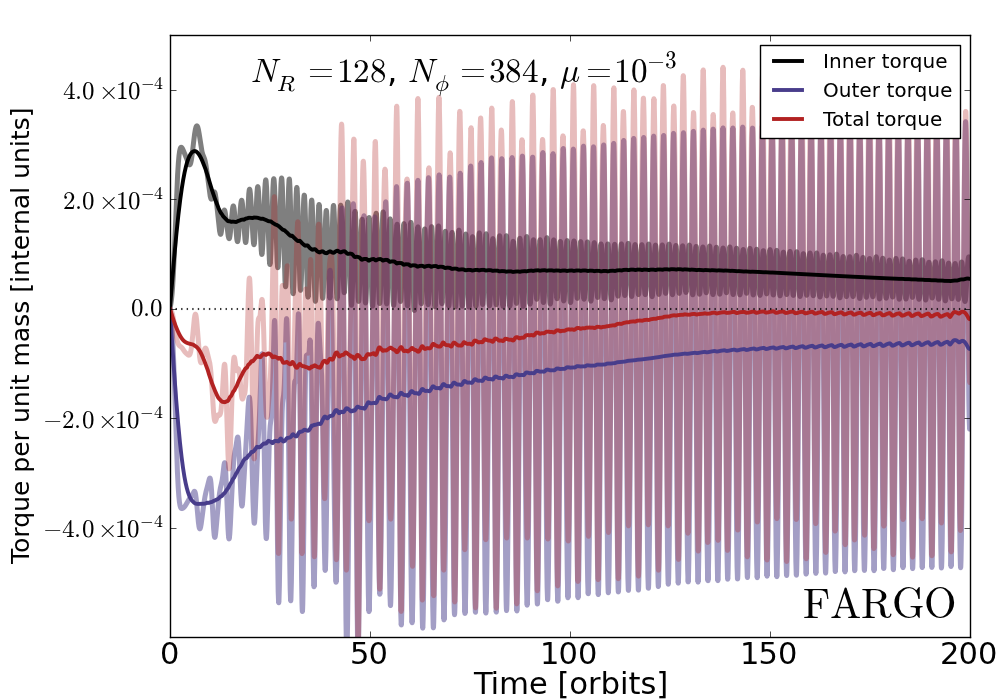}
\caption{Torque evolution for an inviscid disc for $\mu=10^{-4}$ and a temperature profile of $T\propto R^{-1}$
using  {\scriptsize AREPO} (upper panel) and  {\scriptsize FARGO} (lower panel).
The torque averaged over the interval between 175 and 200 orbits is $-1.37\times10^{-5}$ (internal units)
in the {\scriptsize AREPO} run and $-6.81\times10^{-6}$ (internal units) using  {\scriptsize FARGO} .
Although the total torque toward the end of the run is roughly consistent in both calculations
(within a factor of two; see Table~\ref{tab:torque_results}), the inner and outer torques differ by about
a factor of three (outer torque) to six (inner torque) between the two numerical schemes.  
\label{fig:arepo_fargo_torque}}
\end{center}
\end{figure}

{In the numerical experiments described above, we have 
used a globally isothermal equation of state to model the circumstellar disc.
This choice enables an easier comparison to some analytic calculations
\citep[e.g., ][]{tan02} and is simpler computationally, since there is no need
to solve an energy evolution equation. Evidently, this model is oversimplified,
since it is known that protoplanetary discs exhibt radial temperature
gradients \citep[e.g.][]{and09}.  There is a wealth of 
examples in the numerical astrophysics literature where discs are modeled 
with temperature gradient of the form $T\propto R^{-1}$. Although
this specific temperature profile does not necessarily reflect the
structure of real circumstellar discs \citep[see, e.g.][]{chi97,kra11}, it is a popular
approximation because it confers discs with a degree of self-similarity,
since then the aspect ratio is independent of radius. The Stockholm Code Comparison Project 
of \citet{val06} imposed this temperature profile for all of the experiments
in their simulation suite. We have added to our experiments a small set
of runs that include a temperature gradient to facilitate a comparison
with the results of \citet{val06}.}

{We have modified the 
hydrodynamic solver used in the previous sections to include a 
temperature gradient ($T\propto R^{-1}$) such that at each
cell interface, a local sound speed is used in the iterative isothermal
Riemann solver. The additional cost of computing this space-varying
sound speed is minimal, because the centroid of each face is already
calculated  since it is needed to carry out the spatial extrapolation
in the MUSCL-Hancock scheme \citep{spr10a}. Since {\footnotesize AREPO}
is a finite-volume scheme and, by default, it solves the non-radiative 
energy equation of the gas, we have attempted to run a simulation
with the adiabatic Riemann solver but using $\gamma=1+\epsilon$ 
($\epsilon\ll1$) as suggested
by \citet{val06} for finite-volume codes. However, we remain skeptical
about the validity of this approach, unless it is used for planet-disc
interaction simulations well within the linear regime or for very small
planet masses \citep[e.g.][]{duf12}. The reason why this shortcut is
not entirely applicable is that in gap-opening simulations there is significant
radial motion near the coorbital region. It can be shown that using $\gamma=1+\epsilon$
ensures isothermality of gas parcels in time in the Lagrangian sense, i.e.,
gas retains its initial temperature regardless of its spatial trajectory. Therefore,
unless gas motion is very close to azimuthal, using this approximation
can produce results that differ significantly from the strictly isothermal solutions.
Interestingly, low resolution {\footnotesize AREPO} runs using this prescription (not shown)
reproduce the density bumps inside and outside the gaps better than the those with a
strictly isothermal Riemann solver. However, since the resulting sound
speed -- defined as $c_s^2=(1+\epsilon)\,P/\Sigma$ -- is departs in time from
the initial  $c_s^2\propto R^{-1}$, we have obtained seemingly correct
results for the wrong reasons.}

{Part of analysis of the previous sections carried with globally isothermal
gas remains qualitatively unchanged and will not be repeated for the locally
isothermal runs (namely, the morphological analysis of surface density and vortensity).
However, a detailed discussion of the torque calculation for this set of additional runs
 is justified, as it was one of the main objectives of the Stockholm code comparison.}

{The torque per unit mass is calculated in the same manner as in 
Section~\ref{sec:torques}, this time setting $M_p=1$ in Equation~(\ref{eq:torque_formula2}).
 Figure~\ref{fig:arepo_fargo_torque}
shows a comparison of the raw (thick light-colored curves) and smoothed 
(thin dark-colored curves) between {\footnotesize AREPO} (upper panel)
and {\footnotesize FARGO} (lower panel) simulations of a locally isothermal disc
with a Jupiter-mass planet. Despite a rough agreement between these runs during the first
$\sim50$ orbits, after some time the instantaneous torque is stronger in the
 {\footnotesize FARGO}  case. The inner and outer torques are between a factor or two
and five in smaller in the {\footnotesize AREPO} run than in the {\footnotesize FARGO} equivalent.
This can be explained by the weaker density bumps formed in the {\footnotesize AREPO} case,
which get further depleted in time. The difference in amplitude of the high-frequency oscillations
in the torque is also due to these shallower density bumps. As discussed in 
Section~\ref{sec:viscosity_resolution}, a moderate increment in resolution can solve this 
discrepancy, but as we have seen in our globally isothermal simulations, the unstable
configuration of a planet-disc interaction with open gaps cannot provide a testbed for the
convergence of quantities such as the torque. Higher resolution runs do not exhibit a consistent
torque evolution in relation  to that shown in Figure~\ref{fig:arepo_fargo_torque}, sometimes even showing
a reversals in torque sign after 200 orbits. We have confirmed this type of outcome with  {\footnotesize FARGO},
experimenting with 128, 256, 512 and 1024 radial zones, finding that even after an apparent saturation
of the torque is reached at $\sim120$ orbits, edge instabilities can change this quantity at later times.}

{ However, viscous runs do show a more consistent behavior. Table~\ref{tab:torque_results}
lists the total planet-disc tidal torque averaged over 25 orbital periods between 175 and 200 orbits for viscous
runs ($\mu=10^{-5})$ in the Neptune-mass ($\mu=10^{-4}$) and the Jupiter-mass ($\mu=10^{-3}$) cases
for locally isothermal discs (constant aspect ratio $h=0.05$) using {\footnotesize AREPO} and {\footnotesize FARGO} 
at two different resolutions. In addition, we include the simple average of the Stockholm simulation
suite. For each configuration, the two {\footnotesize FARGO}  runs at different resolutions are consistent with each
other to within $2\%$ to $10\%$.  The {\footnotesize AREPO} torques, on the other hand, show a significant increase in
absolute magnitude when increasing the number of radial zones from 128 to 256, still lying at $\sim1.5$-$\sigma$ below
from the Stockholm average, while the {\footnotesize FARGO} torques have values of only $\sim0.6$-$\sigma$ below
the Stockholm averages. As suggested by Figures~\ref{fig:jup-visc_torque} and ~\ref{fig:nep-visc_torque}, a number
of radial zones closer $512$ would be necessary to capture a converged saturated torque with {\footnotesize AREPO}.}

\begin{table}
\caption{Tidal torque per unit planet mass averaged between 175 and 200 planet orbits
in locally isothermal runs (Hill sphere excluded).
\label{tab:torque_results}}
\begin{center}
\begin{tabular}{ l | c | c }
\hline
{Code} & $T/M_p$ & resolution\\
\noalign{\smallskip}
\hline
\noalign{\smallskip}
 \multicolumn{3}{c}{$\mu=10^{-3}$,~$\nu=10^{-5}$} \\
\hline
{\footnotesize AREPO} & $-2.68\times10^{-5}$ & 128 radial zones \\ 
{\footnotesize AREPO} & $-4.16\times10^{-5}$ & 256 radial zones \\ 
{\footnotesize FARGO} &  $-7.19\times10^{-5}$ & $N_R=128$, $N_\phi=384$  \\ 
{\footnotesize FARGO} &  $-7.31\times10^{-5}$ & $N_R=256$, $N_\phi=768$  \\ 
Stockholm suite &  $(-8.34\pm1.88)\times10^{-5}$& -- \\ 
\hline
\noalign{\smallskip}
 \multicolumn{3}{c}{$\mu=10^{-4}$,~$\nu=10^{-5}$} \\
\hline
{\footnotesize AREPO} & $-7.97\times10^{-6}$& 128 radial zones \\ 
{\footnotesize AREPO} & $-1.46\times10^{-5}$ & 256 radial zones \\ 
{\footnotesize FARGO} &  $-2.64\times10^{-5}$& $N_R=128$, $N_\phi=384$  \\ 
{\footnotesize FARGO} &  $-2.29\times10^{-5}$& $N_R=256$, $N_\phi=768$  \\ 
Stockholm suite &   $(-3.72\pm1.48)\times10^{-5}$& -- \\ 
\hline
\end{tabular}
\end{center}
$^{a}$ Simple average and standard deviation
of  the torque results obtained with the 11-12 different codes
included in the Stockholm Code Comparison Project \citep{val06}.\\
\end{table}

\section{Discussion}\label{sec:discussion}
Applying {\footnotesize AREPO} to an idealized version of the
planet-disc interaction problem provides a very important opportunity
to benchmark the moving-mesh method in a setup which we anticipate as
challenging for this type of code. Reassuringly, we have obtained
results that are very consistent with simulation results in the
literature. This invalidates recent suggestions that a code like
{\footnotesize AREPO} would have severe intrinsic limitations for
problems requiring a high-degree of symmetry, especially with a
supersonic shearing flow \citep{duf12}. However, we do agree that in
situations where the flow is very close to axisymmetric,
{\footnotesize AREPO} can at best be competitive with static grid
methods, since here its advantages in the form of reduced advection
errors are wiped out by the additional grid noise due to the moving
mesh. In this regime, a freely moving mesh scheme will naturally be
less efficient, given the additional computational expense of
reconstructing the tessellation at every time-step.

\begin{figure*}
\centering
\includegraphics[width=0.9\textwidth]{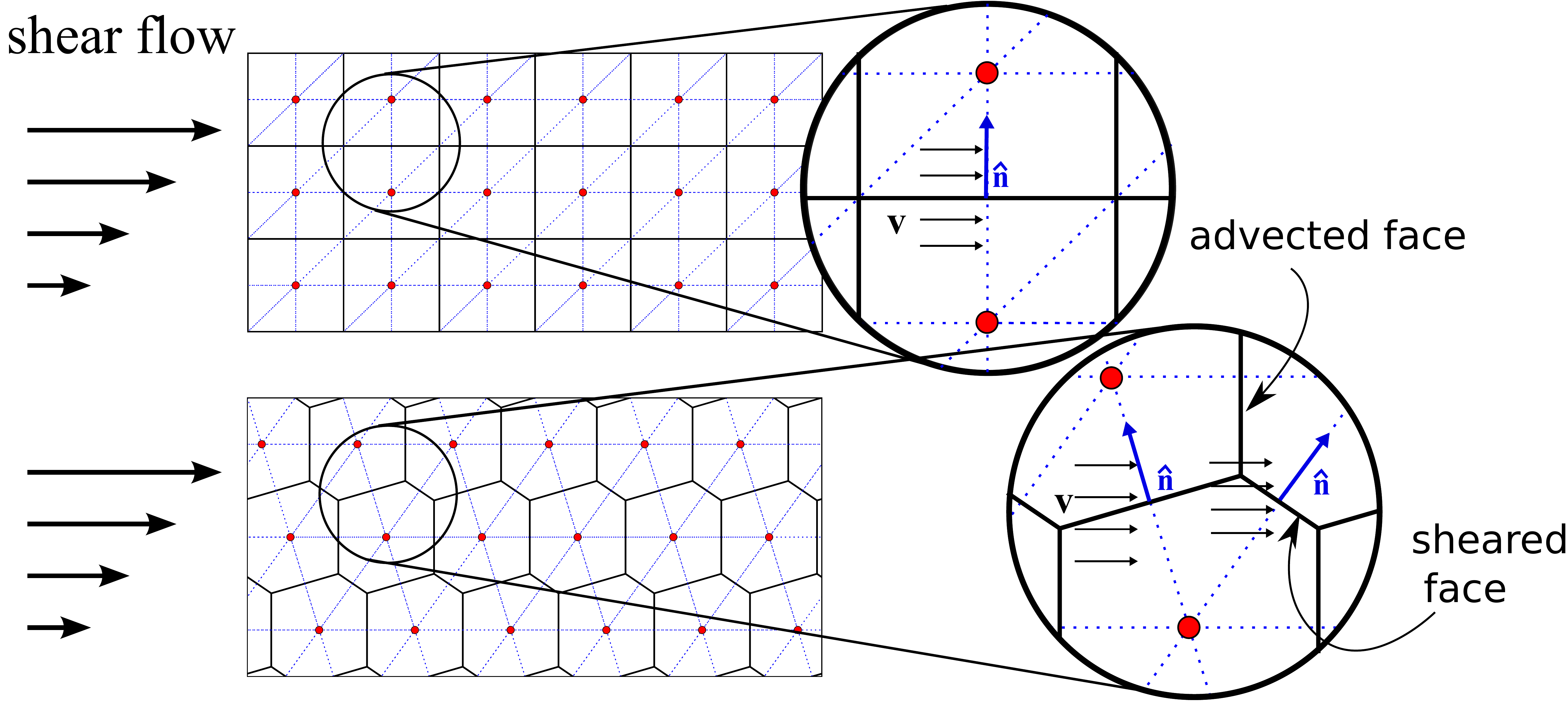}
\caption{Schematic representation of the shearing of an initially
  Cartesian mesh, where the red dots represent the mesh-generating
  points, the solid-black lines are the Voronoi tessellation, and the
  dotted-blue lines correspond to the Delaunay tessellation.  A
  velocity field of the form $\mathbf{v}(x,y)=v_x(y)\hat{\mathbf{x}}$
  will inevitably move the mesh-generating points away from their
  Cartesian-grid positions.  Since the Voronoi tessellation always
  puts the normal to the interface parallel to the line connecting the
  two cell centers, two cells with centers at different
  $y$-coordinates will exchange mass through this ``sheared face"
  since the flow will not be perpendicular to the face normal (lower
  zoomed-in region), as it would have been if the mesh had remained
  Cartesian (top zoomed-in region). On the other hand, the ``advected
  face" remains strictly perpendicular to the flow, since the
  mesh-generating points that share this face are moving in a straight
  line.  Note that in this particular example, all cells have the same
  area in both upper and lower panels. In such highly symmetric cases,
  two interfaces that communicate regions of different $y$-coordinate
  should produce mutually canceling mass fluxes. However, in general,
  cells will have a spurious mass exchange that can act as shear
  viscosity.
\label{fig:mesh_shear}}
\end{figure*}

We have observed that even higher resolution will be needed in future
work to assess the convergence of the torque calculation for
Neptune-mass planets, let alone for Earth-mass planets.  We attribute
the noise component of the net torque evolution observed in the
Neptune case to fluctuations in mass and position of the cells. Future
work should explore new ways to drift the mesh under a point-mass
potential. An interesting possibility is to drift the mesh-generating
points using an integrator of higher order than the leapfrog but with
similar conservative properties, such as the Wisdom-Holman mapping
used for planetary systems \citep{wis91}. In addition, new algorithms
to regularize the mesh should be explored. At this point, mesh
regularization has -- unsurprisingly -- degrees of freedom that do not
necessarily depend on the flow evolution. Moving toward parameter-free
regularization techniques would be highly desirable.

It is known that the so-called ``false diffusion" (or ``numerical
viscosity" or ``advection error") on the grid arises from four
different sources.  The most obvious one is the grid spacing. The
least obvious one is grid noise. The other two are (1) the fluid
velocity {\it with respect to the grid}, and (2) the angle between the
flow direction and the coordinate axis used to discretize the domain
\citep{dev72, pat80}; i.e. the alignment of the grid with the flow.
Although these last two features are interrelated, they are not quite
equivalent.  The first aspect of numerical diffusion lies at the heart
of moving-mesh codes like {\footnotesize AREPO} or advection-subtracted
codes like {\footnotesize FARGO}: eliminating the velocity of the
fluid with respect to the grid reduces the errors since these are
proportional to the velocity field. The second aspect is why
cylindrical grid codes are favored for accretion disc
simulations.{ In these schemes, the equations are first written
in their cylindrical components and then discretized,
and space is discretized by binning in a "cartesian fashion" the axes
of a orthorgonal coordinate system (in this case $R$ and $\phi$). 
 in an orthogonal. This subtlety in distingushing between the discretized 
 equations of motion and discretized coordinates has special
 relevance in a scheme like {\footnotesize AREPO}, in which the
 equations of motion are always in cartesian coordinates, but rotated
 constantly throughout the mesh to match the local geometry and
 orientation of the cell faces. However, by the properties of the tessellation
 geometry}, Voronoi-based schemes can produce {\it some} cell faces
that align along or perpendicular to the flow, but the very nature of
the tessellation forbids this to happen for every interface.

In a Voronoi topology, alignment of the mesh in one direction might
come at the expense of misalignment in another direction. As briefly
discussed by \citet{duf12}, a supersonic shearing flow might not be
ideal for a Voronoi mesh since contiguous cells at different speeds
will share a rapidly rotating interface that will not be parallel to
the direction of the flow at that point, thus eliminating the benefit
that meshes that are aligned with the flow have in reducing numerical
diffusivity. We illustrate this schematically in
Figure~\ref{fig:mesh_shear}.  Consider a velocity field
$\mathbf{v}(x,y)=v_x(y)\hat{\mathbf{x}}$.  Cells moving along the same
line (equal $y$) and at the same speed are simply advected (and so is
the interface between them), and thus exchange a negligible amount of
mass. Cells moving in the same direction but along different lines
(different $y$) and at different speeds will share a face that will be
rotating as a result of this velocity gradient. This will cause a
misalignment between the face and the background velocity field (see
Figure~\ref{fig:mesh_shear}, bottom panel) and thus the cells on both
sides of the interface will exchange mass. In contrast, if the mesh
topology remained to be composed of square cells at all times (see
Figure~\ref{fig:mesh_shear}, top), the projected mass flux onto the
``shearing face" would be zero. Therefore, the locally misaligned mesh
adds noise to the numerical solution. In well-resolved, stationary
flows, the total mass exchange of a given cell should cancel out (some
faces will have positive mass flux, and others negative
flux). However, in complex and time-dependent flows, truncation errors
can deteriorate the face-extrapolated quantities, producing additional numerical diffusion.

Note, however, that for most resolutions used in astrophysical disc
simulations, the change in velocity in the direction of the shear (the
$y$-direction in Figure~\ref{fig:mesh_shear}) will be much smaller
than the bulk velocity of the cells (along the $x$-direction in
Figure~\ref{fig:mesh_shear}), and thus the reduction of advection
error due to this motion is in general more critical than the
reduction of the advection error due to shear. Therefore, although the
rotating face due to shear adds numerical noise, the subtraction of
this cell's bulk motion should still provide a significant improvement
with respect to a fully static grid. Codes like {\footnotesize FARGO}
and {\footnotesize DISCO} aim to solve both issues: bulk motion and
mesh alignment, at the expense of having no resolution flexibility. In
contrast, a mesh comprised of a {\it static} tessellation would fail
to improve numerical diffusion on both fronts, thus being the worst
choice for disc simulations, for this given order of accuracy, along
with a static Cartesian grid.

The unavoidable misalignment of Voronoi faces with respect to the
azimuthal flow led \citet{duf12} to revert to a structured grid
approach in which, although cells are indeed moving, their motion is
restricted to be along the azimuthal direction, in a technique highly
reminiscent of the original {\footnotesize FARGO} scheme. However, as
we have stated above, the shear-induced noise is only secondary to the
bulk motion-induced error. In order to properly bypass this type of
error, the fluid equations must be solved in a frame moving with the
fluid, and this requires a boosted Riemann solver. {Still, an even more
important benefit of a conveniently structured grid is that is always
accompanied by a corresponding choice coordinates (cylindrical
coordinate equations plus cylindrical-coordinate grid.)}

{
There are now two codes in the literature, {\footnotesize AREPO} and  {\footnotesize TESS}, that use Voronoi tessellation.
There are notable difference between the two hydrodynamics schemes. {\footnotesize AREPO} takes 
into account the motion of
the mesh by adding the respective Godunov fluxes at each interface
assuming the mesh is {\it at rest} at any given time, allowing in
effect the subtraction of an advection flux that compensates for the
fact that the face is moving. A crucial detail in the {\footnotesize
  AREPO} scheme is how the Godunov fluxes are obtained in the first
place:
the Riemann problem is solved in a
{\it boosted} frame (i.e., the flow velocity to the left and right of
the interface is always small), and the resulting primitive variables
are boosted back into the lab frame. Only then are the Godunov fluxes
calculated and the quantities updated. On the other hand,
for a tessellation code like that solves the Riemann problem directly
in the lab frame an exact Riemann solver is a necessity, since
the velocity flow to the left and
right of the interface can be very large. For approximate, lab-frame solvers,
even if the posterior
subtraction of the advection flux reduces the mass exchange between
two contiguous cells, the truncation error has already done its damage
and the scheme will be diffusive \footnote{Furthermore, even exact lab-frame 
solvers can be diffusive since they are limited by severe
time-step conditions, thus requiring more operations than the boosted one.}.
As long as the
Riemann problem is properly boosted to the face rest frame,
the use of an approximate Riemann solver is of
secondary importance, as demonstrated by \citet{pak11}, 
}

\section{Summary}\label{sec:summary}
We have presented results for low-resolution simulations of planet
disc interaction using the moving-mesh code {\footnotesize AREPO} for
two different planet-to-star mass ratios exploring the dependence of
the result on resolution and viscosity.

\begin{enumerate}

\item
We have shown that {\footnotesize AREPO} can perform adequately on
problems with a high degree of symmetry like that of planet disc
interactions, even though this is not what a code like this is
expected to excel at.  Although concerns about the numerical noise
associated with faces that are misaligned with the flow is
well-founded, this does not affect the overall performance of the code
significantly.  We argue that the Riemann solver (exact or
approximate) should always be boosted to the frame of the face. From a
practical point of view, this allows for much longer time steps, since
lab frame Riemann solvers must use a CLF criterion based on the lab
frame speed of the flow.

\item 
Among the different sources of noise, we conjecture that grid noise is
the main concern in moving mesh simulations due to its sensitivity to
the development of instabilities.

\item 
We have found that proper convergence of the simulations is a function
of planet mass. This is not surprising since the perturbations exerted
on the disc are proportional to the planet mass (in the linear
regime), and thus it is easier to overcome numerical fluctuations with
larger planetary masses. However, the moving mesh introduces some
degree of uncertainty when separating the torque into different
components and when excluding the contribution from within the Hill
sphere. For low resolution, the Hill sphere is barely resolved,
meaning that cells entering and leaving this region can introduce a
degree of stochasticity that may not be present if cells were not
considered as point particles for torque calculation purposes.
Ultimately, fully consistent migration simulations (e.g. with
self-gravity; see \citealp{cri09a}) are necessary to elucidate the
importance of material within the Hill sphere for the orbital
evolution of the planet in our {\footnotesize AREPO} simulations.

\item 
The quasi-Lagrangian nature of a code like {\footnotesize AREPO} opens
new possibilities to the high-resolution study of planet-disc
interaction.  The possibility of a very flexible increment of
resolution around areas of interest presents an efficient alternative
to uniformly increasing the resolution globally, allowing one to avoid
the computational costs this entails.

\item 
Although the merit and success of {\footnotesize FARGO}
and {\footnotesize FARGO}-like codes is indisputable for this kind of
problem, we believe there is still room for moving-mesh codes as well,
especially for tackling the adaptive mesh refinement difficulties that
arise in cylindrical coordinates, but that are straightforward for
{\footnotesize AREPO}.  For example, in {\footnotesize AREPO} the
increased resolution within the Hill sphere can allow for
high-resolution studies of the circumplanetary environments embedded
in global disc simulations. Furthermore, the increased resolution
along the planetary wakes (see Figure~\ref{fig:planet_wmesh}) enables
the capture of sharper features. Along these lines, future work will
assess the minimum resolution requirement to reproduce the negative
torque density observed at very high (but uniform) resolution by
\citet{don11a} and \citet{duf12}.
\end{enumerate}

\section*{acknowledgements}
{The research presented here was carried out as part of DJM's PhD
  thesis at Harvard University.  The simulations in this paper were
  run on the Odyssey cluster supported by the FAS Science Division
  Research Computing Group at Harvard University. We are thankful to
  Ramesh Narayan, Crist\'oval Petrovich, Mark Vogelsberger,
  Chris Hayward and Dylan Nelson for helpful discussions. 
  DJM would like to thank Dimitar
  Sasselov, Matthew Holman, Ruth Murray-Clay and James Stone for
  insightful feedback and support throughout the development of this
  work. VS acknowledges support by the European Research Council under
  ERC-StG grant EXAGAL-308037.}




 \end{document}